\documentclass[aps,prd,amsmath,amssymb,amsfonts,showpacs,superscriptaddress,floatfix,twocolumn,nofootinbib]{revtex4-1}
\usepackage{amsmath}
\usepackage{amssymb}
\usepackage{graphicx}
\usepackage{bbm}
\usepackage{caption}
\usepackage{subcaption}
\usepackage{epsfig}
\usepackage{braket}
\usepackage{slashed}
\usepackage{bm}
\usepackage[usenames,dvipsnames]{color}
\usepackage{color}
\usepackage{float}
\usepackage{hyperref}

\newcommand{\ta}{{\mathrm{a}}}
\newcommand{\tb}{{\mathrm{b}}}

\newcommand{\umps}[1]{\Psi_u[{#1}]}
\newcommand{\tanv}[3]{\Phi_{#1}[#2,#3]}

\def\be{\begin{equation}}
\def\ee{\end{equation}}

\def\bea{\begin{eqnarray}}
\def\eea{\end{eqnarray}}
\begin{document}

\title{Real-time simulation of the Schwinger effect with Matrix Product States} 

\author{Boye Buyens}
  \affiliation{Department of Physics and Astronomy,
Ghent
 University,
  Krijgslaan 281, S9, 9000 Gent, Belgium}

\author{Jutho Haegeman}
  \affiliation{Department of Physics and Astronomy,
Ghent
 University,
  Krijgslaan 281, S9, 9000 Gent, Belgium}
  
\author{Florian Hebenstreit}
  \affiliation{Albert Einstein Center for Fundamental Physics, Institute for Theoretical Physics, University of Bern, Sidlerstrasse 5, 3012 Bern, Switzerland}
 
\author{Frank Verstraete}
  \affiliation{Department of Physics and Astronomy,
Ghent
 University,
  Krijgslaan 281, S9, 9000 Gent, Belgium}
  \affiliation{Vienna Center for Quantum Science and Technology, Faculty of Physics, University of Vienna, Boltzmanngasse 5, 1090 Vienna, Austria}

 \author{Karel Van Acoleyen}
\affiliation{Department of Physics and Astronomy,
Ghent
 University,
  Krijgslaan 281, S9, 9000 Gent, Belgium}

 \begin{abstract}
\noindent  
\noindent Matrix Product States (MPS) are used for the simulation of the real-time  dynamics induced by an electric quench on the vacuum state of the massive Schwinger model. 
For small quenches it is found that the obtained oscillatory behavior of local observables can be explained from the single-particle excitations of the quenched Hamiltonian. 
For large quenches damped oscillations are found and comparison of the late time behavior with the appropriate Gibbs states seems to give some evidence for the onset of thermalization. 
Finally, the MPS real-time simulations are compared with results from real-time lattice gauge theory which are expected to agree in the limit of large quenches.
\end{abstract}

\maketitle
\section{Introduction.}  
Gauge theories lie at the heart of high energy physics and hence play an essential role in our understanding of nature.
Morever, gauge theories also emerge as low energy effective theories in several condensed matter systems \cite{Benton2016}. 
Lattice gauge theories provide a non-perturbative regularization of such theories that can often be simulated very efficiently by using Quantum Monte Carlo (QMC) methods.
However, several most pressing questions in that regard, e.g., the phase diagram of quantum chromodynamics (QCD) at finite chemical potential or the real-time dynamics of relativistic heavy ion collisions have largely remained out of reach \cite{Bali1999}.

Over the last decade, the Tensor Network States (TNS) approach has become a powerful alternative method to study strongly correlated quantum systems since it does not suffer from the sign problem \cite{Orus2004,Verstraete2004,Verstraete2008}. 
The most famous example of TNS are the Matrix Product States (MPS) \cite{Schollwoeck2011} in one spatial dimension. Ever since the formulation of Density Matrix Renormalization Group \cite{White1992} in terms of MPS, the number of algorithms for quantum many-body systems has increased rapidly. Recently, MPS have also been successfully applied to lattice gauge theories \cite{Banuls2013a,Banuls2016a,Banuls2016b,Banuls2016c,Banuls2016d,Byrnes2003a,Byrnes2003b,Sugihara2005,Rico2014,Kuehn2014,Kuehn2015,Silvi2016,Milsted2015}. 

In this publication we consider $(1+1)$-dimensional quantum electrodynamics (QED), the so-called massive Schwinger model \cite{Schwinger1962}.
Despite being an Abelian gauge theory, it shares several important features with the theory of strong interactions (QCD) such as chiral symmetry breaking or confinement.
Due to the reduced dimensionality this model has become an active playground for testing novel analytical and numerical methods \cite{Schwinger1962,Coleman1975,Coleman1976,Hamer1982,Iso1990,Hosotani1996,Adam1997,Byrnes2003a,Byrnes2003b,Cichy2012,Banuls2013a,Hebenstreit2013,Hebenstreit2014,Kuehn2014,Banuls2016a,Buyens2013,Buyens2014,Buyens2015,Buyens2015b,Buyens2016,Buyens2017}
and for studying intriguing non-equilibrium questions that have been beyond the reach of convential QCD simulations, e.g., jet energy loss and photon production in relativistic heavy ion collisions \cite{Kharzeev2013,Kharzeev2014} or the dynamics of string breaking \cite{Hebenstreit2013a}.
Recently, there have been promising proposals that might allow to quantum simulate the Schwinger model in analog systems of ultracold ions or atoms in optical lattices \cite{Hauke2013,Wiese2013,Martinez2016,Kasper2015,Kasper2016}.

An intriguing effect in the Schwinger model concerns the non-equilibrium dynamics after a quench that is induced by the application of a uniform electric field $E_0 = g\alpha$ onto the ground state $\ket{\Psi_0}$ at time $t=0$.
Physically, this process corresponds to the so-called Schwinger pair creation mechanism \cite{Schwinger1951} in which an external electric field separates virtual electron-positron dipoles to become real electrons and positrons. 
Recently, this process has attracted much interest since high-intensity laser facilities like the Extreme Light Infrastructure (ELI) will for the first time be powerful enough to probe this effect experimentally.
So far, theoretical investigations have mainly been restricted to the regime in which the fermions are treated quantum mechanically whereas the gauge fields are described classically (quantum kinetic theory \cite{Schmidt1998, Kluger1992} or phase-space methods \cite{Hebenstreit2011}), or classical-statistically (real-time lattice techniques \cite{Hebenstreit2013,Kasper2014}). 

In this publication we apply the MPS framework to investigate the non-equilibrium dynamics at the full quantum level.
We perform real-time simulations for small, intermediate and large quenches.
Furthermore, we use MPS computations of ground states, single-particle excitations and Gibbs states to analyse and interpret our results. 
Finally, we explicitly compare the MPS simulations with those obtained using real-time lattice techniques. 

\section{Setup} 
\subsection{Kogut-Susskind Hamiltonian}
The massive Schwinger model describes $(1+1)$-dimensional QED with one fermion flavor that is described by the Lagrangian density
\be \mathcal{L} = \bar{\psi}\left(\gamma^\mu(i\partial_\mu+g A_\mu) - m\right) \psi - \frac{1}{4}F_{\mu\nu}F^{\mu\nu}\,.\label{Lagrangian} \ee
Here, $\psi$ is a two-component fermion field, $A_\mu$ denotes the $U(1)$ gauge field and $F_{\mu\nu}=\partial_\mu A_\nu-\partial_\nu A_\mu$ is the corresponding field strength tensor.

In the following, we employ a lattice regularization \`{a} la Kogut-Susskind \cite{Kogut1975}.
Therefore the two-component fermions are decomposed into their particle and antiparticle components which reside on a staggered lattice.
These staggered fermions are converted to quantum spins $1/2$ by a Jordan-Wigner transformation with the local Hilbert space basis $\{\ket{s_n}_n: s_n \in \{-1,1\} \}$ of $\sigma_z(n)$ at site $n$.
The charge $-g$ `electrons' reside on the odd lattice sites, where spin down ($s=-1$) denotes an occupied site whereas spin up ($s=+1$) corresponds to an unoccupied site.
Conversely, the even sites are related to charge $+g$ `positrons' for which spin down/up corresponds to an unoccupied/occupied sites, respectively.

Moreover, we introduce the compact gauge field $\theta(n) = a g A_1(n)$, which lives on the link that connects neighboring lattice sites, and its conjugate momentum $E(n)$, which correspond to the electric field.
The commutation relation $[\theta(n),E(n')]=ig\delta_{n,n'}$ determines the spectrum of $E(n)$ up to a constant: $E(n)/g = L(n) + \alpha$.
Here, $L(n)$ denotes the angular operator with integer spectrum and $\alpha \in \mathbb{R}$ corresponds to the background electric field. 
Accordingly, the Kogut-Susskind Hamiltonian reads \cite{Kogut1975,Banks1976}
\bea\label{eq:Hamiltonian} \mathcal{H}_{\alpha}&=& \frac{g}{2\sqrt{x}}\Biggl(\sum_{n =1}^{2N} \left[L(n) + \alpha\right]^2 + \frac{\sqrt{x}}{g} m \sum_{n =1}^{2N}(-1)^n\sigma_z(n) \nonumber
\\ &+& x \sum_{n=1}^{2N-1}(\sigma^+ (n)e^{i\theta(n)}\sigma^-(n + 1) + h.c.)\biggl)\, , \eea
where $\sigma^{\pm} = (1/2)(\sigma_x \pm i \sigma_y)$ are the ladder operators. 
Here, we have introduced the parameter $x$ as the inverse lattice spacing in units of $g$: $x \equiv 1/(g^2a^2)$. 
The continuum limit then corresponds to $x\rightarrow \infty$.  
We note that $\mathcal{H}_\alpha$ is only invariant under $\mathcal{T}^2$ (translations over two sites) due to the staggered mass term in the Hamiltonian.
The Hamiltonian is invariant under local gauge transformations that are generated by:
\begin{align} G(n) =  L(n)-L(n-1)-\frac{\sigma_z(n) + (-1)^n}{2}\,.\label{eq:Gauss} \end{align}
If we restrict ourselves to physical (i.e., gauge invariant) operators $O$ for which $[O,G(n)]=0$, the Hilbert space decomposes into dynamically disconnected superselection sectors, which are distinguished by the eigenvalues of $G(n)$.
The sector with $G(n)=0$ at every site $n$ constitutes the physical sector of the Hilbert space.
The condition $G(n)=0$ is referred to as the Gauss law constraint as it is the discretized version of $\partial_z E = j^0$, where $j^0$ is the charge density of dynamical fermions. 

\subsection{MPS for real-time evolution.} Similar as in \cite{Buyens2013,Buyens2017} we block site $n$ and link $n$ into one effective site with local Hilbert space spanned by $\{\ket{\kappa_n} = \ket{s_n,p_n}_n: s_n = -1,1; p_n \in \mathbb{Z} \}$.
In our approach we approximate the states of the lattice system eq.~\eqref{eq:Hamiltonian} by Matrix Product States (MPS) $\ket{\umps{A(1)A(2)}}$ that take the form
\begin{subequations}\label{eq:MPS}
\begin{multline} \sum_{\bm{\kappa}}v_L^\dagger \left(\prod_{n = 1}^{N} A_{\kappa_{2n-1}}(1) A_{\kappa_{2n}}(2)\right)v_R \ket{\kappa_1,\ldots, \kappa_{2N}}.\end{multline}
Here we have $A_\kappa(n) \in \mathbb{C}^{D \times D}$ and $v_L, v_R \in \mathbb{C}^{D \times 1}$. 
The MPS ansatz associates a matrix $A_{\kappa_n}(n)= A_{s_n,p_n}(n)$ with each site $n$ and every local basis state $\ket{\kappa_n}_n =\ket{s_n,p_n}_n$. The indices $\alpha$ and $\beta$ are referred to as virtual indices, and $D$ is called the bond dimension. Note that this ansatz is $\mathcal{T}^2$ invariant. As such we can consider the ansatz directly in the thermodynamic limit ($N \rightarrow + \infty$), bypassing any possible finite size artifacts. In this limit the expectation values of all local observables are independent of the boundary vectors $v_L$ and $v_R$.

The Gauss law constraint \mbox{$G(n)=0$} imposes the following form on the matrices \cite{Buyens2013}
\be
{[A_{s,p}(n)]}_{(q,\alpha_q),(r,\beta_r)} =  {[a_{q,s}(n)]}_{\alpha_q,\beta_r}\delta_{q+(s+(-1)^n)/2,r}\delta_{r,p}
\label{eq:gaugeMPS}\,,\ee
\end{subequations}
where $\alpha_q = 1\ldots D_q$, $\beta_r = 1 \ldots D_r$. 
The variational freedom of the gauge invariant state $\ket{\umps{A(1)A(2)}}$ thus lies within the matrices $a_{q,s}(n) \in \mathbb{C}^{D_q \times D_r}$ and the total bond dimension of the MPS equals $D = \sum_{q \in \mathbb{Z}}D_q$. 

In our simulations, we start from the ground state of the Hamiltonian $\mathcal{H}_{\alpha=0}$ without background field, for which we found a faithful gauge invariant MPS approximation $\ket{\umps{A(1)A(2)}}$ by using the time-dependent variational principle (TDVP) \cite{Buyens2013,Haegeman2011,Haegeman2013,Haegeman2014a,Buyens2017} (see Appendix Sec.~\ref{subsec:TDVPGS}) for a brief review).
At time $t = 0$, we perform a quench and apply a uniform electric field, which is simulated by evolving the ground state with respect to the Hamiltonian $\mathcal{H}_{\alpha\neq0}$ with non-vanishing background field: $\ket{\Psi(t)} = e^{-i\mathcal{H}_{\alpha\neq0} t} \ket{\umps{A(1)A(2)}}$. 
The evolution is performed using the infinite time-evolving block decimation (iTEBD) \cite{Vidal2007} which adapts the bond dimension of this MPS dynamically according to the Schmidt spectrum, see Appendix  Sec.~\ref{subsec:iTEBDRT}. As explained there, the errors introduced by this method are well-controlled and argued to be only of order $10^{-3}$ or smaller. We refer also to \cite{Buyens2013,Buyens2017} for a discussion on the systematics of the iTEBD. 

\subsection{Observables and their discretization}
We focus on the real-time evolution of the following observables
\begin{subequations}\label{eq:EandN}
\be E(t) = \frac{g}{2N}\sum_{n = 1}^{2N}\Braket{  L(n) + \alpha }_t,\ee
\be j^1(t) =\frac{-i\sqrt{x}g}{4N}\sum_{n = 1}^{2N-1}\Braket{\sigma^+(n)e^{i\theta(n)}\sigma^{-}(n+1) - h.c. }_t,\ee
\be \Sigma(t) =  \frac{\sqrt{x}}{2N}\sum_{n=1}^{2N}\Braket{ \frac{\sigma_z(n) + (-1)^n}{2}}_t, \ee
\end{subequations}
with $\braket{\ldots}_t = \braket{\Psi(t)\vert \ldots \vert \Psi(t)}$. Here, $E(t)$ is the expectation value of the total electric field, $j^1 = \Braket{\bar{\psi}\gamma^1\psi}_t$ the current which can also be obtained from the electric field via Amp\`{e}re's law ($\dot{E} = -gj^1$) and $\Sigma(t)$ is the discrete version of the chiral condensate $\Braket{\bar\psi\psi}_t$.
We will use $N(t)=\Sigma(t)-\Sigma(0)$ as a measure for the fermion particle number but notice that only in the non-relativistic limit $m\gg g$ we have a clear notion of electron and positron number.
We will also show the real-time evolution of the half-chain von Neumann entropy, $S(t)=-Tr \rho \log \rho$, where $\rho$ is the density matrix of the half-chain subsystem. Finally, as explained in Appendix Sec.~\ref{appsec:continuum}, notice that $E(t)$, $j^1(t)$, $N(t)$ and $\Delta S(t) = S(t) - S(0)$ are UV finite and already close to the continuum limit $x\rightarrow \infty$ for $x = 100$.
\\
\begin{figure}[t]
\begin{subfigure}[b]{.24\textwidth}
\includegraphics[width=\textwidth]{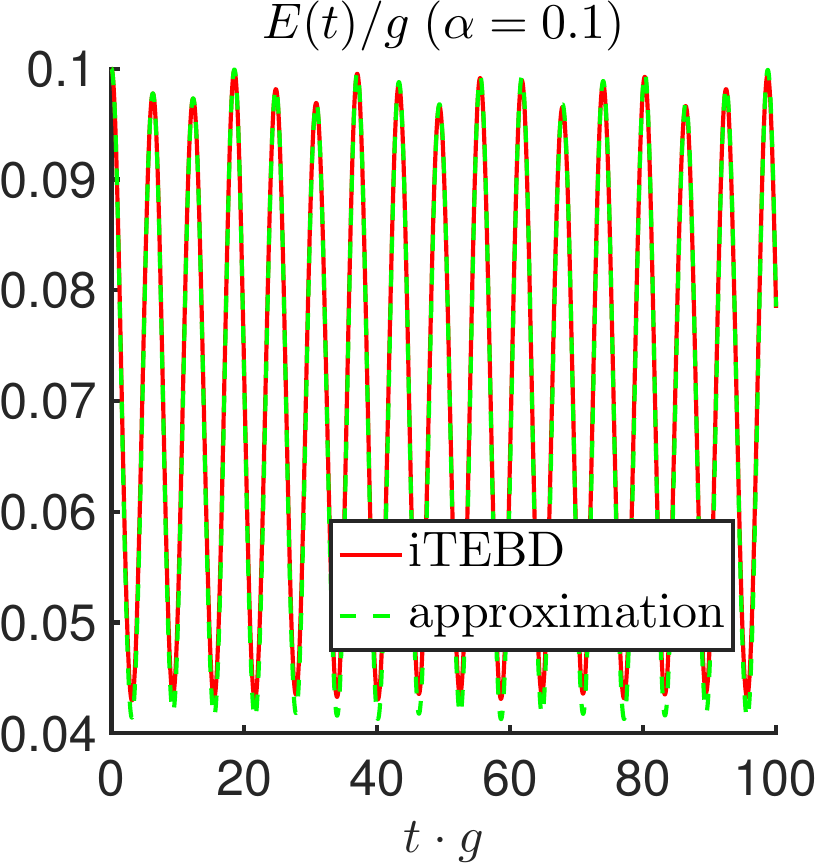}
\caption{\label{fig:CoherentStateAppa}}
\end{subfigure}\hfill
\begin{subfigure}[b]{.24\textwidth}
\includegraphics[width=\textwidth]{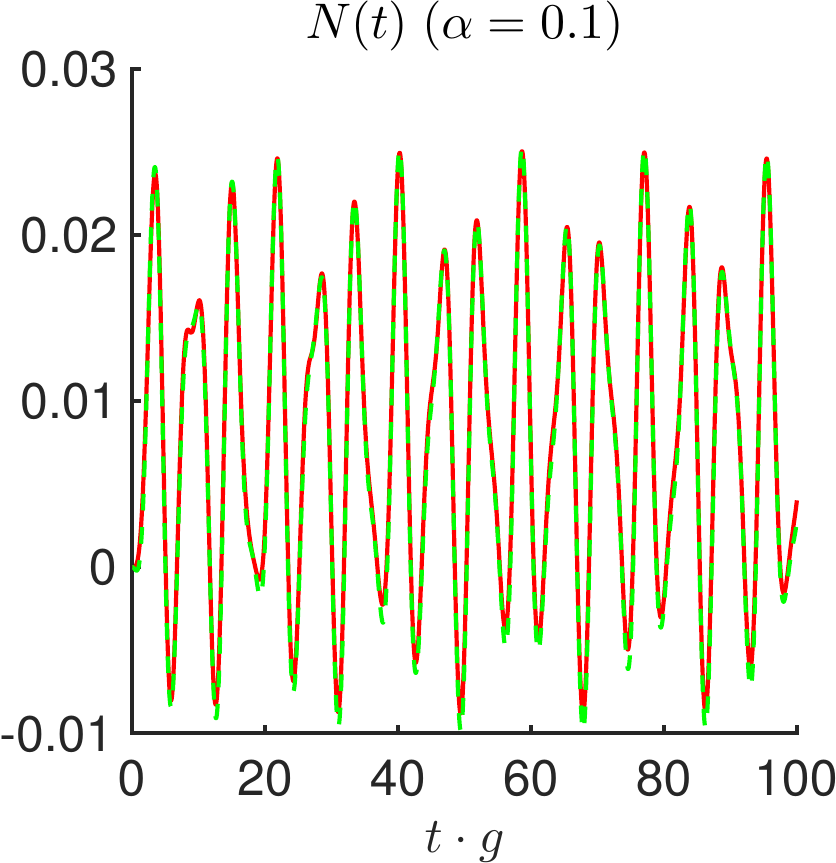}
\caption{\label{fig:CoherentStateAppb}}
\end{subfigure}\vskip\baselineskip
\captionsetup{justification=raggedright}
\caption{\label{fig:CoherentStateApp} $m/g= 0.25, x = 100$. Comparison of iTEBD simulations (full line) with the approximation Eq.~(\ref{eq:coherenstateAppb}) (dashed line). (a): $E(t)/g$ $(\alpha = 0.1)$. (b): $N(t)$ $(\alpha = 0.1)$.}
\end{figure}

\section{Results}
\subsection{Weak-field regime} \label{subsec:weakfieldRegime}
In \cite{Buyens2013} we found that the quasi-period of the oscillations of the electric field in the linear response regime ($\alpha \leq 0.01$) could be traced back to the first single-particle excitation of $\mathcal{H}_0$. However, for $\alpha \gtrsim 0.1$ we observed that the quasi-period grows with $\alpha$, and, hence, cannot be explained by the mass of the same single-particle excitation for each $\alpha$. It turns out that for $\alpha\lesssim 0.25$ the original vacuum $\ket{\Psi_0}$ is well described as a small density coherent state of single-particle excitations of the quenched Hamiltonian $\mathcal{H}_\alpha$. This leads to the oscillatory behavior of Fig.~\ref{fig:CoherentStateApp}. Specifically, as we discuss below, this behavior can be explained quantitatively in terms of the matrix-elements of $\mathcal{H}_0$, and the considered observables $E$ and $N$ in the truncated Hilbert-space consisting of the ground state and the two single-particle excitations of the quenched Hamiltonian $\mathcal{H}_\alpha$. 

As explained in more detail in Appendix Sec.~\ref{appsec:cohstateapp}, for a given $\alpha$, we approximate all observables $\mathcal{O}$ in terms of a series of the creation $\ta_m^\dagger(k)$ and annihilation operators $\ta_m(k)$ of the single-particle excitations $\ket{\mathcal{E}_m(k)}$ of $\mathcal{H}_\alpha$ with energy $\mathcal{E}_m$ and momentum $k$\footnote[4]{Notice that in this approximation we drop the multi-particle scattering states and can therefore consider the creation/annihilation operators  $\ta_m^\dagger(k)$, $\ta_m(k)$ as corresponding to the asymptotic in- or out-states.}; and this up to first order in $\ta_m$ and $\ta_m^\dagger$ \cite{Delfino2016,Delfino2016b}:
\begin{multline}\label{eq:coherenstateAppb} \mathcal{O} \approx \lambda_{\mathcal{O}} \mathbbm{1} + \int dk \int dk'\left(\sum_{m,n}o_{1,m,n}(k,k') \ta_m^\dagger(k)\ta_n(k') \right) \\ + \int dk\;\left( \sum_m o_{2,m}(k) \ta_m(k) + \bar{o}_{2,m}(k)\ta_m^\dagger(k)\right) . \end{multline}
  
\begin{table}[t]
\begin{tabular}{| c||   c | c | }
\hline 
$\alpha$ & $\rho_1\xi$ & $\rho_2\xi$\\
\hline
$0.01$ & $2.6 \times 10^{-5}$ & $ 9.4 \times 10^{-9}$ \\
$0.1$ & $2.7 \times 10^{-3}$ & $8.9\times 10^{-5}$\\
$0.2$ & $1.2 \times 10^{-2}$   & $1.4 \times 10^{-3}$  \\
$0.3$ & $3.5 \times 10^{-2}$ & $8.0 \times 10^{-3}$ \\ 
$0.4$ & $8.9 \times 10^{-2}$ & $2.9 \times 10^{-2}$ \\ 
\hline
\end{tabular}
\captionsetup{justification=raggedright}
\caption{\label{table:valuesdprimem} $m/g = 0.25, x = 100$. Particle densities in units of correlation length for the two single-particle excitations of $\mathcal{H}_\alpha$.}
\end{table}

Here $m=1,2$ labels the two single-particle excitations of $\mathcal{H}_\alpha$ and the integral runs over the momenta $k \in [-\pi,\pi]$. Using the MPS approximations for the ground state and the two single-particle excitations obtained in \cite{Buyens2015,Buyens2015b}, we can extract the coefficients $o_{1,m,n}$ and $o_{2,m}$. For $\mathcal{O}=\mathcal{H}_{0}$ this leads to the approximation of the $\alpha=0$ ground state $\ket{\Psi_0}$, as a coherent state of $\mathcal{H}_{\alpha}$: $\ta_m(k) \ket{\Psi_0} = d'_m \delta(k)\ket{\Psi_0}$ with $d'_m \in \mathbb{C}$. This corresponds to a state with particle densities $\rho_m=\frac{\sqrt{x}}{2\pi}|d'_m|^2$ of the two zero-momentum single-particle excitations on top of the ground state of $\mathcal{H}_\alpha$. In table \ref{table:valuesdprimem} we display the obtained densities for different $\alpha$ in units of the correlation length $\xi=1/\mathcal{E}_1(0)$. One would expect our single-particle approximation to hold as long as $\xi\rho_1,\xi\rho_2\ll 1$ which is in the line with our results. The approximation on the evolution for $E(t)$ and $N(t)$ is obtained by extracting the coefficients in eq. \eqref{eq:coherenstateAppb} for the appropriate operators (Eq.~\ref{eq:EandN}), and by considering the proper time-evolution $\ta_m(t)=\ta_m e^{-i\mathcal{E}_mt}$. As can be observed in Fig.~\ref{fig:CoherentStateApp}, the approximation works very well for $\alpha=0.1$, which lies already well beyond the linear response regime. For $\alpha \gtrsim 0.2$ our approximation still predicts the right quasi-periods, but overestimates the amplitudes of the minima of $E(t)$ and the amplitudes of the maxima of $N(t)$ by approximately $20 \%$. These discrepancies become larger when $\alpha$ increases and, eventually, when $\alpha \gtrsim 0.4$ this approximation also fails in predicting the right quasi-periods (see Appendix Sec.~\ref{appsec:cohstateapp}, in particular Figs.~\ref{fig:appLRTEF} and \ref{fig:appLRTCC}). 

Finally, let us mention the holographic approach of \cite{daSilva2016} and the studies of other models with confinement \cite{Kormos2016,Rakovszky2016} that also obtained an oscillatory behavior of local observables, bearing some resemblance with our results.

\subsection{Strong-field regime} 
\begin{figure}[t]
\begin{subfigure}[b]{.24\textwidth}
\includegraphics[width=\textwidth]{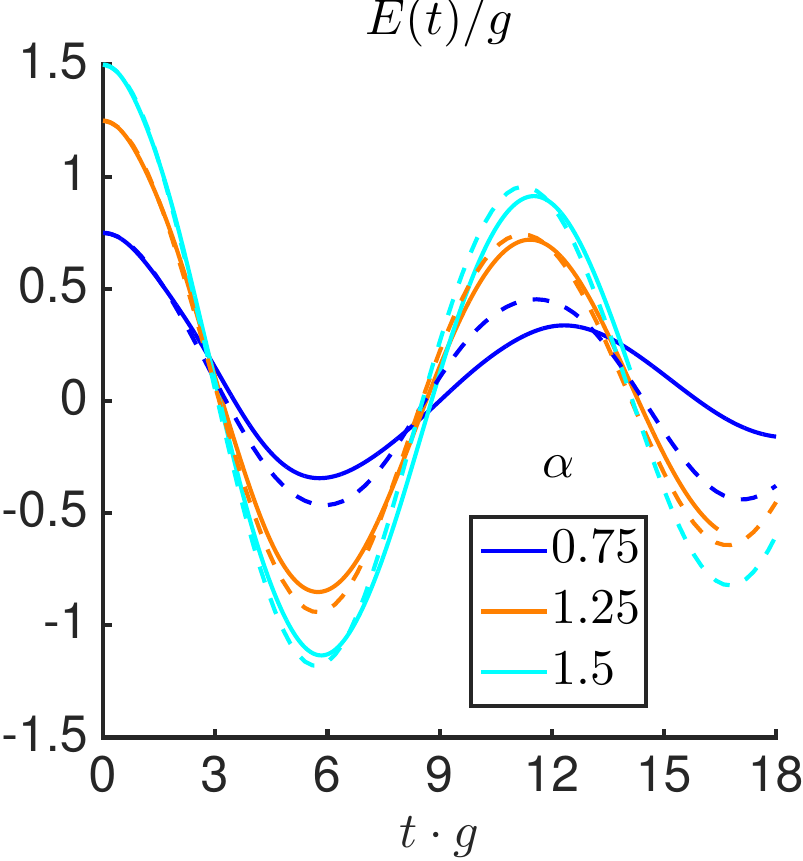}
\caption{\label{fig:EFsemiClassical}}
\end{subfigure}\hfill
\begin{subfigure}[b]{.24\textwidth}
\includegraphics[width=\textwidth]{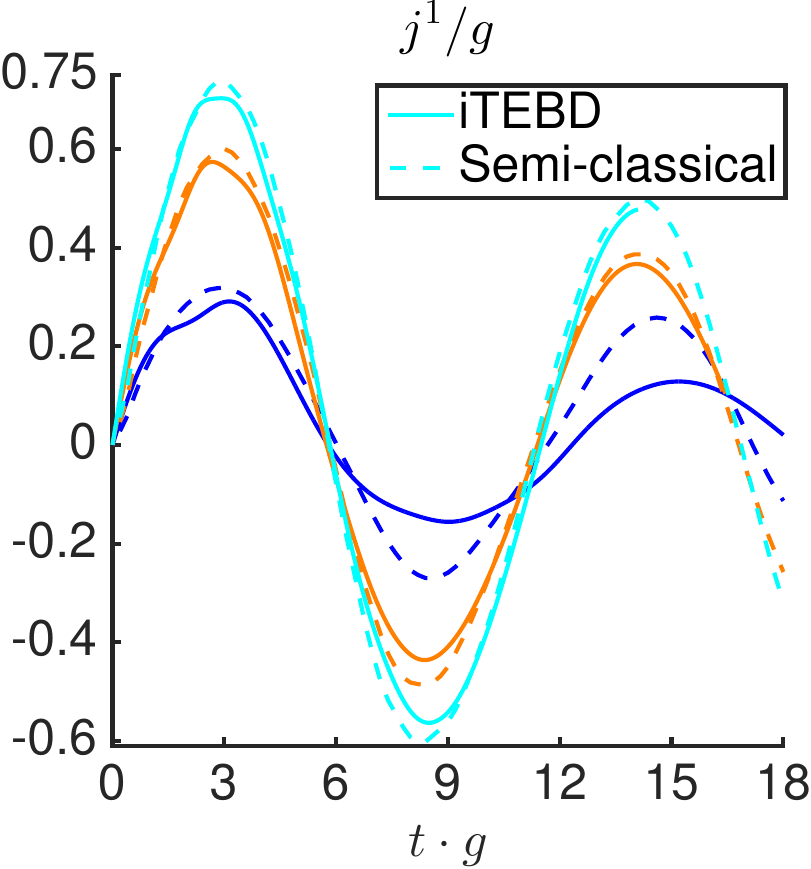}
\caption{\label{fig:CurrentsemiClassical}}
\end{subfigure}\vskip\baselineskip
\begin{subfigure}[b]{.24\textwidth}
\includegraphics[width=\textwidth]{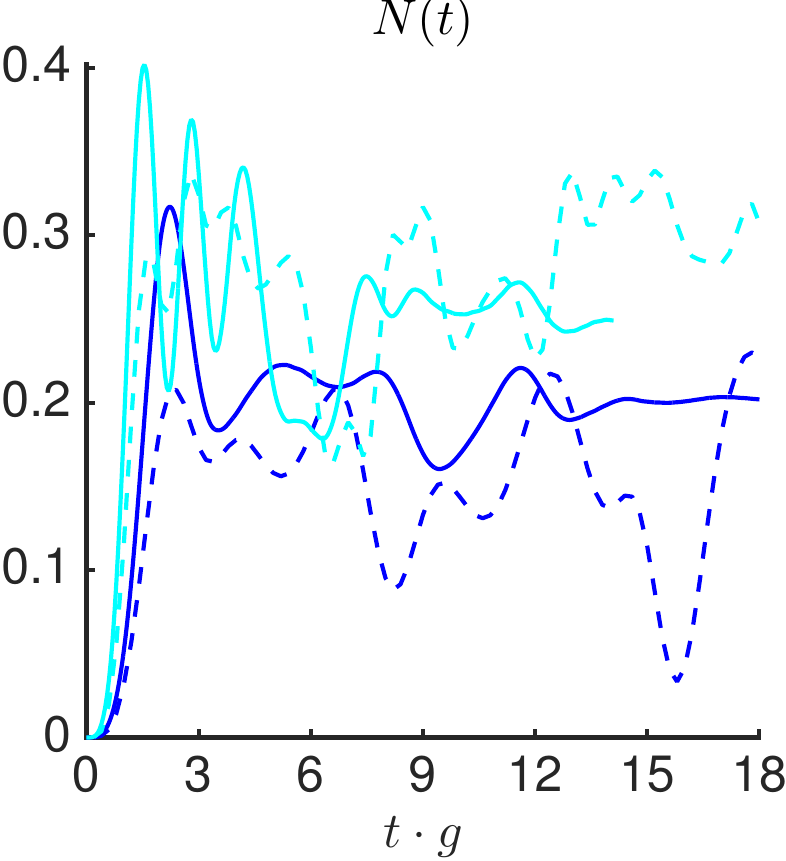}
\caption{\label{fig:PNsemiClassical}}
\end{subfigure}\hfill
\begin{subfigure}[b]{.24\textwidth}
\includegraphics[width=\textwidth]{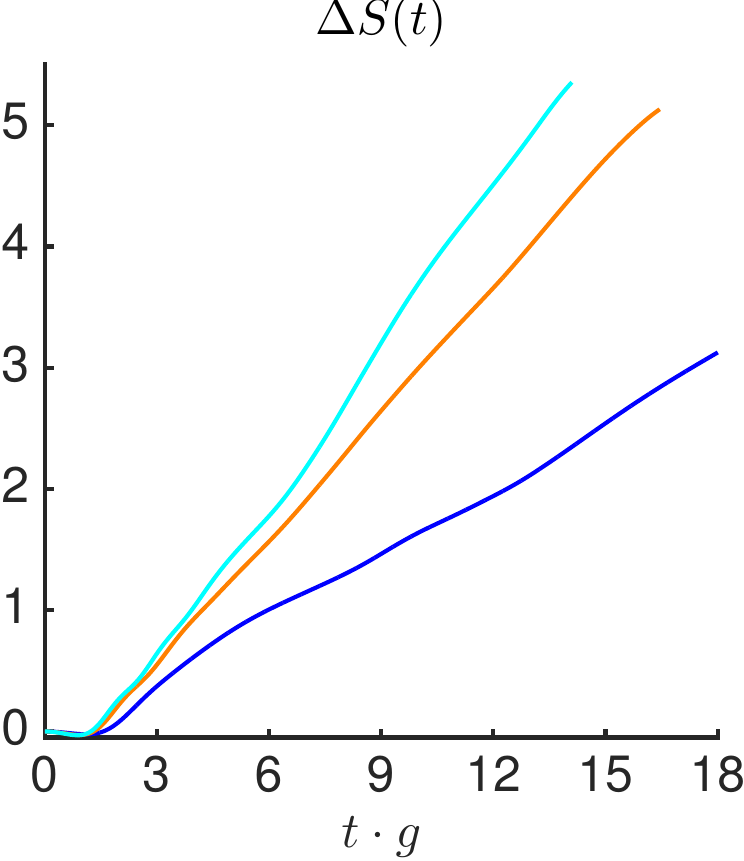}
\caption{\label{fig:EntrsemiClassical}}
\end{subfigure}\vskip\baselineskip
\captionsetup{justification=raggedright}
\caption{\label{fig:semiclassical} Results for $m/g= 0.25, x = 100$. Comparison of full quantum simulations (full line) with real-time lattice simulations (dashed line) (a): electric field $E(t)/g$. (b): current $j^1(t)/g$. (c) particle number $N(t)$. (d) entropy excess  $\Delta S(t)$.}
\end{figure}

Let us now consider larger quenches: $\alpha \geq 0.75$. 
In Fig.~\ref{fig:semiclassical} we compare the full quantum simulations (full line) with results obtained from real-time lattice gauge theory simulations \cite{Hebenstreit2013} (dashed line). 
The latter should give reliable results as long as the classicality conditions are fulfilled, i.e., anti-commutator expectation values for typical gauge field modes are much larger than the corresponding commutators.
This regime is characterized by non-perturbatively large field amplitudes \cite{Kasper2014}.

Focussing first on the electric field $E(t)$ and the current $j^1(t)$\footnote[3]{Notice that the time evolution of the electric field and the current are connected through Amp\`{e}re's law: $j^1(t)= -\dot{E}(t)/g$ which holds at the operator level. We computed both quantities independently for the MPS simulations and found agreement with Amp\`{e}re's law up to $10^{-3}$ . For the real-time lattice simulations we derived $j^1(t)$ from $E(t)$ using Amp\`{e}re's law.}  one can observe good agreement between the MPS and real-time lattice simulations.
The agreement further improves for growing $\alpha$ which is a nice cross-check for these two different techniques. However, for the particle number $N(t)$ we find sizeable deviations.
We attribute this discrepancy to differences in the initial states: the MPS simulation starts from the full ground state of the Hamiltonian $\mathcal{H}_{\alpha=0}$ and hence incorporates interactions of the fermions with the fluctuating gauge field.
On the other hand, the real-time lattice simulations are initialized in the bare Dirac vacuum that does not account for these interactions.

In a semi-classical picture the behavior of $E(t)$, $j^1(t)$ and $N(t)$ can be attributed to the nontrivial interplay between fermion and gauge field dynamics (backreaction) \cite{Kluger1992,Hebenstreit2013}: 
the electric field creates electron-positron pairs out of the vacuum and then accelerates them almost to the speed of light.
This process costs energy, the electric field therefore decreases due to energy conservation so that particle creation terminates and the current saturates. After this initial creation of electron-positron pairs, which essentially occurs during the first oscillation of the electric field, we enter a regime of plasma oscillations, for which the onset at $tg \gtrsim 3$ can be observed in Figs.~\ref{fig:EFsemiClassical} and \ref{fig:CurrentsemiClassical}. Also the behavior of the entanglement entropy $\Delta S(t)$ fits nicely with the semi-classical picture \cite{Calabrese2005}: after the local production of entangled electron-positron pairs, the pairs will separate, entangling the system over even larger distances. From Figs.~\ref{fig:PNsemiClassical} and \ref{fig:EntrsemiClassical} one can indeed observe that the entropy starts increasing linearly after the initial period of pair production. 

Even for large quenches we expect that the classicality conditions that underlie the real-time lattice technique are briefly violated during the times at which $E(t)$ crosses zero. We can indeed observe in Fig.~\ref{fig:EFsemiClassical} that the full quantum MPS results start deviating from the real-time lattice results after the first transit through zero. In particular the MPS simulations predict a stronger damping. We interpret this damping as the onset of equilibration. It is accepted that a state which is brought out of equilibrium relaxes and equilibrates locally at late times \cite{Linden2009}. In fact, it is believed that, under some generic conditions, the state thermalizes to a Gibbs state of the quenched Hamiltonian at a certain temperature \cite{Eisert2015,Rigol2008,Deutsch1991,Srednicki1994,Tasaki1998,Rigol2012,Rigol2009,Steinigeweg2014,Beugeling2014,Polkovnikov2011,Riera2012}. There are however some exceptions, such as when the state as a whole is not thermal even if some local quantities already indicate thermalization \cite{Berges2004,Banuls2011,Mueller2013}, when the system is integrable and it converges towards a so-called generalized Gibbs ensemble \cite{Caux2011,Caux2012,Caux2013,Gogolin2011,Cramer2008,Cassidy2011,Altland2012,Vidmar2016}, pre-thermalization \cite{Marcuzzi2013,Essler2014,Geiger2014,Abanin2015} or many-body localization \cite{Lagendijk2009,Gogolin2011,Nandkishore2015,Bauer2013,Serbyn2015}. 

\begin{figure}[t]
\begin{subfigure}[b]{.24\textwidth}
\includegraphics[width=\textwidth]{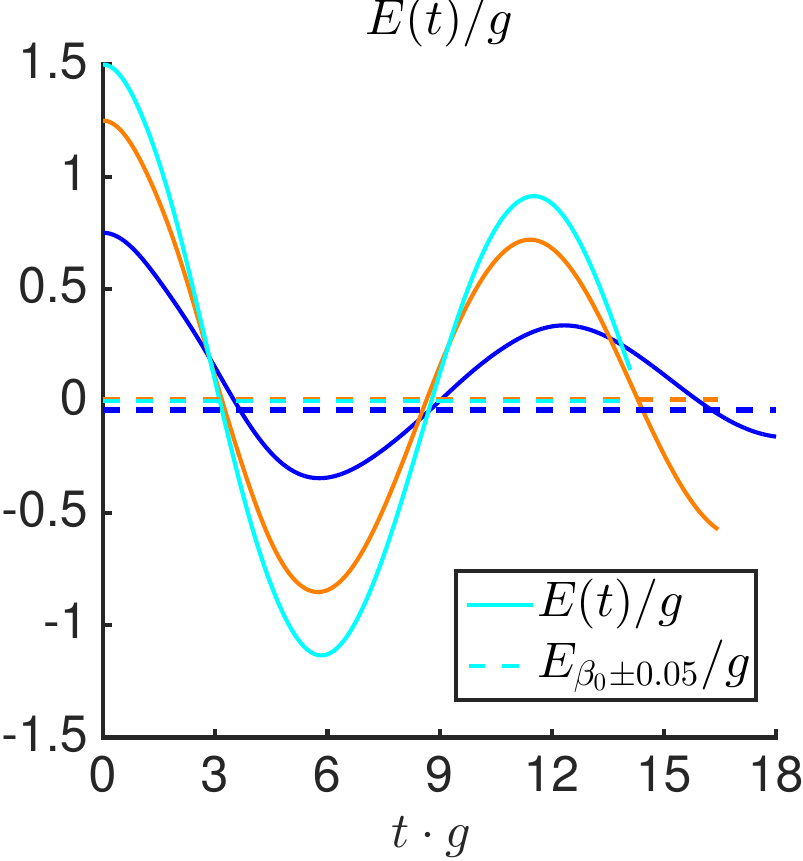}
\caption{\label{fig:EFcheckTherm}}
\end{subfigure}\hfill
\begin{subfigure}[b]{.24\textwidth}
\includegraphics[width=\textwidth]{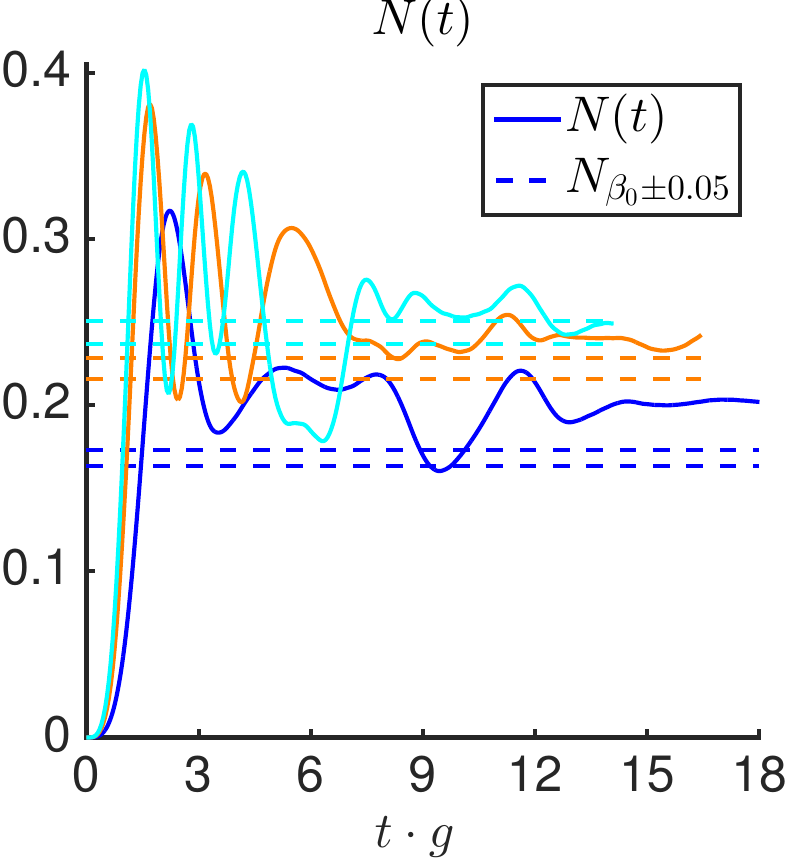}
\caption{\label{fig:PNcheckTherml}}
\end{subfigure}\vskip\baselineskip
\captionsetup{justification=raggedright}
\caption{\label{fig:thermalization} Results for $m/g= 0.25, x = 100$. Comparison real-time simulations (full line) with predicted asymptotic value in thermal equilibrium (dashed line) (a): $E(t)/g$. (b) $N(t)/g$. }
\end{figure}

Under the assumption that the state would thermalize, we can determine its inverse temperature $\beta_0$ from energy conservation and by using our results from finite temperature simulations \cite{Buyens2016} (see Appendix Sec.~\ref{subsec:determineGibbsState}). In fig.~\ref{fig:thermalization} we compare $E(t)$ and $N(t)$ (full line) with its predicted thermal values $E_{\beta_0}$ and $N_{\beta_0}$ (dashed line). Note that our finite temperature simulations only enable us to determine $\beta_0$ numerically up to $\Delta\beta = 0.05$, therefore we show the intervals $E_{\beta_0\pm 0.05}$ and $N_{\beta_0 \pm 0.05}$. 
Although the electric field seems to oscillate around $E_{\beta_0}$, the amplitudes of the oscillations are still too large for a definite conclusion. On the other hand, one might be more tempted to say that $N(t)$ is close to its thermal value for $\alpha = 1.25$ and $\alpha = 1.5$, although one should be cautious here as well. 

To reach a definite conclusion, we would have to push the MPS simulations further in time. Unfortunately, the linear growth of entanglement, see Fig.~\ref{fig:EntrsemiClassical}, requires the variational freedom of the MPS representation to grow exponentially in time (see Appendix Sec.~\ref{subsec:iTEBDRT}, in particular Fig.~\ref{fig:evolutionDmax}). This precludes computations at large $t g$ and hence constrains the maximum time up to which we can reliably track the state.

\section{Conclusion} 
We demonstrated the potential of MPS to solve the real-time simulation of gauge field theories near the continuum limit, based on the paradigmatic example of an electric quench in the massive Schwinger model. 
For small quenches the real-time dynamics can be explained by using the single-particle excitations of the quenched Hamiltonian.
For large quenches $\alpha=\mathcal{O}(1)$, which is related to the phenomenon of Schwinger pair production, we compared the MPS simulations with results from real-time lattice gauge theory simulations and found good agreement between those methods.
In this regime, we further investigated whether the state thermalizes at late times by using finite temperature simulations.
While we found evidence that supports the onset of thermalization, the increase of entanglement prevented us to reach a decisive conclusion yet.

The MPS method provides a unique means to benchmark quantum simulators of the massive Schwinger model or related models using ultracold ions or atoms in optical lattices \cite{Hauke2013,Wiese2013,Martinez2016,Kasper2015,Kasper2016}.
On the other hand, it is a major goal to extend this type of real-time simulation technique to more than one spatial dimension using projected entangled pair states (PEPS) \cite{Verstraete2004}. The major progress on PEPS algorithms in the last decade \cite{Murg2007,Corboz2009,Jordan2008,Corboz2010,Kraus2010,Corboz2014,Vanderstraeten2015b,Phien2015,Corboz2016,Vanderstraeten2016b} in combination with recent promising PEPS and TNS results for higher-dimensional gauge theories \cite{Tagliacozzo2014,Haegeman2015,Zohar2015,Milsted2016,Zohar2016} makes us confident that this will be realized in the foreseeable future.

\section*{Acknowledgments} 
We acknowledge interesting discussions with Mari-Carmen Ba\~{n}uls, David Dudal and Esperanza Lopez. 
This work is supported by an Odysseus grant from the FWO, a PhD-grant from the FWO (B.B), a post-doc grant from the FWO (J.H.), the FWF grants FoQuS and Vicom, the ERC grant QUERG, the EU grant SIQS and a grant from the European Union's Seventh Framework Programme (FP7/2007-2013)/ERC under grant agreement 339220 (F.H.). 

\appendix
\numberwithin{equation}{section}
\renewcommand\theequation{\Alph{section}.\arabic{equation}}
\section{MPS for the Schwinger model}\label{app:MPSSchwingerModel}
\noindent In this section we explain the Matrix Product States (MPS) methods that are used for the Schwinger model. More specifically we discuss:
\begin{enumerate}
\item How the time-dependent variational principle (TDVP) is used to find the optimal translational invariant MPS approximation for the ground state in the thermodynamic limit (see Sec.~\ref{subsec:TDVPGS}).
\item How we approximate the single-particle excitations using MPS (see Sec.~ \ref{subsec:RRforExc}).
\item How we perform real-time evolution within the manifold of MPS using the infinite time-evolving block decimation algorithm (iTEBD) (see Sec.~\ref{subsec:iTEBDRT}).
\item How we use the MPS approximations for the ground state and the single-particle excitations to approximate the real-time evolution in the weak-field regime (see Sec.~\ref{appsec:cohstateapp}).
\item How we determine the temperature of to the equilibrium state given that the state brought out of equilibrium by the quench thermalizes (see Sec.~\ref{subsec:determineGibbsState}). 
\end{enumerate}

More details and results can also be found in our earlier papers \cite{Buyens2013,Buyens2014,Buyens2015,Buyens2015b,Buyens2016,Buyens2017}
\subsection{Ground-state ansatz}\label{subsec:TDVPGS}
\noindent Consider the Kogut-Susskind Hamiltonian Eq.~(\ref{eq:Hamiltonian}) of the Schwinger model:
\begin{multline} \label{eq:Hamiltonianapp} \mathcal{H}_{\alpha}= \frac{g}{2\sqrt{x}}\Biggl(\sum_{n=1}^{2N}[L(n) + \alpha]^2 + \frac{\sqrt{x}}{g} m \sum_{n =1}^{2N}(-1)^n\sigma_z(n)  + \\ 
x \sum_{n=1}^{2N-1}(\sigma^+ (n)e^{i\theta(n)}\sigma^-(n + 1) + h.c.)\biggl).\end{multline}
We block site $n$ and link $n$ into one effective site with local Hilbert space spanned by $\{\ket{s_n,p_n}_n: s_n = -1,1; p_n \in \mathbb{Z} \}$. 
Writing $ \kappa_n = (s_n,p_n)$ and
$$\bm{\kappa} = \bigl((s_1,p_1),(s_2,p_2),\ldots,(s_{2N},p_{2N})\bigl) = (\kappa_1,\ldots,\kappa_{2N})$$
a general state on this system of $2N$ sites takes the form
$$\ket{\Psi} = \sum_{\bm{\kappa}} C_{\kappa_1,\ldots,\kappa_{2N}} \ket{\bm{\kappa}} $$
with basis coefficients $C^{\kappa_1,\ldots,\kappa_{2N}}$. 

A MPS $\ket{\umps{A(1)A(2)}}$ assumes now a special form for these coefficients:
$$C_{\kappa_1,\ldots,\kappa_{2N}} = v_L^\dagger \left(\prod_{n = 1}^{N} A_{\kappa_{2n-1}}(1) A_{\kappa_{2n}}(2)\right)v_R,$$
i.e.,
\begin{subequations}\label{eq:MPSapp}
\be\ket{\umps{A(1)A(2)}} = \sum_{\bm{\kappa}}v_L^\dagger \left(\prod_{n = 1}^{N} A_{\kappa_{2n-1}}(1) A_{\kappa_{2n}}(2)\right)v_R \ket{\bm{\kappa}}.\ee
Here we have $A_\kappa(n) \in \mathbb{C}^{D(n) \times D(n+1)}$ and $v_L, v_R \in \mathbb{C}^{D(1) \times 1}$. The MPS ansatz associates with each site $n$ and every local basis state $\ket{\kappa_n}_n =\ket{s_n,p_n}_n$ a matrix $A_{\kappa_n}(n)= A_{s_n,p_n}(n)$. The indices $\alpha$ and $\beta$ are referred to as virtual indices, and $D(n)$ are called the bond dimensions. Note that here $A_\kappa(n)$ only depends on the parity of $n$, in accordance with the $\mathcal{T}^2$ symmetry of the Hamiltonian. As such we can consider the ansatz directly in the thermodynamic limit ($N \rightarrow + \infty$), bypassing any possible finite size artifacts. In this limit the expectation values of all local observables are independent of the boundary vectors $v_L$ and $v_R$.

As explained in \cite{Buyens2013}, to parameterize gauge invariant MPS, i.e. states that obey $G(n)\ket{\Psi(A)}=0$ for every $n$, 
$$G(n) = L(n) - L(n-1) + \frac{\sigma_z(n) + (-1)^n}{2},$$
it is convenient to give the virtual indices a multiple index structure $\alpha\rightarrow (q,\alpha_q); \beta \rightarrow (r,\beta_r)$, where $q$ resp. $r$ labels the eigenvalues of $L(n-1)$ resp. $L(n)$. One can verify that the condition $G(n)=0$ then imposes the following form on the matrices:
\be
{[A_{s,p}(n)]}_{(q,\alpha_q),(r,\beta_r)} =  {[a_{q,s}(n)]}_{\alpha_q,\beta_r}\delta_{q+(s+(-1)^n)/2,r}\delta_{r,p}
\label{eq:gaugeMPSapp},\ee
\end{subequations}
where $\alpha_q = 1\ldots D_q(n)$, $\beta_r = 1 \ldots D_r(n+1)$. The formal total bond dimensions of this MPS are $D(n) = \sum_q D_q(n)$, but notice that, as (\ref{eq:gaugeMPSapp}) takes a very specific form, the true variational freedom lies within the matrices $a_{q,s}(n) \in \mathbb{C}^{D_q(n) \times D_r(n+1)}$. 

\noindent To find the optimal ground state of $\mathcal{H}_\alpha$ within the class of gauge invariant MPS Eq.~(\ref{eq:MPSapp}) we apply the time-dependent variational principle (TDVP) \cite{Haegeman2011,Haegeman2013,Haegeman2014a} to the Schr\"{o}dinger equation
$$\partial_\tau \ket{\umps{A(1)A(2)}} = -\mathcal{H}_{\alpha} \ket{\umps{A(1)A(2)}} $$
in imaginary time $d\tau = -idt$. When $\tau \rightarrow + \infty$ we indeed find the optimal approximation $ \ket{\umps{A(1)A(2)}}$ for the ground state of $\mathcal{H}_{\alpha}$. As the Schmidt decomposition of $\ket{\umps{A(1)A(2)}}$ with respect to the bipartition of the lattice consisting of the two regions $\mathcal{A}_1(n) = \mathbb{Z}[1, \ldots, n]$ and $\mathcal{A}_2(n) = \mathbb{Z}[n+1,\ldots, 2N]$ equals
\be \label{eq:MPSschmidtGaugeapp} \ket{\umps{A(1)A(2)}} = \sum_{q} \sum_{\alpha_q=1}^{D_q} \sqrt{\lambda_{q,\alpha_q}(n)} \ket{\psi_{q,\alpha_q}^{\mathcal{A}_1(n)}}\ket{\psi_{q,\alpha_q}^{\mathcal{A}_2(n)}}, \ee
it follows that to obtain a faithful approximation for the ground state one has to choose $D_q$ such that the discarded Schmidt values for each charge sector are {\em sufficiently} small. In particular we could take $D_q=0$ for $|q|>3$ which is explained by the first term in the Hamiltonian Eq.~(\ref{eq:Hamiltonianapp}). A proper justification of truncating the charge sectors is provided in \cite{Buyens2017}. We refer to \cite{Buyens2015,Buyens2017} for the details on the TDVP.

\subsection{MPS approximation for single-particle excitations}\label{subsec:RRforExc}
\noindent Once we have an MPS approximation $\ket{\umps{A(1)A(2)}}$ for the ground state of $\mathcal{H}_{\alpha}$, see Sec.~\ref{subsec:TDVPGS}, we use the method of \cite{Haegeman2012,Haegeman2013} to approximate the single-particle excitations. The ansatz for the single-particle excitations with momentum $k$ that we will use is:
\begin{subequations}\label{eq:excAnsatzt2}
\begin{multline} \ket{\tanv{k}{B}{A(1)A(2)}} = \sum_{m  = 1}^{N} e^{2ikn/\sqrt{x}}  \sum_{\{\kappa_n\}} \\ v_L^\dagger \left(\prod_{n = 1}^m A_{\kappa_{2n-1}}(1)A_{\kappa_{2n}}(2)\right)  B_{\kappa_{2n-1},\kappa_{2n}} \\ 
\left(\prod_{n = m+1}^N A_{\kappa_{2n-1}}(1)A_{\kappa_{2n}}(2)\right)v_R \ket{\bm{\kappa}},\end{multline}
where $A(1)$ and $A(2)$ correspond to the ground state $\ket{\umps{A(1)A(2)}}$ of $\mathcal{H}_{\alpha}$ and gauge invariance is imposed by
\begin{multline} [B_{s_1,p_1,s_2,p_2}]_{(q,\alpha_q);(r,\beta_r)}  \\ 
=  [b_{q,s_1,s_2}]_{\alpha_q,\beta_r}\delta_{p_1,q + (s_1 -1)/2}\delta_{p_2,q + (s_1+s_2)/2}\delta_{r,p_2}.\end{multline}
\end{subequations}
where $\kappa_n = (s_{n},p_{n})$ and $b_{q,s_1,s_2} \in \mathbb{C}^{D_q \times D_r}$. The algorithm to find the optimal approximation $\ket{\tanv{k}{B}{A(1)A(2)}}$ for the excited states is discussed in \cite{Haegeman2013,Buyens2013,Buyens2017}: one has to find $b_{q,s_1,s_2}$ such that
$$\frac{\braket{\tanv{k}{\overline{B}}{\overline{A(1)A(2)}} \vert \mathcal{H}_{\alpha} \vert \tanv{k}{B}{A(1)A(2)}}}{\braket{\tanv{k}{\overline{B}}{\overline{A(1)A(2)}}\vert {\tanv{k}{B}{A(1)A(2)}}}} $$
is minimized with respect to $\overline{b}_{q,s_1,s_2}$. This boils down in a generalized eigenvalue equation for $b_{q,s_1,s_2}$ where the smallest eigenvalues correspond to the energies of the single-particle excitations. Only the ones who are stable against variation of the bond dimensions $D_q$ are physical. We refer to \cite{Buyens2013,Buyens2015b,Buyens2017} for the details. \\
\\In \cite{Buyens2015,Buyens2017} we found for $m/g = 0.25$ and $\alpha \lesssim 0.47$ two single-particle excitations with masses $\mathcal{E}_1$ and $\mathcal{E}_2$. The energies at non-zero momentum are in the continuum limit determined by the Lorentz dispersion relations: $\mathcal{E}_m(k) = \sqrt{k^2 + \mathcal{E}_m^2}$. The corresponding MPS approximations at non-zero lattice spacing $a = 1/g\sqrt{x}$ are $\ket{\tanv{k}{B^{(m,k)}}{A(1)A(2)}}$ with
\begin{multline}[B_{s_1,p_1,s_2,p_2}^{(m,k)}]_{(q,\alpha_q);(r,\beta_r)} \\ =  [b_{q,s_1,s_2}^{(m,k)}]_{\alpha_q,\beta_r}\delta_{p_1,q + (s_1 -1)/2}\delta_{p_2,q + (s_1+s_2)/2}\delta_{r,p_2} \end{multline}
and are normalized such that \cite{Haegeman2013}
\begin{subequations} \label{eq:normSP}
\begin{multline} \braket{\tanv{k'}{\overline{B^{(n,k')}}}{\overline{A(1)A(2)}}\vert {\tanv{k}{B^{(m,k)}}{A(1)A(2)}}} 
\\ = 2\pi \delta_{n,m} \delta(k - k') \end{multline}
\be \braket{\umps{\overline{A(1)A(2)}} \vert {\tanv{k}{B^{(m,k)}}{A(1)A(2)}}} = 0.  \ee
\end{subequations}
The delta-Dirac function originates from the infinite lattice length and has to be read as
\be\label{eq:diracreg0}\delta(k-k') = \lim_{N \rightarrow + \infty } \frac{2N}{2\pi}\delta_{k,k'}\ee
where $2N$ ($N \rightarrow + \infty$) is the number of sites on the lattice. 

For a local observable $\mathcal{O} = \sum_{n = 1}^{2N-1} \mathcal{T}^{n-1} o \mathcal{T}^{-n+1}$, where $o$ is a Hermitian operator which acts only non-trivial on sites $1$ and $2$, we first subtract the ground state contribution such that 
$$\braket{\umps{\overline{A(1)A(2)}} \vert \mathcal{O} \vert \umps{A(1)A(2)}} = 0 .$$
With this renormalization we have that
\begin{subequations}\label{eq:overlapSP}
 \begin{multline} \bra{\tanv{k}{\overline{B}}{\overline{A(1)A(2)}}} \mathcal{O} \ket{\tanv{k'}{B}{A(1)A(2)}} 
\\ = 2\pi\delta(k-k')  O_{eff}^1[\overline{B},B']\end{multline}
\begin{multline}  \braket{\umps{\overline{A(1)A(2)}} \vert \mathcal{O}\vert  \tanv{k}{B}{A(1)A(2)}}  \\
= 2\pi \delta(k) O_{eff}^2[\overline{A(1)A(2)},B]\end{multline}
\end{subequations}
where $O_{eff}^1[\overline{B},B']$ and $O_{eff}^2[\overline{A(1)A(2)},B]$ are finite quantities that can be computed efficiently, see \cite{Haegeman2013}. The delta-Dirac distributions have to be regularized according to Eq.~(\ref{eq:diracreg0}).

\subsection{iTEBD for real-time evolution}\label{subsec:iTEBDRT}
\noindent To evolve a state approximated by a MPS Eq.~(\ref{eq:MPSapp}) at $t = 0$, i.e. to find
$$\ket{\Psi(t)} = e^{-i\mathcal{H}_\alpha t} \ket{\umps{A(1)A(2)}},$$
we used the infinite time-evolving block decimation (iTEBD) \cite{Vidal2007}. At the core of this method lies the Trotter decomposition \cite{Hatano2005} which decomposes $e^{- i dt \mathcal{H}}$ into a product of local operators, the so-called Trotter gates. Specifically, we did a fourth order Trotter decomposition of $e^{-i \mathcal{H}_\alpha dt}$ for small steps $dt$ and projected afterwards $\ket{\Psi(t+dt)} =  e^{-i \mathcal{H}_\alpha dt}\ket{\umps{A(1)A(2)}}$ to a MPS $\ket{\umps{\tilde{A}(1)\tilde{A}(2)}}$ with smaller bond dimensions $D_q$ . Similar as for the ground state, $D_q$ is chosen by discarding the Schmidt values smaller than a preset tolerance $\epsilon^2$ in Eq.~(\ref{eq:MPSschmidtGaugeapp}). In this way the virtual dimensions are adapted dynamically. We refer to \cite{Buyens2013} for the details on the implementation of the iTEBD.  

\begin{figure}
\null\hfill
\begin{subfigure}[b]{.24\textwidth}
\includegraphics[width=\textwidth]{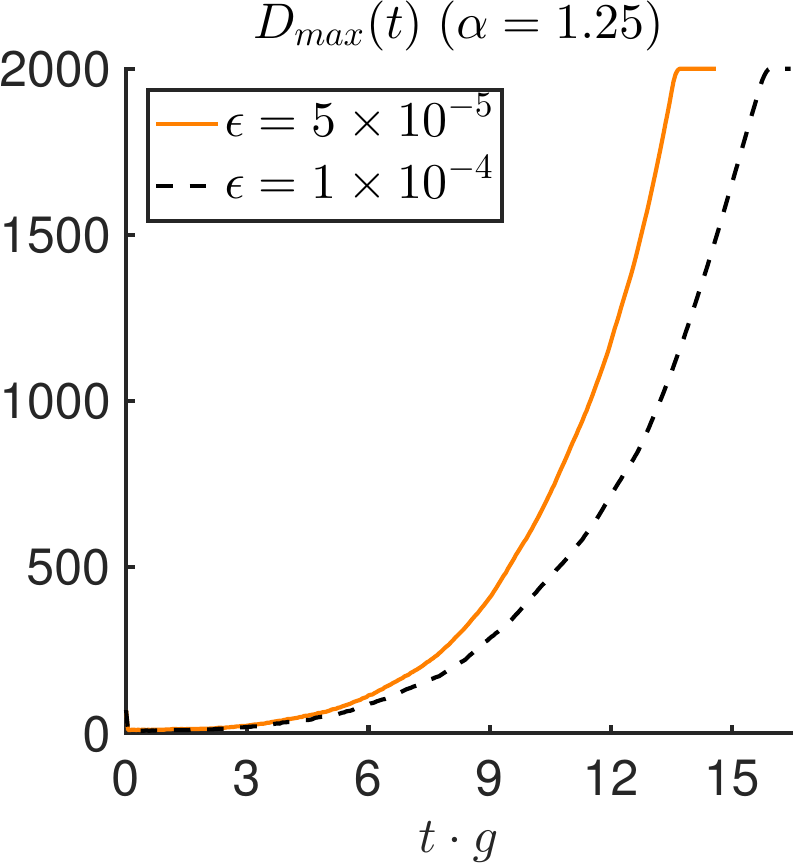}
\caption{\label{fig:evolutionDmaxa}}
\end{subfigure}\hfill
\begin{subfigure}[b]{.24\textwidth}
\includegraphics[width=\textwidth]{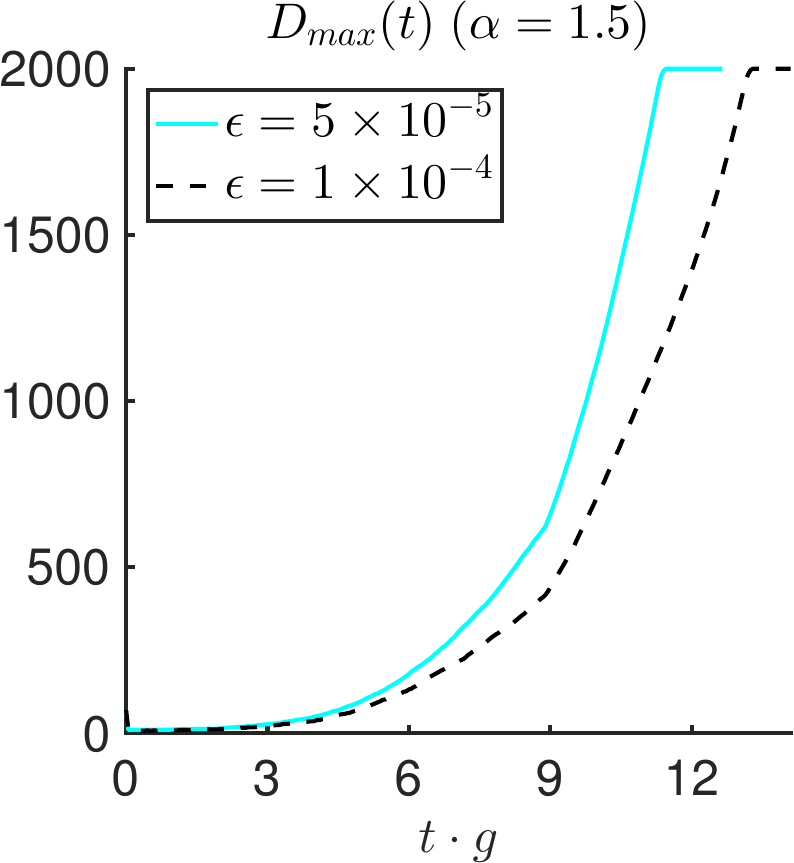}
\caption{\label{fig:evolutionDmaxb}}
\end{subfigure}\hfill\null
\vskip\baselineskip
\captionsetup{justification=raggedright}
\caption{\label{fig:evolutionDmax} $m/g= 0.25$. Evolution of the maximum of the bond dimension over the charge sectors for fixed values of $\epsilon$: $\epsilon = 5 \times 10^{-5}$ (full line) and $\epsilon = 1 \times 10^{-4}$ (dashed line) (a) $\alpha = 0.125$. (b) $\alpha = 1.5$.  }
\end{figure}

Taking a non-zero value for $\epsilon$ yields a truncation in the entanglement spectrum and hence a truncation in turn determines the required bond dimensions $D_q$ for every charge sector. For instance, in Fig.~\ref{fig:evolutionDmax} we show how the maximum of the bond dimension over the charge sectors $D_{max}  = \max_q D_q$ varies with time for a given value of $\epsilon$. It is this growth of the required bond dimensions, which can be traced back to the growth of entanglement, that makes the computations more costly at later times. Note that to save computational resources we imposed that $D_{max} \leq 2000$. 

\begin{figure}
\null\hfill
\begin{subfigure}[b]{.24\textwidth}
\includegraphics[width=\textwidth]{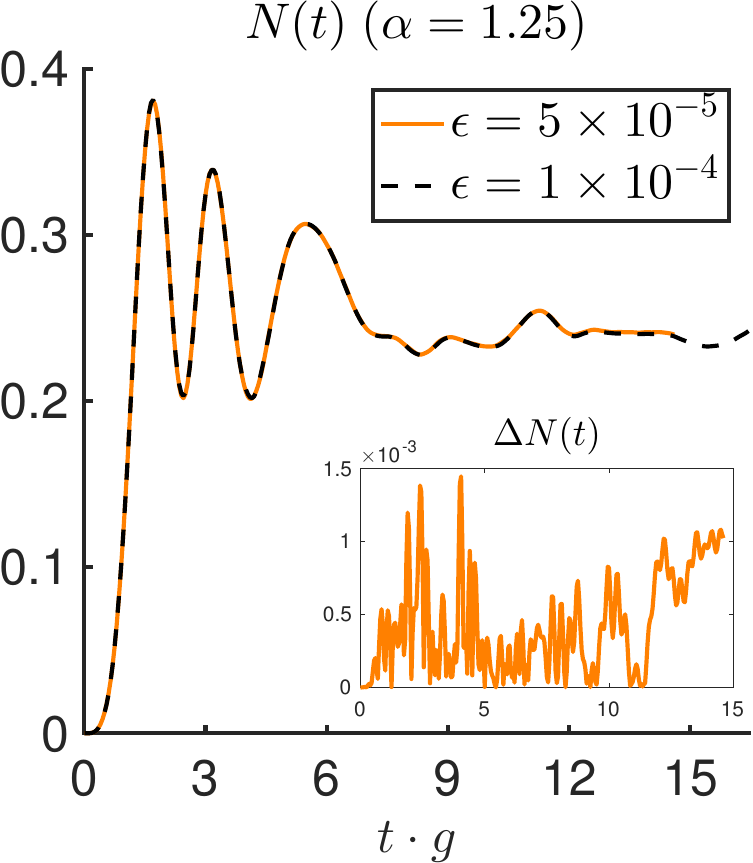}
\caption{\label{fig:Ndiffepsa}}
\end{subfigure}\hfill
\begin{subfigure}[b]{.24\textwidth}
\includegraphics[width=\textwidth]{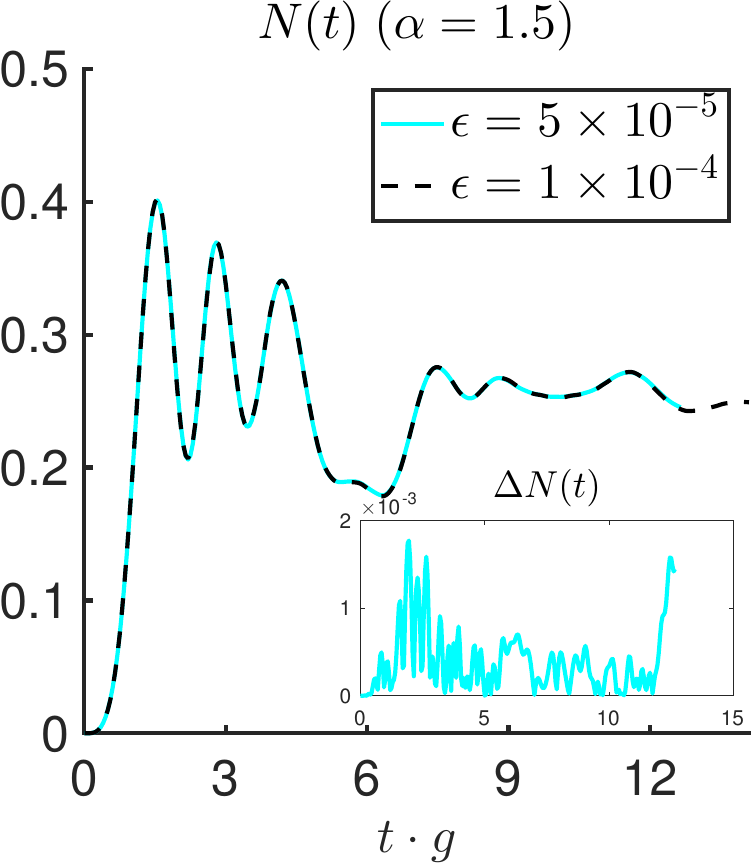}
\caption{\label{ig:Ndiffepsb}}
\end{subfigure}\hfill\null
\vskip\baselineskip
\null\hfill
\begin{subfigure}[b]{.24\textwidth}
\includegraphics[width=\textwidth]{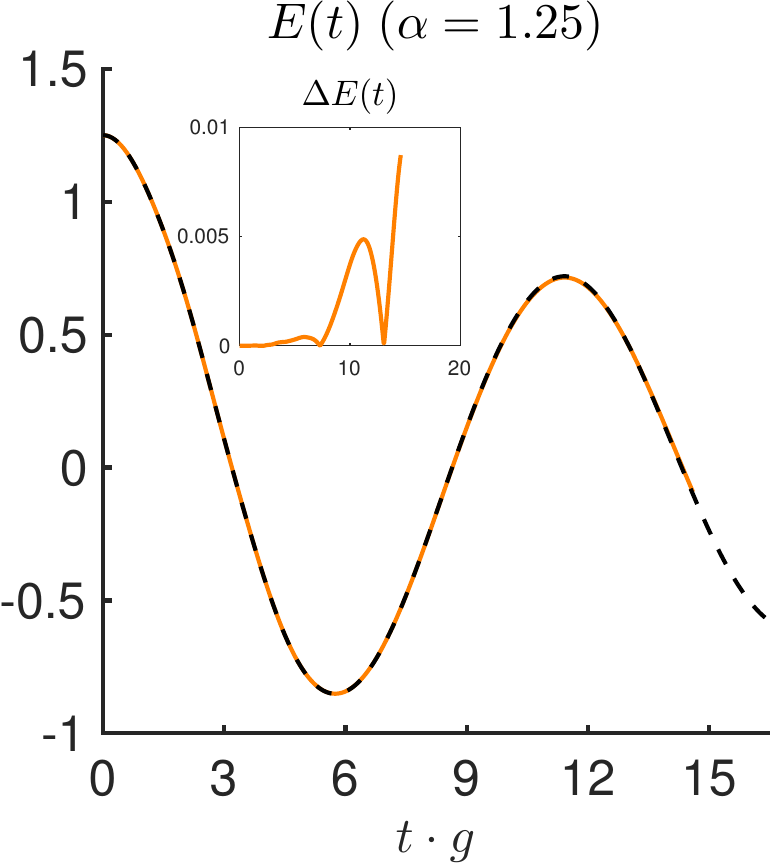}
\caption{\label{fig:EFdiffepsa}}
\end{subfigure}\hfill
\begin{subfigure}[b]{.24\textwidth}
\includegraphics[width=\textwidth]{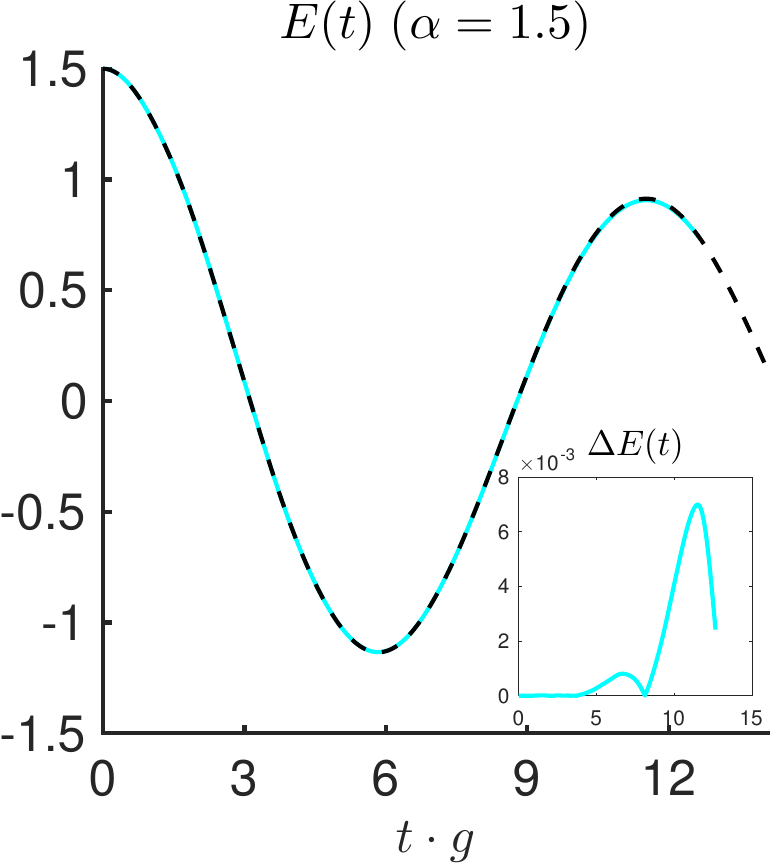}
\caption{\label{fig:EFdiffepsb}}
\end{subfigure}\hfill\null
\vskip\baselineskip
\captionsetup{justification=raggedright}
\caption{\label{fig:Qdiffeps}$m/g= 0.25$. Evolution of the electric field and particle number for different values of $\epsilon$: $\epsilon = 5 \times 10^{-5}$ (full line) and $\epsilon = 1 \times 10^{-4}$ (dashed line). Inset: difference in magnitude of the considered quantity for the simulation with $\epsilon = 5 \times 10^{-5}$ and the simulation with $\epsilon = 1 \times 10^{-4}$.   (a) $N(t)$ ($\alpha = 1.25$). (b) $N(t)$ ($\alpha = 1.5$). (c) $E(t)$ ($\alpha = 1.25$). (d) $E(t)$ ($\alpha = 1.5$). }
\end{figure}

As explained in \cite{Buyens2013} the simulation should be exact as $\epsilon \rightarrow 0$. Therefore the convergence in $\epsilon$ can be used to control the truncation error for a certain observable. In order to have a rough idea about the error for taking non-zero $\epsilon$ we compare the results for the simulation for the two smallest values of $\epsilon$. We illustrate this in Fig.~\ref{fig:Qdiffeps} for the electric field expectation value and the particle number $N(t)$ where we compare the simulations for $\epsilon = 5 \times 10^{-5}$ (full line) with the simulations for $\epsilon = 1 \times 10^{-4}$. As can be observed from the inset, where we plot the differences in magnitude of the electric field,
$$ \Delta E(t) = \vert E_{\epsilon = 5 \times 10^{-5}}(t) - E_{\epsilon = 1 \times 10^{-4}}(t)\vert $$ 
and the particle number,
$$\Delta N(t) = \vert N_{\epsilon = 5 \times 10^{-5}}(t) - N_{\epsilon = 1 \times 10^{-4}}(t) \vert,$$
the results are in agreement with each other up to at most $8 \times 10^{-3}$. For other (smaller) values of $\alpha$ we found that this error was even smaller. Therefore we can trust that our results are reliable up to at least $1\%$.

\subsection{Weak-field regime approximation}\label{appsec:cohstateapp}
\noindent If $\mathcal{H}_{\alpha_0}$ is the Hamiltonian in an electric background field $\alpha_0$
\begin{multline} \mathcal{H}_{\alpha_0}= \frac{g}{2\sqrt{x}}\Biggl(\sum_{n=1}^{2N}[L(n) + \alpha_0]^2 \\
+ \frac{\sqrt{x}}{g} m \sum_{n =1}^{2N}(-1)^n\bigl(\sigma_z(n) + (-1)^n\bigl)  \\
+ x \sum_{n=1}^{2N-1}(\sigma^+ (n)e^{i\theta(n)}\sigma^-(n + 1) + h.c.)\biggl).\end{multline}
and $\mathcal{H}_{\alpha}$ is the Hamiltonian in an electric background field $\alpha$, then we can write (up to an irrelevant constant)
$$ \mathcal{H}_{\alpha} = \mathcal{H}_{\alpha_0} + \epsilon \mathcal{V} $$
where 
$$ \mathcal{V} = \frac{g}{\sqrt{x}}\sum_{n=1}^{2N}L(n) $$
and $\epsilon = \alpha - \alpha_0$. Consider now the annihilation and creation operators $\ta_m$ and $\ta_m^\dagger$ of the single-particle excitations with energy $\mathcal{E}_m(k)$ and momentum $k$ of $\mathcal{H}_\alpha$. In principle, they can obey either the canonical commutation relations for bosons or fermions, but as we will see later we need to impose boson statistics:
\begin{multline}\label{eq:commRelA} [\ta_n(k'), \ta_m^\dagger(k)]  =  \delta(k'-k)\delta_{m,n}, 
\\ [\ta_m(k'),\ta_n(k')] = 0, [\ta_n^\dagger(k'),\ta_m^\dagger (k)] = 0. \end{multline}

Using the TDVP, see Sec.~\ref{subsec:TDVPGS}, we have a MPS approximation $\ket{\umps{A(1)A(2)}}$ for the ground state of $\mathcal{H}_\alpha$ and by using the method discussed in Sec.~\ref{subsec:RRforExc}, we have a MPS approximation $\ket{\tanv{k}{B^{(m,k)}}{A}}$ for the $m$-th single-particle excitation with momentum $k$ and energy $\mathcal{E}_{m}(k)$. They are normalized as
\begin{subequations}\label{eq:Norm}
\be \braket{\umps{\overline{A(1)A(2)}} \vert \umps{A(1)A(2)}} = 1,\ee
\be \braket{\umps{\overline{A(1)A(2)}} \vert \tanv{k}{B^{(m,k)}}{A(1)A(2)}} = 0, \ee
and
\begin{multline} \braket{\tanv{k'}{\overline{B^{(n,k')}}}{\overline{A(1)A(2)}}\vert \tanv{k}{B^{(m,k)}}{A(1)A(2)}} \\ = 2\pi \delta(k - k')\delta_{n,m} \end{multline}
\end{subequations}
The delta-Dirac functions originate from the infinite lattice length and have to be read as, see Eq.~(\ref{eq:diracreg0}),
\be\label{eq:diracreg}\delta(k-k') = \lim_{N \rightarrow + \infty } \frac{2N}{2\pi}\delta_{k,k'}\ee
where $2N$ ($N \rightarrow + \infty$) is the number of sites on the lattice. Within this approximation we have that
\begin{subequations} 
\be \mathcal{H}_\alpha \ket{\umps{A(1)A(2)}} = 0, \ee
\begin{multline} \mathcal{H}_{\alpha} \ket{\tanv{k}{B^{(m,k)}}{A(1)A(2)}} \\ = \mathcal{E}_m(k) \ket{\tanv{k}{B^{(m,k)}}{A(1)A(2)}},\end{multline}
\end{subequations}
and 
\begin{subequations}\label{eq:creationoperator} 
\be \ta_m^\dagger(k) \ket{\umps{A(1)A(2)}} = \frac{1}{\sqrt{2\pi}}\ket{\tanv{k}{B^{(m,k)}}{A(1)A(2)}},\ee 
\be \ta_m(k) \ket{\umps{A(1)A(2)}} = 0. \ee
\end{subequations}

We now want to express the ground state $\ket{\Psi(0)}$ of $\mathcal{H}_{\alpha_0}$ in terms of the ground state $\ket{\umps{A(1)A(2)}}$ and the single-particle excitations $\ket{\tanv{k}{B}{A(1)A(2)}}$ of $\mathcal{H}_{\alpha}$. We will expand $\mathcal{H}_{\alpha_0}$ in a series of powers of $(\ta_m(k),\ta_m^\dagger(k))$:
\begin{multline} \label{eq:halpha0app}  
\mathcal{H}_{\alpha_0} \approx \lambda_0 \mathbbm{1} + \int dk \left( \sum_m c_m(k)\ta_m(k) + \sum_m\bar{c}_m(k)\ta_m^\dagger (k) \right) \\ 
+ \int dk \int dk' \left(\; \sum_{m,n}\mu_{m,n}(k,k') \ta_m(k)^\dagger \ta_n(k')\right) + \ldots \end{multline}
where $\lambda_0, c_m(k), \mu_{m,n}(k) \in \mathbb{C}$. The integrals over $k$ and $k'$ run over all the momenta $k,k' \in [-\pi,\pi[$. Note that we only displayed the operators that are non-trivial within the single-particle subspace. Indeed, in higher-order terms there appear products of the form $a_{m_1}(k_1)\ldots a_{m_n}(k_n)$ or of the form $a_{m_1}^\dagger (k_1)\ldots a_{m_n}^\dagger (k_n)$ for $n \geq 2$ and, as such, these operators become trivial when projected onto the single-particle subspace. As we have only MPS approximations for the ground state and the single-particle excitations we need to restrict ourselves to the terms that are displayed in Eq.~(\ref{eq:halpha0app}). Physically this means that we ignore the contributions of multi-particle eigenstates  of $\mathcal{H}_\alpha$. 

Because $\mathcal{H}_{\alpha_0}$ is Hermitian, $\mu_{m,n}$ should also be a Hermitian operator:
$$\mu_{m,n}(k,k') = \overline{\mu_{n,m}(k',k)}. $$ 
Using the ground state $\ket{\umps{A(1)A(2)}}$ and the single-particle excitations $\ket{\tanv{k}{B^{(m,k)}}{A(1)A(2)}}$ of $\mathcal{H}_{\alpha}$ it follows from Eq.~(\ref{eq:creationoperator}) that 
$$ \lambda_0 =  \bra{\umps{\overline{A(1)A(2)}}} \mathcal{H}_{\alpha_0} \ket{\umps{A(1)A(2)}}. $$ 
As the energy is only determined up to a constant we can renormalize $\mathcal{H}_{\alpha_0}$ such that 
$$ \lambda_0 =  \bra{\umps{\overline{A(1)A(2)}}} \mathcal{H}_{\alpha_0} \ket{\umps{A(1)A(2)}} = 0. $$ 
With this convention, it follows from Eq.~(\ref{eq:creationoperator}) that we can compute the coefficients $\mu_{m,n}$ and $c_m$: 
\begin{subequations}\label{eq3}
\begin{multline} \mu_{m,n}(k,k') = \\ \frac{1}{2\pi}\braket{\tanv{k}{\overline{B^{(m,k)}}}{\overline{A(1)A(2)}} \vert \mathcal{H}_{\alpha_0} \vert \tanv{k'}{B^{(n,k')}}{A(1)A(2)}}\end{multline}
\begin{multline} c_m(k) = \\ \frac{1}{\sqrt{2\pi}} \braket{\umps{\overline{A(1)A(2)}} \vert \mathcal{H}_{\alpha_0} \vert \tanv{k}{B^{(m,k)}}{A(1)A(2)}}\end{multline}
\end{subequations}
and as the states are normalized according to Eq.~(\ref{eq:normSP}), it follows from Eq.~(\ref{eq:overlapSP}) that:
\begin{subequations}\label{eq:overlaphalpha0}
\be c_m(k) = \sqrt{2\pi} \delta(k) H_{eff}^2[\overline{A(1)A(2)},B^{(m,k)}]\ee
\be \mu_{m,n}(k,k') = \delta(k-k')  H_{eff}^1[\overline{B^{(m,k)}},B^{(n,k')}]\ee
\end{subequations}
where $H_{eff}^1[\overline{B^{(m,k')}},B^{(n,k)}]$ and $H_{eff}^2[\overline{A(1)A(2)},B^{(m,k)}]$ are finite quantities that we can compute efficiently (see \cite{Haegeman2013} for the details).

Using Eq.~(\ref{eq3}) and (\ref{eq:overlaphalpha0}) we rewrite $\mathcal{H}_{\alpha_0}$, Eq.~(\ref{eq:halpha0app}), now as
\begin{multline} \mathcal{H}_{\alpha_0} = \int dk \; \left(\sum_m c_m(k) \ta_m(k) + \sum_m \bar{c}_m(k) \ta_m^\dagger (k)\right. \\ 
\left.+ \sum_{m,n} M_{m,n}(k) \ta_m^\dagger (k) \ta_n(k)  \right) \end{multline}
where
\begin{subequations}\label{eq5}
\be M_{m,n}(k) =  H_{eff}^1[\overline{B^{(m,k)}},B^{(n,k)}],\ee
\be c_m(k) = \sqrt{2\pi} H_{eff}^2[\overline{A(1)A(2)},B^{(m,0)}]\delta(k).\ee
\end{subequations}
$\mathcal{H}_{\alpha_0}$ is now diagonalized by the following transformations:
$$  \tb_r(k) = \sum_{m}\left( U_{r,m}(k)\ta_m(k) + \frac{U_{r,m}(k)}{\mathcal{E}_{r}(k)}\bar{c}_m(k) \right)$$
where $U(k)$ is the unitary transformation which diagonalize $M(k)$ and $\mathcal{E}(k)$ is the diagonal matrix containing the eigenvalues of $M(k)$, i.e. $M(k) = U(k)^\dagger \mathcal{E}(k)U(k)$. In vector notation we can write this transformation as
\be\label{BT}\vec{\tb}(k) = U(k)\vec{\ta}(k) + \mathcal{E}^{-1}(k)U(k)\vec{\bar{c}}(k) \ee 
or
$$ \vec{\ta}(k) = U^\dagger(k) \vec{\tb}(k) - U^\dagger(k)\mathcal{E}^{-1}(k)U(k)\vec{\bar{c}}(k) . $$
One easily verifies now that 
\begin{multline} \mathcal{H}_{\alpha_0} = \int dk\; \left(\sum_{r}\mathcal{E}_r(k)\tb_r^\dagger(k)\tb_r(k) \right. \\
\left.- \sum_{m,n}[M^{-1}]_{m,n}(k)c_m(k)\bar{c}_n(k)\right).\end{multline}
Some remarks are in order here
\begin{itemize}
\item[i.] The last term in $\mathcal{H}_{\alpha_0}$ is a constant (divergent) term and can be omitted. This therm is only necessary if we are doing computations in the eigenbasis of $\mathcal{H}_{\alpha}$ because it is this term that assures us that
$$\braket{\umps{\overline{A(1)A(2)}} \vert \mathcal{H}_{\alpha} \vert \umps{A(1)A(2)}} = 0.$$
\item[ii.] In the Hamiltonian $\mathcal{H}_{\alpha_0}$ there appear terms of the form $c_m(k)c_n(k)$ which is ill-defined as $c_m(k) \varpropto \delta(k)$. One can regularize this by replacing the Dirac-functions by $\delta(k) \rightarrow \delta_{k,0}2N/(2\pi)$  and the $dk$ by $dk \rightarrow 2\pi/2N$ ($2N$ the number of sites on the lattice, $2N \rightarrow + \infty$). 
\item[iii.] $\mathcal{E}_r(k)$ should be positive, otherwise the quadratic expansion of $\mathcal{H}_{\alpha_0}$ in the creation and annihilation operators $a_n^\dagger(k)$ and $\ta_n(k)$ is certainly not a valid approximation anymore.
\end{itemize}

Now we have diagonalized $\mathcal{H}_{\alpha_0}$, the ground state $\ket{\Psi(0)}$ of $\mathcal{H}_{\alpha_0}$ is found as the state for which
\begin{subequations}\label{eq:vacHalpha0}
\be \tb_r(k) \ket{\Psi(0)} = 0, \forall k \in [-\pi, \pi[ \mbox{ and } \forall r,\ee
or 
\be \ta_m(k) \ket{\Psi(0)} = d_m(k)\ket{\Psi(0)}\ee
where
\be d_m(k) = - \sum_r [M(k)^{-1}]_{m,r}\bar{c}_r(k)\ee
\end{subequations}
as follows from Eq.~(\ref{BT}). Note that if $k \neq 0$ that $d_m(k) = 0$, so for non-zero momenta (in this approach) $\mathcal{H}_{\alpha_0}$ and $\mathcal{H}_{\alpha}$ have the same vacuum. This can be interpreted as the fact that a translation invariant quench cannot create particles with non-zero momentum out of the vacuum. Again, $d_m(k)$ involves a Delta-dirac distribution,
\be \label{eq:defdprimem} d_m(k) = \delta(k)d'_m, d'_m \in \mathbb{C} \ee 
which can be regularized as in Eq.~(\ref{eq:diracreg}). In order that the approximation Eq.~(\ref{eq:halpha0app}) remains valid we must have that $\vert d'_m \vert^2 \ll \vert d'_m \vert$, i.e. that $\vert d'_m \vert  \ll 1$.

Note that Eq.~(\ref{eq:vacHalpha0}) implies that $\ket{\Psi(0)}$ is a coherent state, i.e. an eigenvalue of $\ta_m(k)$. This is only possible for non-zero $d'_m$ if the creation and annihilation operators obey boson statistics. This means that within our approximation the single-particle excitations must behave as bosons, see Eq.~(\ref{eq:commRelA}). In this approximation the vacuum $\ket{\Psi(0)}$ of $\mathcal{H}_{\alpha_0}$ is interpreted as the vacuum of $\mathcal{H}_\alpha$ with on top of it a small density of zero-momentum single-particle excitations. This number of single-particles per site can be computed and equals
$$\frac{1}{N}\int dk \braket{\Psi(0) \vert \ta_m^\dagger (k) \ta_m(k) \vert \Psi(0)} = \frac{1}{2\pi} \sum_m \vert d'_m \vert^2 $$
where $N \rightarrow + \infty$ is the number of sites and we regularized $dk = 2\pi/N$ and the Dirac-delta distribution according to Eq.~(\ref{eq:diracreg}). 

\begin{figure}[t]
\null\hfill
\begin{subfigure}[b]{.24\textwidth}
\includegraphics[width=\textwidth]{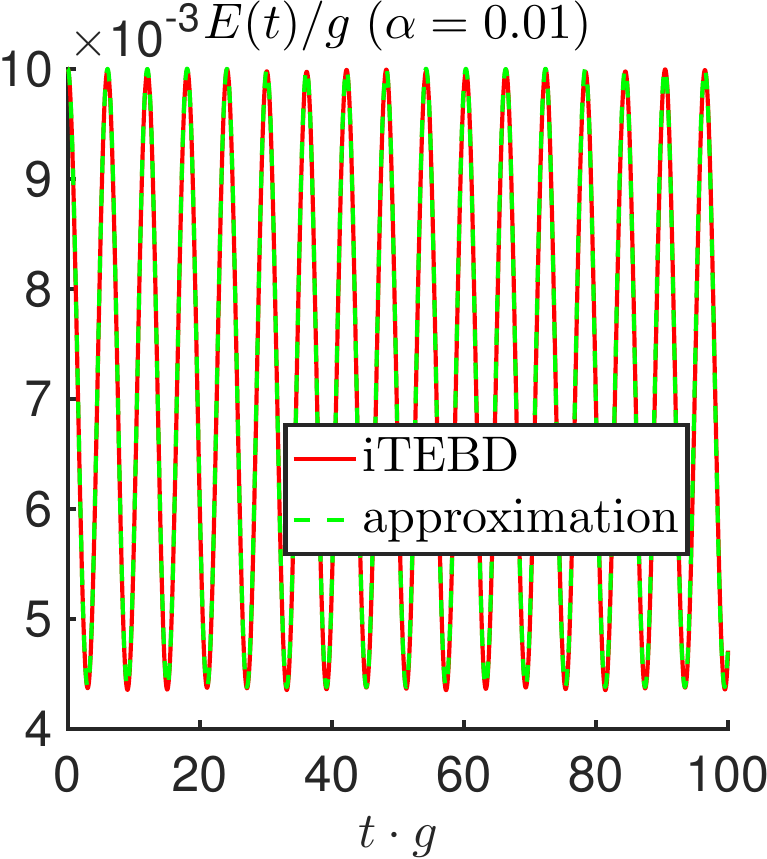}
\caption{\label{fig:appLRTEFa}}
\end{subfigure}\hfill
\begin{subfigure}[b]{.24\textwidth}
\includegraphics[width=\textwidth]{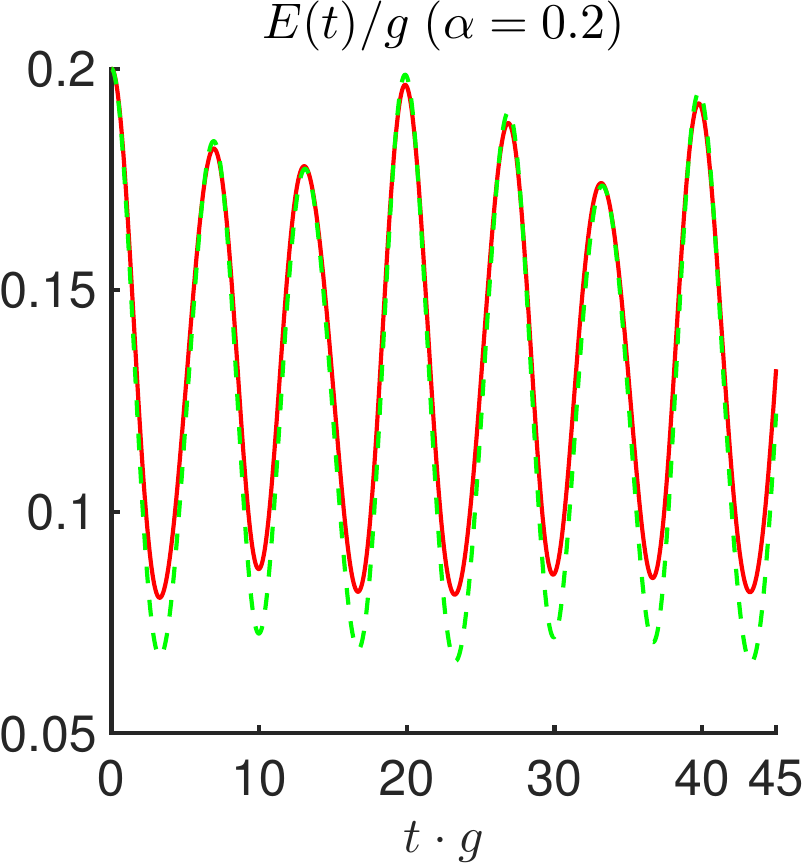}
\caption{\label{fig:appLRTEFb}}
\end{subfigure}
\hfill\null
\vskip\baselineskip
\null\hfill
\begin{subfigure}[b]{.24\textwidth}
\includegraphics[width=\textwidth]{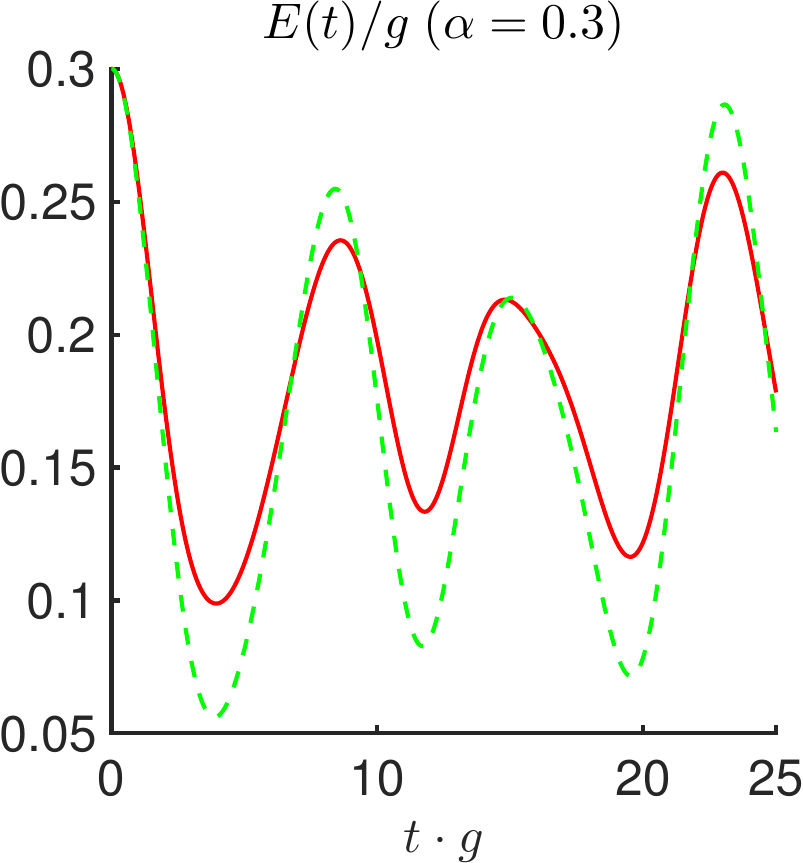}
\caption{\label{fig:appLRTEFc}}
\end{subfigure}\hfill
\begin{subfigure}[b]{.24\textwidth}
\includegraphics[width=\textwidth]{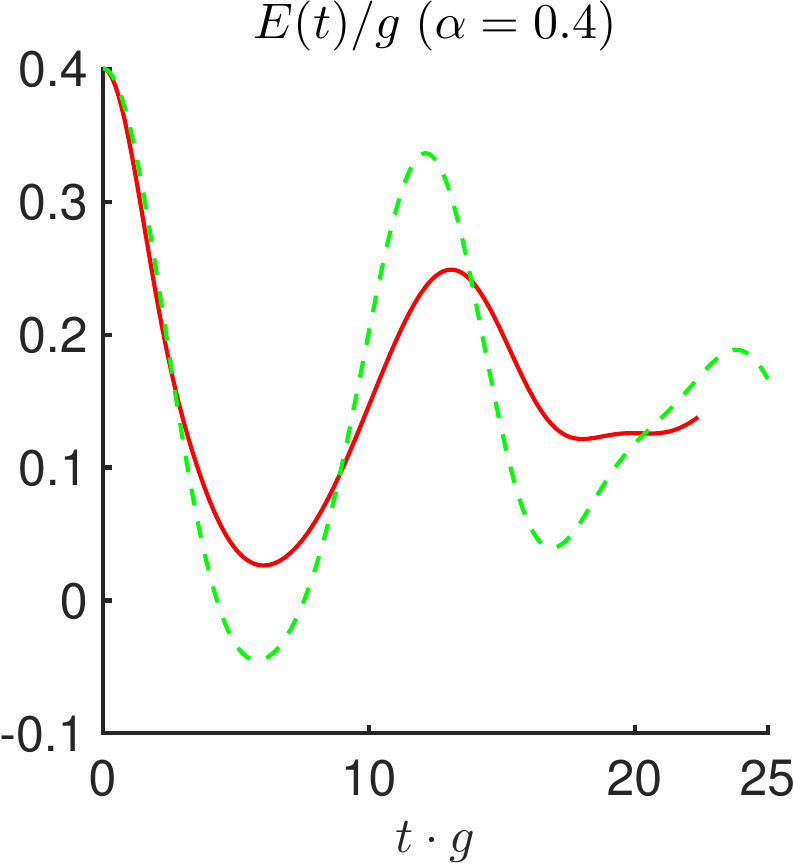}
\caption{\label{fig:appLRTEFd}}
\end{subfigure}
\hfill\null
\vskip\baselineskip
\captionsetup{justification=raggedright}
\caption{\label{fig:appLRTEF} $m/g= 0.25, x = 100$. Comparison of iTEBD simulations (full line) with the approximation Eq.~(\ref{eq:appcoherenstateAppb}) (dashed line) for the electric field $E(t)$. (a): $\alpha = 0.01$. (b): $\alpha = 0.2$. (c) $\alpha = 0.3$. (d): $\alpha = 0.4$.}
\end{figure}

Assume now we want to compute expectation values with respect to $\ket{\Psi(0)}$ of a translation invariant observable 
$$\mathcal{O} = \sum_{n = 1}^{2N} \mathcal{T}^{n-1} o \mathcal{T}^{-n+1}$$ 
where $o$ has only support on sites $1$ and $2$. Then we expand this operator similar as $\mathcal{H}_{\alpha_0}$ quadratically in the annihilation and creation operators of $\mathcal{H}_{\alpha}$:
\begin{multline} \mathcal{O} \approx \int dk\;\left( \sum_m o_{2,m}(k) \ta_m(k) + \bar{o}_{2,m}(k)\ta_m^\dagger(k)\right)  \\  
+ \int dk \int dk'\;\left(\sum_{m,n}o_{1,m,n}(k,k') \ta_m^\dagger(k)\ta_n(k') \right)  \end{multline}
where we renormalized $\mathcal{O}$ such that $\braket{\umps{\overline{A(1)A(2)}} \vert \mathcal{O}\vert\umps{A(1)A(2)}} = 0$. The coefficients can be extracted similar to Eq.~(\ref{eq:overlaphalpha0}):
\begin{multline} o_{1,m,n}(k,k')  \\  
=  \frac{1}{2\pi}\bra{\tanv{k}{\overline{B^{(m,k)}}}{\overline{A(1)A(2)}}} \mathcal{O} \ket{\tanv{k'}{B^{(n,k')}}{A(1)A(2)}} \\ 
 =  \delta(k-k')  O_{eff}^1[\overline{B^{(m,k)}},B^{(n,k')}] \end{multline}
\begin{multline}  o_{2,m}(k)  = \\ 
\frac{1}{\sqrt{2\pi}} \braket{\umps{\overline{A(1)A(2)}} \vert \mathcal{O} \vert \tanv{k}{B^{(m,k)}}{A(1)A(2)}} \\ 
 = \sqrt{2\pi} \delta(k) O_{eff}^2[\overline{A(1)A(2)},B^{(m,k)}] \end{multline}
where $O_{eff}^1$ and $O_{eff}^2$ are finite quantities which we can compute efficiently.

Hence, we find
\begin{multline} \label{eq:opHeisenbergPict} \mathcal{O}  \approx \sum_m \left(o_{2,m}\ta_m(0) + \bar{o}_{2,m}\ta_m^\dagger(0)\right) \\ 
+ \int dk \left(\sum_{m,n}o_{1,m,n}(k) \ta_m^\dagger(k)\ta_n(k) \right)  \end{multline}
with 
\begin{subequations} 
\be o_{1,m,n}(k) =  O_{eff}^1[\overline{B^{(m,k)}},B^{(n,k)}] \ee
and 
\be o_{2,m} = \sqrt{2\pi} O_{eff}^2[\overline{A(1)A(2)},B^{(m,0)}].\ee
\end{subequations}

\begin{figure}[t]
\null\hfill
\begin{subfigure}[b]{.24\textwidth}
\includegraphics[width=\textwidth]{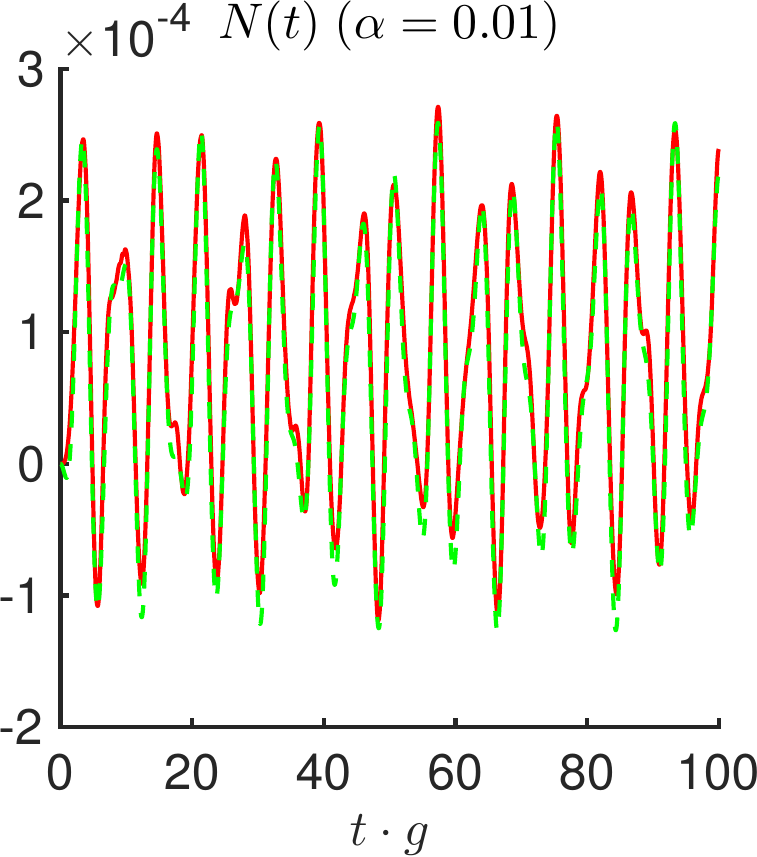}
\caption{\label{fig:appLRTCCa}}
\end{subfigure}\hfill
\begin{subfigure}[b]{.24\textwidth}
\includegraphics[width=\textwidth]{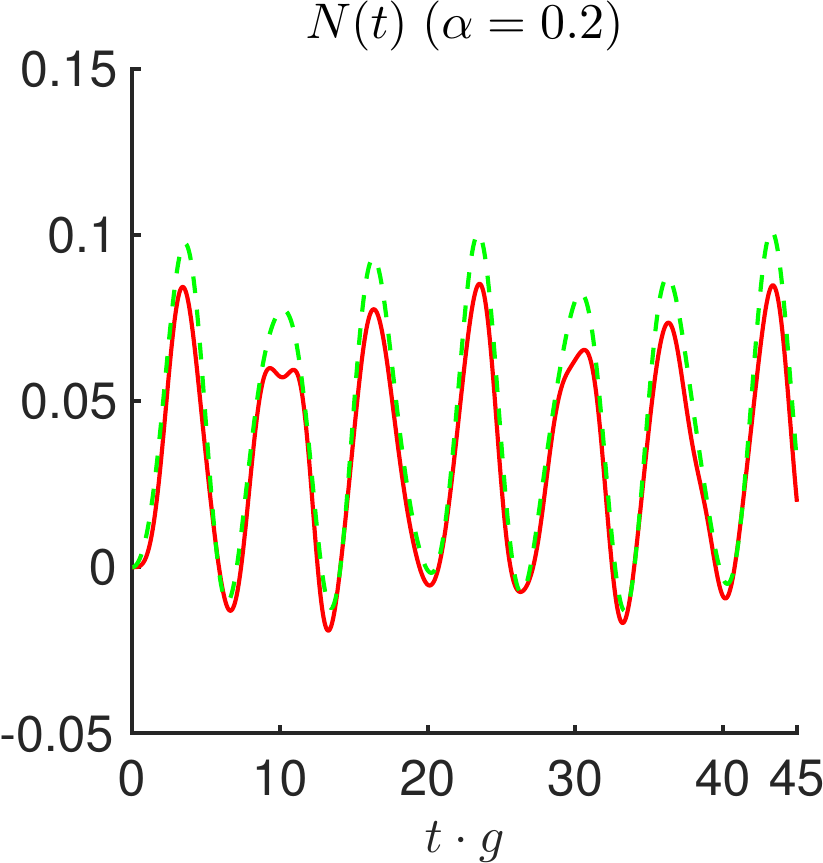}
\caption{\label{fig:appLRTCCb}}
\end{subfigure}
\hfill\null
\vskip\baselineskip
\null\hfill
\begin{subfigure}[b]{.24\textwidth}
\includegraphics[width=\textwidth]{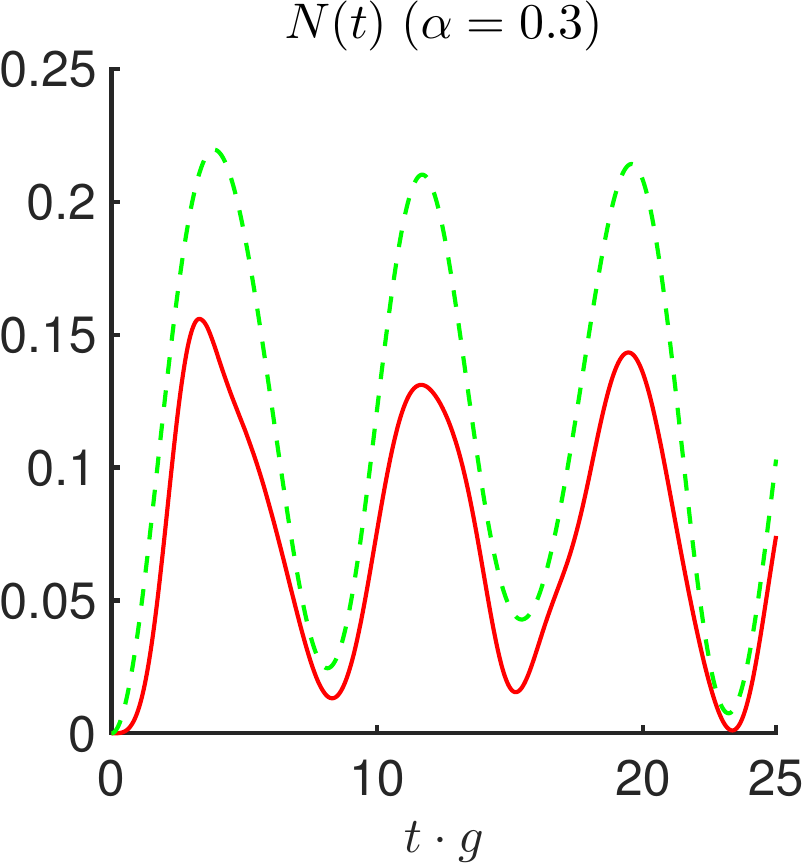}
\caption{\label{fig:appLRTCCc}}
\end{subfigure}\hfill
\begin{subfigure}[b]{.24\textwidth}
\includegraphics[width=\textwidth]{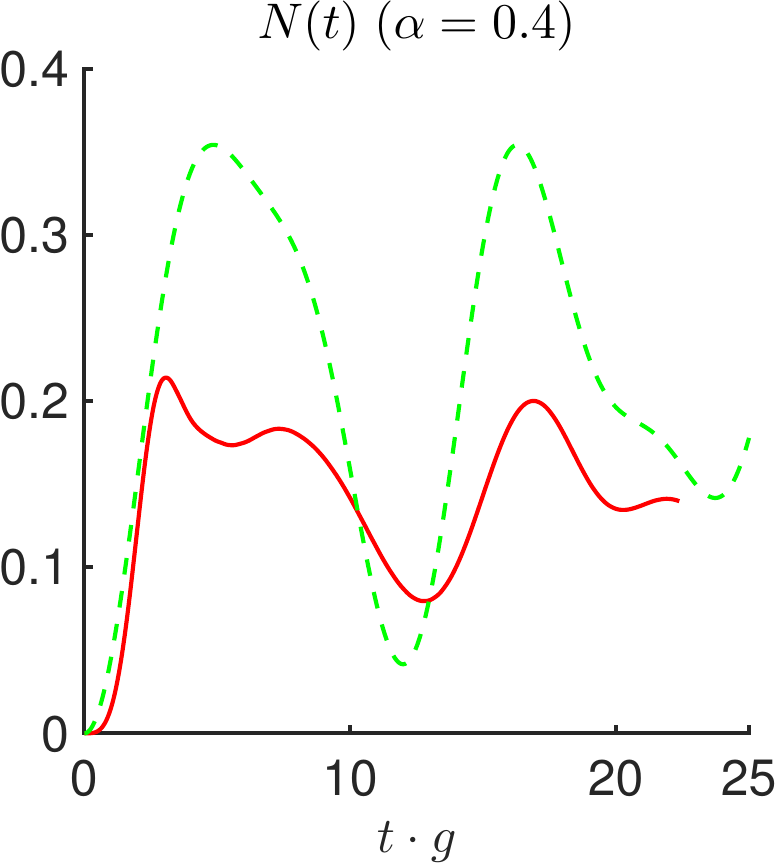}
\caption{\label{fig:appLRTCCd}}
\end{subfigure}
\hfill\null
\vskip\baselineskip
\captionsetup{justification=raggedright}
\caption{\label{fig:appLRTCC} $m/g= 0.25, x = 100$. Comparison of iTEBD simulations (full line) with the approximation Eq.~(\ref{eq:appcoherenstateAppb}) (dashed line) for $N(t)$. (a): $\alpha = 0.01$. (b): $\alpha = 0.2$. (c) $\alpha = 0.3$. (d): $\alpha = 0.4$.}
\end{figure}

To perform real-time evolution with $\mathcal{H}_{\alpha}$ we will work in the Heisenberg picture. The creation- and annihilation operator $\ta_m^\dagger(k)$ and $\ta_m(k)$ satisfy the following differential equation
\begin{subequations}\label{eq:evolveAlinresp}
\be\dot{\ta}_m(k) = i[\mathcal{H}_{\alpha},\ta_m(k)],  \ee 
\be \dot{\ta}_m^\dagger(k) = i[\mathcal{H}_{\alpha},\ta_m^\dagger(k)] .\ee
\end{subequations}
If we restrict the Hilbert space to the vacuum and the single-particle excitations we find that
\begin{subequations}
\be [\mathcal{H}_{\alpha},\ta_m(k)] = -\mathcal{E}_m(k)\ta_m(k) ,\ee 
\be [\mathcal{H}_{\alpha},\ta_m^\dagger (k)] = \mathcal{E}_m(k)\ta_m^\dagger (k). \ee
\end{subequations}

It follows that within this approximation:
$$\ta_m(k,t) = e^{-i\mathcal{E}_m(k) t}\ta_m(k)\mbox{ and } \ta_m^\dagger(k,t) = e^{i\mathcal{E}_m(k) t}\ta_m^\dagger(k).$$
In the Heisenberg picture Eq.~(\ref{eq:opHeisenbergPict}) becomes
\begin{multline}  \mathcal{O}(t) = \sum_m \left(o_{2,m}\ta_m(0,t) + \bar{o}_{2,m}\ta_m^\dagger(0,t)\right) \\ + \int dk \left(\sum_{m,n}o_{1,m,n}(k) \ta_m^\dagger(k,t)a_n(k,t) \right) \end{multline}
and the expectation value with respect to $\ket{\Psi(0)}$, the vacuum of $\mathcal{H}_{\alpha_0}$, see Eq.~(\ref{eq:vacHalpha0}), then reads
\begin{multline}  \braket{\Psi(0) \vert \mathcal{O}(t) \vert \Psi(0)} = 
\\ \sum_m o_{2,m} d_m(0) e^{-i\mathcal{E}_m(0) t} 
+ \sum_m \bar{o}_{2,m} \bar{d}_m(0) e^{i\mathcal{E}_m(0) t} \\ 
+ \int dk \left(\sum_{m,n}o_{1,m,n}(k)e^{i(\mathcal{E}_m(k) - \mathcal{E}_n(k))t}\bar{d}_m(k){d}_n(k)\right) \end{multline}
where we used eqs. (\ref{eq:vacHalpha0}) and (\ref{eq:evolveAlinresp}). As already noted before, $d_m(k)$ involves a delta-Dirac contribution: $d_m(k) = \delta(k)d'_m$. The expression $\braket{0 \vert \mathcal{O}(t) \vert 0} $ is regularized by $\delta(k) \rightarrow \delta_{k,0}2N/(2\pi)$ and $dk = 2\pi/2N$. This yields the following results:
\begin{multline} \braket{\Psi(0) \vert \mathcal{O}(t) \vert \Psi(0)} = \\ 
\frac{2N}{2\pi}\left[\sum_m o_{2,m} d'_m e^{-i\mathcal{E}_m(0) t} + \sum_m \bar{o}_{2,m} \bar{d}'_m e^{i\mathcal{E}_m(0) t} \right. \\
\left. +  \left(\sum_{m,n}o_{1,m,n}(0)e^{i(\mathcal{E}_m(0) - \mathcal{E}_n(0))t}\bar{d}'_m{d}'_n\right)\right].\end{multline}
Because $\mathcal{O} = \sum_{n = 1}^{2N-1} T^{n-1} o T^{-n+1}$, $\braket{0 \vert \mathcal{O}(t) \vert 0}$ will scale with the number of lattice sites ($2N$). It follows that
\begin{multline} \label{eq:appcoherenstateAppb}  \frac{1}{2N} \braket{\Psi(0) \vert \mathcal{O}(t) \vert \Psi(0)} = 
\\ \frac{1}{2\pi}\left[\sum_m o_{2,m} d'_m e^{-i\mathcal{E}_m(0) t} + \sum_m \bar{o}_{2,m} \bar{d}'_m e^{i\mathcal{E}_m(0) t}\right. 
\\ \left.+  \left(\sum_{m,n}o_{1,m,n}(0)e^{i(\mathcal{E}_m(0) - \mathcal{E}_n(0))t}\bar{d}'_m{d}'_n\right)\right] \end{multline}
is the expectation value per site and is finite.
\\
\\Within this approximation all coefficients appearing above can be computed from the MPS approximations $\ket{\umps{A(1)A(2)}}$ and $\ket{\tanv{k}{B^{(m,k)}}{A(1)A(2)}}$ for the ground state and the single-particle excitations of $\mathcal{H}_\alpha$. In our case for $m/g = 0.25$ and $x = 100$ we have for the values of $\alpha$ we considered here two single-particle excitations. Hence, the sum over $m$ runs from 1 to 2. We expect the above approximation to be true as long as the contribution of the multi-particle excitations of $\mathcal{H}_\alpha$ is negligible. Physically this means that the ground state $\ket{\Psi(0)}$ of $\mathcal{H}_{\alpha_0}$ is a coherent state of the creation and annihilation operators of $\mathcal{H}_\alpha$. This can be interpreted as the fact that $\ket{\Psi(0)}$ is constructed from the ground state of $\mathcal{H}_{\alpha}$ with a small density of single-particles of $\mathcal{H}_\alpha$ on top of it. We can indeed expect that for small values of $\alpha$ and at early times that this is the case. In Figs.~\ref{fig:appLRTEF} and \ref{fig:appLRTCC} we compare the real-time simulations with iTEBD (full line) with this approximation Eq.~(\ref{eq:appcoherenstateAppb}) and find agreement for $\alpha \lesssim 0.2$ while for $\alpha = 0.4$ the difference between both results is quit large. A discussion is provided in Sec.~\ref{subsec:weakfieldRegime}.

\subsection{Predicting the asymptotic thermal values of real-time evolution}\label{subsec:determineGibbsState}
\noindent In \cite{Buyens2016} we succeeded to approximate the Gibbs state $\rho(\beta)$ at temperature $T = 1/\beta$ by using Matrix Product Operators (MPO) with
$$\rho(\beta) = \frac{\mathcal{P}e^{-\beta \mathcal{H}_\alpha}}{\mbox{tr}\left(\mathcal{P}e^{-\beta \mathcal{H}_\alpha}\right)} $$
where $\mathcal{P}$ is the orthogonal projector onto the $(G(n) = 0)$-subspace. If the state $\ket{\Psi(t)} = e^{- i\mathcal{H}_\alpha t}\ket{\Psi(0)}$ would eventually equilibrate to a Gibbs state then we can estimate its inverse temperature $\beta_0$ from the requirement that 
$$\braket{\Psi(0) \vert \mathcal{H}_{\alpha} \vert \Psi(0)} = \frac{\mbox{tr}\left(\mathcal{H}_{\alpha} \mathcal{P}e^{-\beta_0 \mathcal{H}_\alpha}\right)} {\mbox{tr}\left( \mathcal{P}e^{-\beta_0 \mathcal{H}_\alpha}\right)}, $$
as follows from energy conservation during real-time evolution. 

\begin{figure}[t]
\null\hfill
\begin{subfigure}[b]{.24\textwidth}
\includegraphics[width=\textwidth]{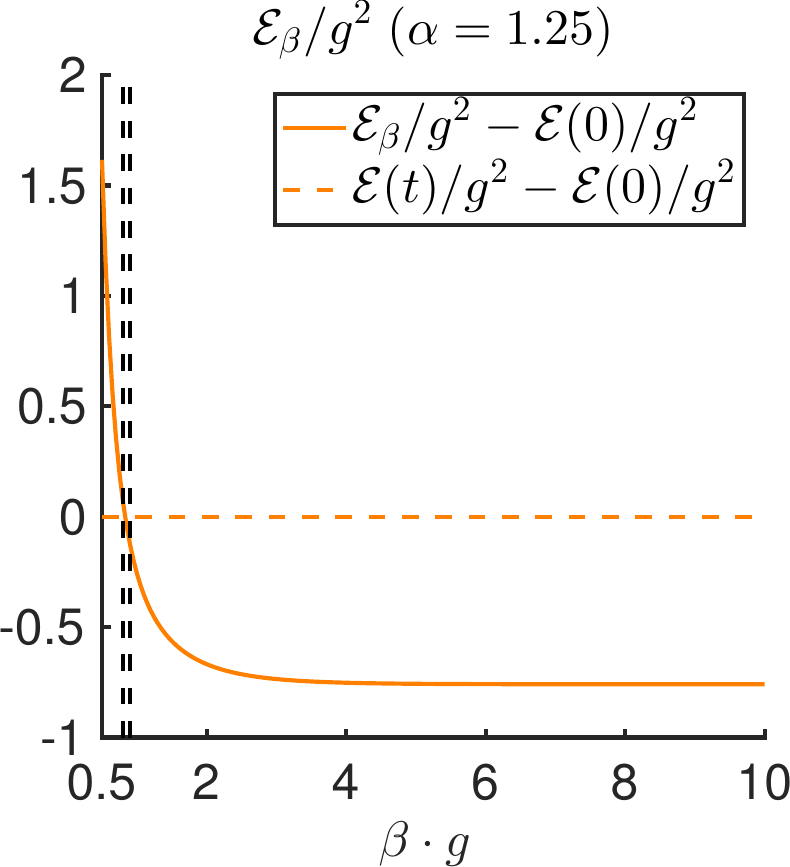}
\caption{\label{fig:FindTempa}}
\end{subfigure}\hfill
\begin{subfigure}[b]{.24\textwidth}
\includegraphics[width=\textwidth]{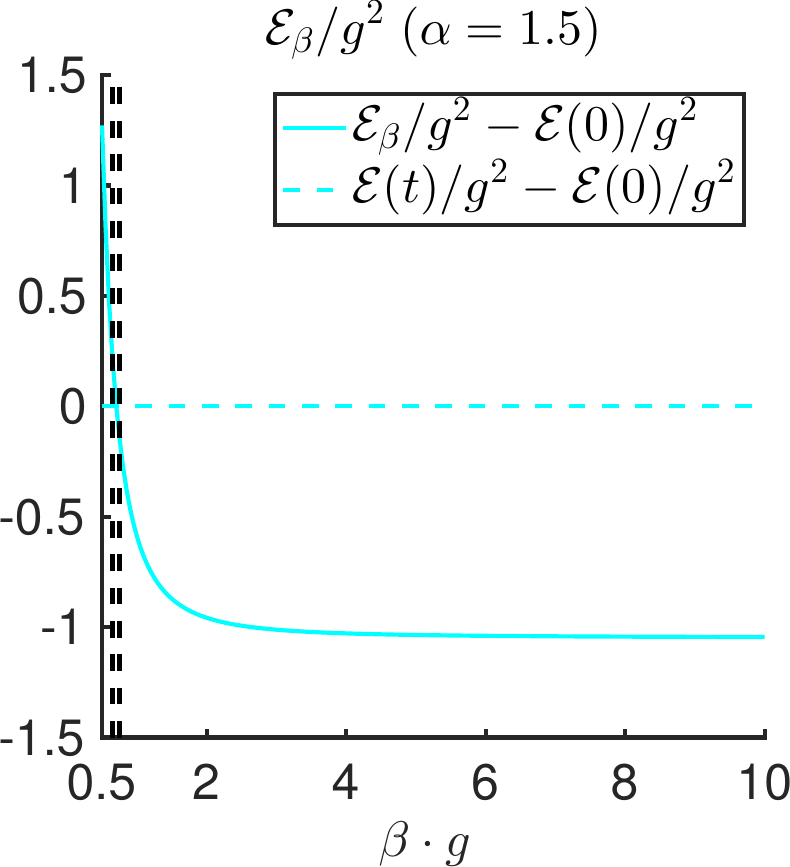}
\caption{\label{fig:FindTempb}}
\end{subfigure}\hfill\null
\vskip\baselineskip
\captionsetup{justification=raggedright}
\caption{\label{fig:findTemp} Results for $m/g= 0.25, x = 100$. Determination of the temperature of the asymptotic state in thermal equilibrium by finding the intersection of the conserved energy $\mathcal{E}_t $(dashed line) with the energy of the Gibbs state $\mathcal{E}_\beta$ (full line). (a): $\alpha = 1.25$. (b): $\alpha = 1.5$. }
\end{figure}

In Figs.~\ref{fig:FindTempa} and \ref{fig:FindTempb} we show the energy per unit of length $\mathcal{E}_\beta$ of the Gibbs state $\rho(\beta)$ as a function of $\beta$ and the (conserved) energy per unit of length $\mathcal{E}(t)$ of the state $\ket{\Psi(t)}$. We subtracted from both quantities the energy per unit of length of $\ket{\Psi(0)}$. The intersection between the curves determines the value of $\beta_0$. Because we simulated the thermal evolution with steps $d\beta = 0.05$ we can only determine $\beta_0$ up to $0.05/g$. For $\alpha = 0.75$ we find $\beta_0g = 1.35 (\pm 0.05)$, $\alpha = 1.25$ we find $\beta_0g = 0.85 (\pm 0.05)$ and for $\alpha = 1.5$ we find $\beta_0 g = 0.70 (\pm 0.05)$.

\section{Scaling to the continuum limit of the real-time results}\label{appsec:continuum}
In this paper we consider the following quantities:
\begin{itemize}
\item[(a)] The electric field:
\begin{multline} \label{eq:ElectricFieldRT} E(t) = \braket{\Psi(t) \vert E \vert \Psi(t) } \\ 
= \frac{1}{2N}\sum_{n=1}^{2N}\braket{\Psi(t) \vert L(n) + \alpha \vert \Psi(t) }\end{multline}
\item[(b)] The current:
\begin{multline} \label{eq:CurrentRT}  j^1(t)  = \braket{\Psi(t) \vert j^1 \vert \Psi(t) } \\
= \frac{-i\sqrt{x}g}{2N}\sum_{n = 1}^{2N}\Braket{\Psi(t) \vert \sigma^+(n)e^{i\theta(n)}\sigma^{-}(n+1) - h.c. \vert\Psi(t)},\end{multline}
\item[(c)] The particle number $N(t) = (\Sigma(t) - \Sigma(0))/g$ with 
\begin{multline} \Sigma(t) = \braket{\Psi(t) \vert \bar{\psi}(0)\psi(0) \vert\Psi(t)} \\ 
= \Braket{\Psi(t) \left\vert g\frac{\sqrt{x}}{2N}\sum_{n=1}^{2N} \frac{\sigma_z(n) + (-1)^n}{2}\right\vert \Psi(t)} \end{multline}
which counts in the weak coupling limit ($m/g \gg 1$) the number of electrons and positrons per unit of length that are created out of the vacuum or destroyed in the vacuum due to turning on the electric background field $\alpha$ at $t = 0$.
\item[(d)] From the Schmidt spectrum $\{\lambda_{\alpha_q}^q\}$ associated to a cut between an even and an odd site, we can compute the half chain entanglement entropy
\be S = -\sum_{q\in \mathbb{Z}} \sum_{\alpha = 1}^{D_q}\lambda_{\alpha_q}^q \log\left(\lambda_{\alpha_q}^q\right). \ee 
As we will show below, a UV quantity is obtained by considering the renormalized half chain entanglement entropy $$\Delta S(t) = S(t) - S(0). $$
\end{itemize}

\begin{figure}
\null\hfill
\begin{subfigure}[b]{.24\textwidth}
\includegraphics[width=\textwidth]{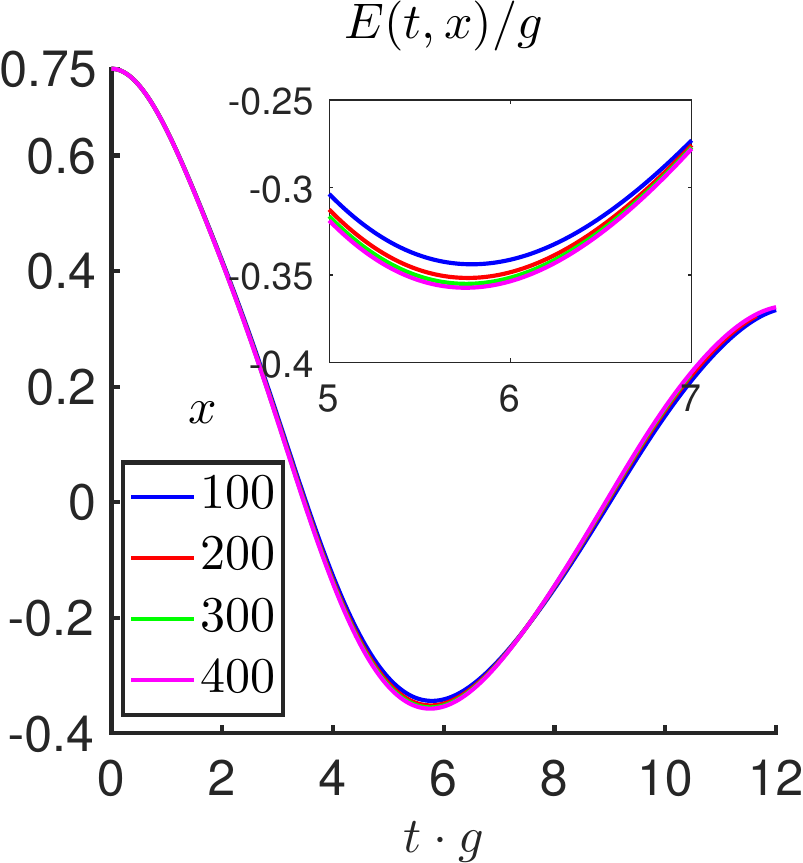}
\caption{\label{fig:RTEFScaling}}
\end{subfigure}\hfill
\begin{subfigure}[b]{.24\textwidth}
\includegraphics[width=\textwidth]{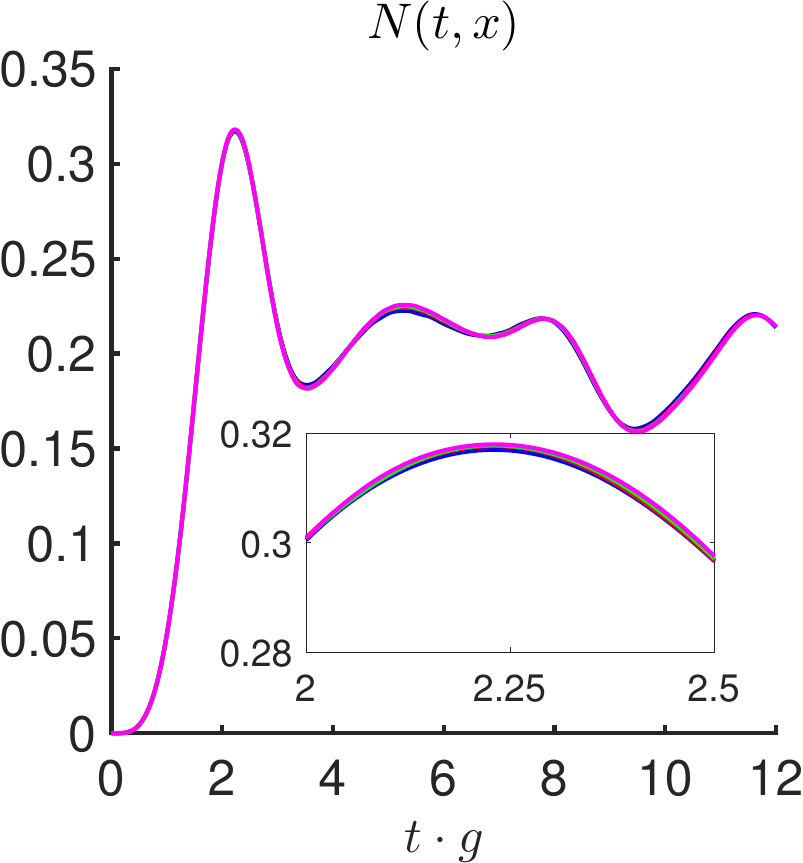}
\caption{\label{fig:RTCCScaling}}
\end{subfigure}\hfill\null
\vskip\baselineskip
\null\hfill
\begin{subfigure}[b]{.24\textwidth}
\includegraphics[width=\textwidth]{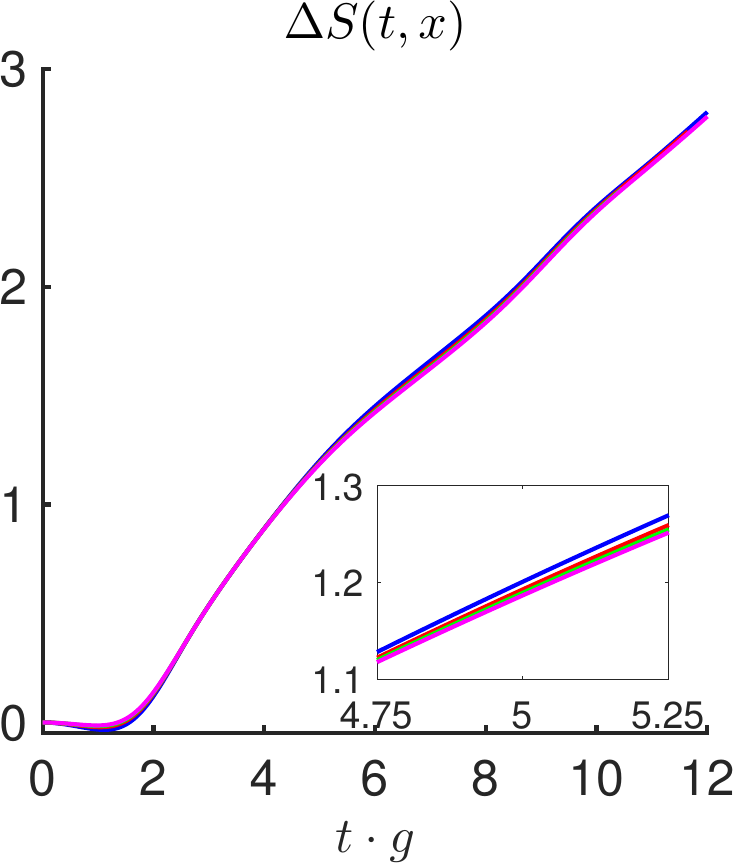}
\caption{\label{fig:RTEntropyScaling}}
\end{subfigure}\hfill
\begin{subfigure}[b]{.24\textwidth}
\includegraphics[width=\textwidth]{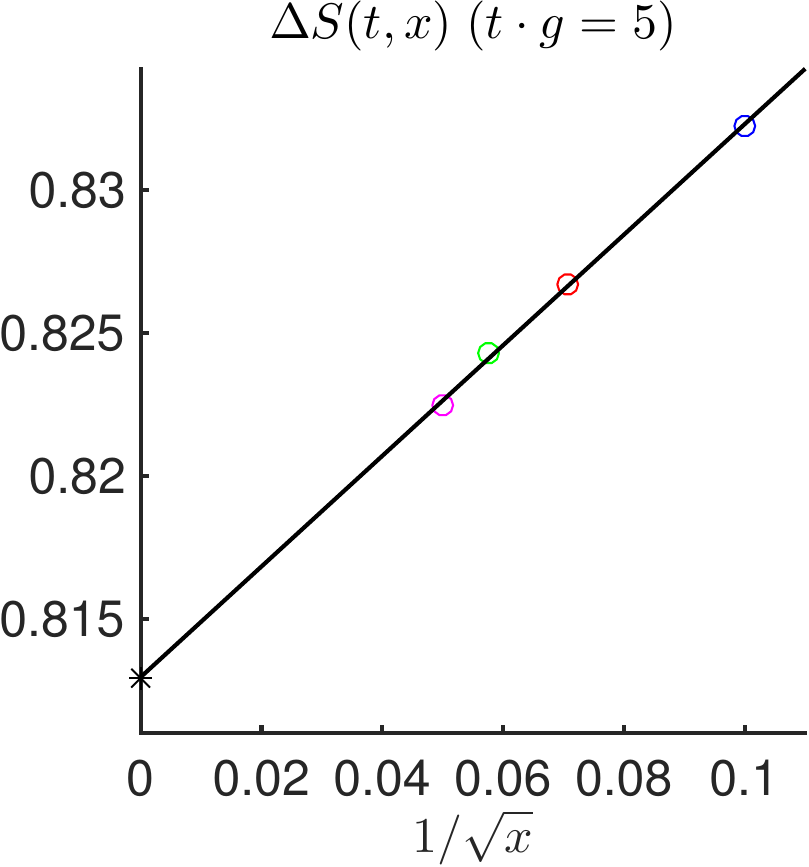}
\caption{\label{fig:RTEntropyContExtr}}
\end{subfigure}\hfill\null
\vskip\baselineskip
\captionsetup{justification=raggedright}
\caption{\label{fig:RTscaling} $m/g= 0.25, \alpha = 0.75$. Scaling of the quantities to $x \rightarrow + \infty$. (a) Electric field $E(t,x)$. (b) Particle number $N(t,x)$. (c) Renormalized entropy $\Delta S(t,x)$. (d) Polynomial extrapolation in $1/\sqrt{x}$ of the renormalized entropy to $x \rightarrow + \infty$. }
\end{figure}

The fact that these quantities are UV finite is corroborated by Fig.~\ref{fig:RTscaling} where we show the evolution of the electric field $E(t,x)$, the particle number $N(t,x)$ and the renormalized entropy $\Delta S(t,x)$ as a function of time for $x = 100,$ $200,$ $300,$ $400$. Note that we here explicitly denote the $x-$dependence of the quantities. We observe that for all these quantities the graphs are almost on top of each other, see Figs.~\ref{fig:RTscaling} (a) - (c). One can also obtain a continuum estimate for these quantities by a polynomial extrapolation, see Fig.~\ref{fig:RTEntropyContExtr} where we perform a polynomial extrapolation for $\Delta S(t)$ for $tg = 5$. It is also clear from this example that we can already expect at $x = 100$ to be close to the continuum limit. (For the current $j^1(t)$ this follows from Amp\`{e}re's law: $\dot{E} = - gj^1$.) This justifies that we restrict ourselves to $x = 100$ for the discussion on the continuum results. . 

\newpage

\bibliography{PaperRT}

\begin{thebibliography}{109}
\expandafter\ifx\csname natexlab\endcsname\relax\def\natexlab#1{#1}\fi
\expandafter\ifx\csname bibnamefont\endcsname\relax
  \def\bibnamefont#1{#1}\fi
\expandafter\ifx\csname bibfnamefont\endcsname\relax
  \def\bibfnamefont#1{#1}\fi
\expandafter\ifx\csname citenamefont\endcsname\relax
  \def\citenamefont#1{#1}\fi
\expandafter\ifx\csname url\endcsname\relax
  \def\url#1{\texttt{#1}}\fi
\expandafter\ifx\csname urlprefix\endcsname\relax\def\urlprefix{URL }\fi
\providecommand{\bibinfo}[2]{#2}
\providecommand{\eprint}[2][]{\url{#2}}

\bibitem[{\citenamefont{{Benton} et~al.}(2016)\citenamefont{{Benton},
  {Jaubert}, {Yan}, and {Shannon}}}]{Benton2016}
\bibinfo{author}{\bibfnamefont{O.}~\bibnamefont{{Benton}}},
  \bibinfo{author}{\bibfnamefont{L.~D.~C.} \bibnamefont{{Jaubert}}},
  \bibinfo{author}{\bibfnamefont{H.}~\bibnamefont{{Yan}}}, \bibnamefont{and}
  \bibinfo{author}{\bibfnamefont{N.}~\bibnamefont{{Shannon}}},
  \bibinfo{journal}{Nature Communications} \textbf{\bibinfo{volume}{7}},
  \bibinfo{eid}{11572} (\bibinfo{year}{2016}), \eprint{1510.01007}.

\bibitem[{\citenamefont{{Bali}}(1999)}]{Bali1999}
\bibinfo{author}{\bibfnamefont{G.~S.} \bibnamefont{{Bali}}},
  \bibinfo{journal}{Fizika B} \textbf{\bibinfo{volume}{8}},
  \bibinfo{pages}{229} (\bibinfo{year}{1999}), \eprint{hep-lat/9901023}.

\bibitem[{\citenamefont{{Or{\'u}s}}(2014)}]{Orus2004}
\bibinfo{author}{\bibfnamefont{R.}~\bibnamefont{{Or{\'u}s}}},
  \bibinfo{journal}{Annals of Physics} \textbf{\bibinfo{volume}{349}},
  \bibinfo{pages}{117} (\bibinfo{year}{2014}), \eprint{1306.2164}.

\bibitem[{\citenamefont{{Verstraete} and {Cirac}}(2004)}]{Verstraete2004}
\bibinfo{author}{\bibfnamefont{F.}~\bibnamefont{{Verstraete}}}
  \bibnamefont{and} \bibinfo{author}{\bibfnamefont{J.~I.}
  \bibnamefont{{Cirac}}}, \bibinfo{journal}{eprint arXiv:cond-mat/0407066}
  (\bibinfo{year}{2004}), \eprint{cond-mat/0407066}.

\bibitem[{\citenamefont{{Verstraete} et~al.}(2008)\citenamefont{{Verstraete},
  {Murg}, and {Cirac}}}]{Verstraete2008}
\bibinfo{author}{\bibfnamefont{F.}~\bibnamefont{{Verstraete}}},
  \bibinfo{author}{\bibfnamefont{V.}~\bibnamefont{{Murg}}}, \bibnamefont{and}
  \bibinfo{author}{\bibfnamefont{J.~I.} \bibnamefont{{Cirac}}},
  \bibinfo{journal}{Advances in Physics} \textbf{\bibinfo{volume}{57}},
  \bibinfo{pages}{143} (\bibinfo{year}{2008}), \eprint{0907.2796}.

\bibitem[{\citenamefont{{Schollw{\"o}ck}}(2011)}]{Schollwoeck2011}
\bibinfo{author}{\bibfnamefont{U.}~\bibnamefont{{Schollw{\"o}ck}}},
  \bibinfo{journal}{Annals of Physics} \textbf{\bibinfo{volume}{326}},
  \bibinfo{pages}{96} (\bibinfo{year}{2011}), \eprint{1008.3477}.

\bibitem[{\citenamefont{White}(1992)}]{White1992}
\bibinfo{author}{\bibfnamefont{S.~R.} \bibnamefont{White}},
  \bibinfo{journal}{Phys. Rev. Lett.} \textbf{\bibinfo{volume}{69}},
  \bibinfo{pages}{2863} (\bibinfo{year}{1992}).

\bibitem[{\citenamefont{{Ba{\~n}uls} et~al.}(2013)\citenamefont{{Ba{\~n}uls},
  {Cichy}, {Cirac}, and {Jansen}}}]{Banuls2013a}
\bibinfo{author}{\bibfnamefont{M.~C.} \bibnamefont{{Ba{\~n}uls}}},
  \bibinfo{author}{\bibfnamefont{K.}~\bibnamefont{{Cichy}}},
  \bibinfo{author}{\bibfnamefont{J.~I.} \bibnamefont{{Cirac}}},
  \bibnamefont{and} \bibinfo{author}{\bibfnamefont{K.}~\bibnamefont{{Jansen}}},
  \bibinfo{journal}{Journal of High Energy Physics}
  \textbf{\bibinfo{volume}{11}}, \bibinfo{eid}{158} (\bibinfo{year}{2013}),
  \eprint{1305.3765}.

\bibitem[{\citenamefont{{Ba{\~n}uls}
  et~al.}(2016{\natexlab{a}})\citenamefont{{Ba{\~n}uls}, {Cichy}, {Jansen}, and
  {Saito}}}]{Banuls2016a}
\bibinfo{author}{\bibfnamefont{M.~C.} \bibnamefont{{Ba{\~n}uls}}},
  \bibinfo{author}{\bibfnamefont{K.}~\bibnamefont{{Cichy}}},
  \bibinfo{author}{\bibfnamefont{K.}~\bibnamefont{{Jansen}}}, \bibnamefont{and}
  \bibinfo{author}{\bibfnamefont{H.}~\bibnamefont{{Saito}}},
  \bibinfo{journal}{\prd} \textbf{\bibinfo{volume}{93}}, \bibinfo{eid}{094512}
  (\bibinfo{year}{2016}{\natexlab{a}}), \eprint{1603.05002}.

\bibitem[{\citenamefont{{Ba{\~n}uls}
  et~al.}(2016{\natexlab{b}})\citenamefont{{Ba{\~n}uls}, {Cichy}, {Cirac},
  {Jansen}, and {K{\"u}hn}}}]{Banuls2016b}
\bibinfo{author}{\bibfnamefont{M.~C.} \bibnamefont{{Ba{\~n}uls}}},
  \bibinfo{author}{\bibfnamefont{K.}~\bibnamefont{{Cichy}}},
  \bibinfo{author}{\bibfnamefont{J.~I.} \bibnamefont{{Cirac}}},
  \bibinfo{author}{\bibfnamefont{K.}~\bibnamefont{{Jansen}}}, \bibnamefont{and}
  \bibinfo{author}{\bibfnamefont{S.}~\bibnamefont{{K{\"u}hn}}},
  \bibinfo{journal}{ArXiv e-prints}  (\bibinfo{year}{2016}{\natexlab{b}}),
  \eprint{1611.00705}.

\bibitem[{\citenamefont{{Ba{\~n}uls}
  et~al.}(2016{\natexlab{c}})\citenamefont{{Ba{\~n}uls}, {Cichy}, {Cirac},
  {Jansen}, {K{\"u}hn}, and {Saito}}}]{Banuls2016c}
\bibinfo{author}{\bibfnamefont{M.~C.} \bibnamefont{{Ba{\~n}uls}}},
  \bibinfo{author}{\bibfnamefont{K.}~\bibnamefont{{Cichy}}},
  \bibinfo{author}{\bibfnamefont{J.~I.} \bibnamefont{{Cirac}}},
  \bibinfo{author}{\bibfnamefont{K.}~\bibnamefont{{Jansen}}},
  \bibinfo{author}{\bibfnamefont{S.}~\bibnamefont{{K{\"u}hn}}},
  \bibnamefont{and} \bibinfo{author}{\bibfnamefont{H.}~\bibnamefont{{Saito}}},
  \bibinfo{journal}{ArXiv e-prints}  (\bibinfo{year}{2016}{\natexlab{c}}),
  \eprint{1611.01458}.

\bibitem[{\citenamefont{{Ba{\~n}uls}
  et~al.}(2016{\natexlab{d}})\citenamefont{{Ba{\~n}uls}, {Cichy}, {Cirac},
  {Jansen}, {K{\"u}hn}, and {Saito}}}]{Banuls2016d}
\bibinfo{author}{\bibfnamefont{M.~C.} \bibnamefont{{Ba{\~n}uls}}},
  \bibinfo{author}{\bibfnamefont{K.}~\bibnamefont{{Cichy}}},
  \bibinfo{author}{\bibfnamefont{J.~I.} \bibnamefont{{Cirac}}},
  \bibinfo{author}{\bibfnamefont{K.}~\bibnamefont{{Jansen}}},
  \bibinfo{author}{\bibfnamefont{S.}~\bibnamefont{{K{\"u}hn}}},
  \bibnamefont{and} \bibinfo{author}{\bibfnamefont{H.}~\bibnamefont{{Saito}}},
  \bibinfo{journal}{ArXiv e-prints}  (\bibinfo{year}{2016}{\natexlab{d}}),
  \eprint{1611.04791}.

\bibitem[{\citenamefont{{Byrnes}
  et~al.}(2002{\natexlab{a}})\citenamefont{{Byrnes}, {Sriganesh}, {Bursill},
  and {Hamer}}}]{Byrnes2003a}
\bibinfo{author}{\bibfnamefont{T.~M.~R.} \bibnamefont{{Byrnes}}},
  \bibinfo{author}{\bibfnamefont{P.}~\bibnamefont{{Sriganesh}}},
  \bibinfo{author}{\bibfnamefont{R.~J.} \bibnamefont{{Bursill}}},
  \bibnamefont{and} \bibinfo{author}{\bibfnamefont{C.~J.}
  \bibnamefont{{Hamer}}}, \bibinfo{journal}{Nuclear Physics B Proceedings
  Supplements} \textbf{\bibinfo{volume}{109}}, \bibinfo{pages}{202}
  (\bibinfo{year}{2002}{\natexlab{a}}), \eprint{hep-lat/0201007}.

\bibitem[{\citenamefont{{Byrnes}
  et~al.}(2002{\natexlab{b}})\citenamefont{{Byrnes}, {Sriganesh}, {Bursill},
  and {Hamer}}}]{Byrnes2003b}
\bibinfo{author}{\bibfnamefont{T.~M.} \bibnamefont{{Byrnes}}},
  \bibinfo{author}{\bibfnamefont{P.}~\bibnamefont{{Sriganesh}}},
  \bibinfo{author}{\bibfnamefont{R.~J.} \bibnamefont{{Bursill}}},
  \bibnamefont{and} \bibinfo{author}{\bibfnamefont{C.~J.}
  \bibnamefont{{Hamer}}}, \bibinfo{journal}{\prd}
  \textbf{\bibinfo{volume}{66}}, \bibinfo{eid}{013002}
  (\bibinfo{year}{2002}{\natexlab{b}}), \eprint{hep-lat/0202014}.

\bibitem[{\citenamefont{{Sugihara}}(2005)}]{Sugihara2005}
\bibinfo{author}{\bibfnamefont{T.}~\bibnamefont{{Sugihara}}},
  \bibinfo{journal}{Journal of High Energy Physics}
  \textbf{\bibinfo{volume}{7}}, \bibinfo{eid}{022} (\bibinfo{year}{2005}),
  \eprint{hep-lat/0506009}.

\bibitem[{\citenamefont{{Rico} et~al.}(2014)\citenamefont{{Rico}, {Pichler},
  {Dalmonte}, {Zoller}, and {Montangero}}}]{Rico2014}
\bibinfo{author}{\bibfnamefont{E.}~\bibnamefont{{Rico}}},
  \bibinfo{author}{\bibfnamefont{T.}~\bibnamefont{{Pichler}}},
  \bibinfo{author}{\bibfnamefont{M.}~\bibnamefont{{Dalmonte}}},
  \bibinfo{author}{\bibfnamefont{P.}~\bibnamefont{{Zoller}}}, \bibnamefont{and}
  \bibinfo{author}{\bibfnamefont{S.}~\bibnamefont{{Montangero}}},
  \bibinfo{journal}{Physical Review Letters} \textbf{\bibinfo{volume}{112}},
  \bibinfo{eid}{201601} (\bibinfo{year}{2014}), \eprint{1312.3127}.

\bibitem[{\citenamefont{{K{\"u}hn} et~al.}(2014)\citenamefont{{K{\"u}hn},
  {Cirac}, and {Ba{\~n}uls}}}]{Kuehn2014}
\bibinfo{author}{\bibfnamefont{S.}~\bibnamefont{{K{\"u}hn}}},
  \bibinfo{author}{\bibfnamefont{J.~I.} \bibnamefont{{Cirac}}},
  \bibnamefont{and} \bibinfo{author}{\bibfnamefont{M.-C.}
  \bibnamefont{{Ba{\~n}uls}}}, \bibinfo{journal}{\pra}
  \textbf{\bibinfo{volume}{90}}, \bibinfo{eid}{042305} (\bibinfo{year}{2014}),
  \eprint{1407.4995}.

\bibitem[{\citenamefont{{K{\"u}hn} et~al.}(2015)\citenamefont{{K{\"u}hn},
  {Zohar}, {Cirac}, and {Ba{\~n}uls}}}]{Kuehn2015}
\bibinfo{author}{\bibfnamefont{S.}~\bibnamefont{{K{\"u}hn}}},
  \bibinfo{author}{\bibfnamefont{E.}~\bibnamefont{{Zohar}}},
  \bibinfo{author}{\bibfnamefont{J.~I.} \bibnamefont{{Cirac}}},
  \bibnamefont{and} \bibinfo{author}{\bibfnamefont{M.~C.}
  \bibnamefont{{Ba{\~n}uls}}}, \bibinfo{journal}{Journal of High Energy
  Physics} \textbf{\bibinfo{volume}{7}}, \bibinfo{eid}{130}
  (\bibinfo{year}{2015}), \eprint{1505.04441}.

\bibitem[{\citenamefont{{Silvi} et~al.}(2016)\citenamefont{{Silvi}, {Rico},
  {Dalmonte}, {Tschirsich}, and {Montangero}}}]{Silvi2016}
\bibinfo{author}{\bibfnamefont{P.}~\bibnamefont{{Silvi}}},
  \bibinfo{author}{\bibfnamefont{E.}~\bibnamefont{{Rico}}},
  \bibinfo{author}{\bibfnamefont{M.}~\bibnamefont{{Dalmonte}}},
  \bibinfo{author}{\bibfnamefont{F.}~\bibnamefont{{Tschirsich}}},
  \bibnamefont{and}
  \bibinfo{author}{\bibfnamefont{S.}~\bibnamefont{{Montangero}}},
  \bibinfo{journal}{ArXiv e-prints}  (\bibinfo{year}{2016}),
  \eprint{1606.05510}.

\bibitem[{\citenamefont{{Milsted}}(2016)}]{Milsted2015}
\bibinfo{author}{\bibfnamefont{A.}~\bibnamefont{{Milsted}}},
  \bibinfo{journal}{\prd} \textbf{\bibinfo{volume}{93}}, \bibinfo{eid}{085012}
  (\bibinfo{year}{2016}), \eprint{1507.06624}.

\bibitem[{\citenamefont{Schwinger}(1962)}]{Schwinger1962}
\bibinfo{author}{\bibfnamefont{J.}~\bibnamefont{Schwinger}},
  \bibinfo{journal}{Phys. Rev.} \textbf{\bibinfo{volume}{128}},
  \bibinfo{pages}{2425} (\bibinfo{year}{1962}).

\bibitem[{\citenamefont{Coleman et~al.}(1975)\citenamefont{Coleman, Jackiw, and
  Susskind}}]{Coleman1975}
\bibinfo{author}{\bibfnamefont{S.}~\bibnamefont{Coleman}},
  \bibinfo{author}{\bibfnamefont{R.}~\bibnamefont{Jackiw}}, \bibnamefont{and}
  \bibinfo{author}{\bibfnamefont{L.}~\bibnamefont{Susskind}},
  \bibinfo{journal}{Annals of Physics} \textbf{\bibinfo{volume}{93}},
  \bibinfo{pages}{267 } (\bibinfo{year}{1975}).

\bibitem[{\citenamefont{Coleman}(1976)}]{Coleman1976}
\bibinfo{author}{\bibfnamefont{S.}~\bibnamefont{Coleman}},
  \bibinfo{journal}{Annals of Physics} \textbf{\bibinfo{volume}{101}},
  \bibinfo{pages}{239 } (\bibinfo{year}{1976}).

\bibitem[{\citenamefont{Hamer et~al.}(1982)\citenamefont{Hamer, Kogut,
  Crewther, and Mazzolini}}]{Hamer1982}
\bibinfo{author}{\bibfnamefont{C.}~\bibnamefont{Hamer}},
  \bibinfo{author}{\bibfnamefont{J.}~\bibnamefont{Kogut}},
  \bibinfo{author}{\bibfnamefont{D.}~\bibnamefont{Crewther}}, \bibnamefont{and}
  \bibinfo{author}{\bibfnamefont{M.}~\bibnamefont{Mazzolini}},
  \bibinfo{journal}{Nuclear Physics B} \textbf{\bibinfo{volume}{208}},
  \bibinfo{pages}{413 } (\bibinfo{year}{1982}).

\bibitem[{\citenamefont{Iso and Murayama}(1990)}]{Iso1990}
\bibinfo{author}{\bibfnamefont{S.}~\bibnamefont{Iso}} \bibnamefont{and}
  \bibinfo{author}{\bibfnamefont{H.}~\bibnamefont{Murayama}},
  \bibinfo{journal}{Progress of Theoretical Physics}
  \textbf{\bibinfo{volume}{84}}, \bibinfo{pages}{142} (\bibinfo{year}{1990}).

\bibitem[{\citenamefont{{Hosotani} and
  {Rodr{\'{\i}}guez}}(1996)}]{Hosotani1996}
\bibinfo{author}{\bibfnamefont{Y.}~\bibnamefont{{Hosotani}}} \bibnamefont{and}
  \bibinfo{author}{\bibfnamefont{R.}~\bibnamefont{{Rodr{\'{\i}}guez}}},
  \bibinfo{journal}{Physics Letters B} \textbf{\bibinfo{volume}{389}},
  \bibinfo{pages}{121} (\bibinfo{year}{1996}), \eprint{hep-th/9608176}.

\bibitem[{\citenamefont{{Adam}}(1997)}]{Adam1997}
\bibinfo{author}{\bibfnamefont{C.}~\bibnamefont{{Adam}}},
  \bibinfo{journal}{Annals of Physics} \textbf{\bibinfo{volume}{259}},
  \bibinfo{pages}{1} (\bibinfo{year}{1997}), \eprint{hep-th/9704064}.

\bibitem[{\citenamefont{{Cichy} et~al.}(2013)\citenamefont{{Cichy},
  {Kujawa-Cichy}, and {Szyniszewski}}}]{Cichy2012}
\bibinfo{author}{\bibfnamefont{K.}~\bibnamefont{{Cichy}}},
  \bibinfo{author}{\bibfnamefont{A.}~\bibnamefont{{Kujawa-Cichy}}},
  \bibnamefont{and}
  \bibinfo{author}{\bibfnamefont{M.}~\bibnamefont{{Szyniszewski}}},
  \bibinfo{journal}{Computer Physics Communications}
  \textbf{\bibinfo{volume}{184}}, \bibinfo{pages}{1666} (\bibinfo{year}{2013}),
  \eprint{1211.6393}.

\bibitem[{\citenamefont{{Hebenstreit}
  et~al.}(2013{\natexlab{a}})\citenamefont{{Hebenstreit}, {Berges}, and
  {Gelfand}}}]{Hebenstreit2013}
\bibinfo{author}{\bibfnamefont{F.}~\bibnamefont{{Hebenstreit}}},
  \bibinfo{author}{\bibfnamefont{J.}~\bibnamefont{{Berges}}}, \bibnamefont{and}
  \bibinfo{author}{\bibfnamefont{D.}~\bibnamefont{{Gelfand}}},
  \bibinfo{journal}{\prd} \textbf{\bibinfo{volume}{87}}, \bibinfo{eid}{105006}
  (\bibinfo{year}{2013}{\natexlab{a}}), \eprint{1302.5537}.

\bibitem[{\citenamefont{{Hebenstreit} and {Berges}}(2014)}]{Hebenstreit2014}
\bibinfo{author}{\bibfnamefont{F.}~\bibnamefont{{Hebenstreit}}}
  \bibnamefont{and} \bibinfo{author}{\bibfnamefont{J.}~\bibnamefont{{Berges}}},
  \bibinfo{journal}{\prd} \textbf{\bibinfo{volume}{90}}, \bibinfo{eid}{045034}
  (\bibinfo{year}{2014}), \eprint{1406.4273}.

\bibitem[{\citenamefont{{Buyens}
  et~al.}(2014{\natexlab{a}})\citenamefont{{Buyens}, {Haegeman}, {Van
  Acoleyen}, {Verschelde}, and {Verstraete}}}]{Buyens2013}
\bibinfo{author}{\bibfnamefont{B.}~\bibnamefont{{Buyens}}},
  \bibinfo{author}{\bibfnamefont{J.}~\bibnamefont{{Haegeman}}},
  \bibinfo{author}{\bibfnamefont{K.}~\bibnamefont{{Van Acoleyen}}},
  \bibinfo{author}{\bibfnamefont{H.}~\bibnamefont{{Verschelde}}},
  \bibnamefont{and}
  \bibinfo{author}{\bibfnamefont{F.}~\bibnamefont{{Verstraete}}},
  \bibinfo{journal}{Physical Review Letters} \textbf{\bibinfo{volume}{113}},
  \bibinfo{eid}{091601} (\bibinfo{year}{2014}{\natexlab{a}}),
  \eprint{1312.6654}.

\bibitem[{\citenamefont{{Buyens}
  et~al.}(2014{\natexlab{b}})\citenamefont{{Buyens}, {Van Acoleyen},
  {Haegeman}, and {Verstraete}}}]{Buyens2014}
\bibinfo{author}{\bibfnamefont{B.}~\bibnamefont{{Buyens}}},
  \bibinfo{author}{\bibfnamefont{K.}~\bibnamefont{{Van Acoleyen}}},
  \bibinfo{author}{\bibfnamefont{J.}~\bibnamefont{{Haegeman}}},
  \bibnamefont{and}
  \bibinfo{author}{\bibfnamefont{F.}~\bibnamefont{{Verstraete}}},
  \bibinfo{journal}{ArXiv e-prints}  (\bibinfo{year}{2014}{\natexlab{b}}),
  \eprint{1411.0020}.

\bibitem[{\citenamefont{{Buyens}
  et~al.}(2016{\natexlab{a}})\citenamefont{{Buyens}, {Haegeman}, {Verschelde},
  {Verstraete}, and {Van Acoleyen}}}]{Buyens2015}
\bibinfo{author}{\bibfnamefont{B.}~\bibnamefont{{Buyens}}},
  \bibinfo{author}{\bibfnamefont{J.}~\bibnamefont{{Haegeman}}},
  \bibinfo{author}{\bibfnamefont{H.}~\bibnamefont{{Verschelde}}},
  \bibinfo{author}{\bibfnamefont{F.}~\bibnamefont{{Verstraete}}},
  \bibnamefont{and} \bibinfo{author}{\bibfnamefont{K.}~\bibnamefont{{Van
  Acoleyen}}}, \bibinfo{journal}{Phys. Rev. X} \textbf{\bibinfo{volume}{6}},
  \bibinfo{pages}{041040} (\bibinfo{year}{2016}{\natexlab{a}}),
  \eprint{1509.00246}.

\bibitem[{\citenamefont{{Buyens} et~al.}(2015)\citenamefont{{Buyens},
  {Haegeman}, {Verstraete}, and {Van Acoleyen}}}]{Buyens2015b}
\bibinfo{author}{\bibfnamefont{B.}~\bibnamefont{{Buyens}}},
  \bibinfo{author}{\bibfnamefont{J.}~\bibnamefont{{Haegeman}}},
  \bibinfo{author}{\bibfnamefont{F.}~\bibnamefont{{Verstraete}}},
  \bibnamefont{and} \bibinfo{author}{\bibfnamefont{K.}~\bibnamefont{{Van
  Acoleyen}}}, \bibinfo{journal}{ArXiv e-prints}  (\bibinfo{year}{2015}),
  \eprint{1511.04288}.

\bibitem[{\citenamefont{{Buyens}
  et~al.}(2016{\natexlab{b}})\citenamefont{{Buyens}, {Verstraete}, and {Van
  Acoleyen}}}]{Buyens2016}
\bibinfo{author}{\bibfnamefont{B.}~\bibnamefont{{Buyens}}},
  \bibinfo{author}{\bibfnamefont{F.}~\bibnamefont{{Verstraete}}},
  \bibnamefont{and} \bibinfo{author}{\bibfnamefont{K.}~\bibnamefont{{Van
  Acoleyen}}}, \bibinfo{journal}{\prd} \textbf{\bibinfo{volume}{94}},
  \bibinfo{eid}{085018} (\bibinfo{year}{2016}{\natexlab{b}}),
  \eprint{1606.03385}.

\bibitem[{\citenamefont{{Buyens} et~al.}(2017)\citenamefont{{Buyens},
  {Montangero}, {Haegeman}, {Verstraete}, and {Van Acoleyen}}}]{Buyens2017}
\bibinfo{author}{\bibfnamefont{B.}~\bibnamefont{{Buyens}}},
  \bibinfo{author}{\bibfnamefont{S.}~\bibnamefont{{Montangero}}},
  \bibinfo{author}{\bibfnamefont{J.}~\bibnamefont{{Haegeman}}},
  \bibinfo{author}{\bibfnamefont{F.}~\bibnamefont{{Verstraete}}},
  \bibnamefont{and} \bibinfo{author}{\bibfnamefont{K.}~\bibnamefont{{Van
  Acoleyen}}}, \bibinfo{journal}{\prd} \textbf{\bibinfo{volume}{95}},
  \bibinfo{eid}{094509} (\bibinfo{year}{2017}), \eprint{1702.08838}.

\bibitem[{\citenamefont{{Kharzeev} and {Loshaj}}(2013)}]{Kharzeev2013}
\bibinfo{author}{\bibfnamefont{D.~E.} \bibnamefont{{Kharzeev}}}
  \bibnamefont{and} \bibinfo{author}{\bibfnamefont{F.}~\bibnamefont{{Loshaj}}},
  \bibinfo{journal}{\prd} \textbf{\bibinfo{volume}{87}}, \bibinfo{eid}{077501}
  (\bibinfo{year}{2013}), \eprint{1212.5857}.

\bibitem[{\citenamefont{{Kharzeev} and {Loshaj}}(2014)}]{Kharzeev2014}
\bibinfo{author}{\bibfnamefont{D.~E.} \bibnamefont{{Kharzeev}}}
  \bibnamefont{and} \bibinfo{author}{\bibfnamefont{F.}~\bibnamefont{{Loshaj}}},
  \bibinfo{journal}{\prd} \textbf{\bibinfo{volume}{89}}, \bibinfo{eid}{074053}
  (\bibinfo{year}{2014}), \eprint{1308.2716}.

\bibitem[{\citenamefont{{Hebenstreit}
  et~al.}(2013{\natexlab{b}})\citenamefont{{Hebenstreit}, {Berges}, and
  {Gelfand}}}]{Hebenstreit2013a}
\bibinfo{author}{\bibfnamefont{F.}~\bibnamefont{{Hebenstreit}}},
  \bibinfo{author}{\bibfnamefont{J.}~\bibnamefont{{Berges}}}, \bibnamefont{and}
  \bibinfo{author}{\bibfnamefont{D.}~\bibnamefont{{Gelfand}}},
  \bibinfo{journal}{Physical Review Letters} \textbf{\bibinfo{volume}{111}},
  \bibinfo{eid}{201601} (\bibinfo{year}{2013}{\natexlab{b}}),
  \eprint{1307.4619}.

\bibitem[{\citenamefont{{Hauke} et~al.}(2013)\citenamefont{{Hauke}, {Marcos},
  {Dalmonte}, and {Zoller}}}]{Hauke2013}
\bibinfo{author}{\bibfnamefont{P.}~\bibnamefont{{Hauke}}},
  \bibinfo{author}{\bibfnamefont{D.}~\bibnamefont{{Marcos}}},
  \bibinfo{author}{\bibfnamefont{M.}~\bibnamefont{{Dalmonte}}},
  \bibnamefont{and} \bibinfo{author}{\bibfnamefont{P.}~\bibnamefont{{Zoller}}},
  \bibinfo{journal}{Physical Review X} \textbf{\bibinfo{volume}{3}},
  \bibinfo{eid}{041018} (\bibinfo{year}{2013}), \eprint{1306.2162}.

\bibitem[{\citenamefont{{Wiese}}(2013)}]{Wiese2013}
\bibinfo{author}{\bibfnamefont{U.-J.} \bibnamefont{{Wiese}}},
  \bibinfo{journal}{Annalen der Physik} \textbf{\bibinfo{volume}{525}},
  \bibinfo{pages}{777} (\bibinfo{year}{2013}), \eprint{1305.1602}.

\bibitem[{\citenamefont{{Martinez} et~al.}(2016)\citenamefont{{Martinez},
  {Muschik}, {Schindler}, {Nigg}, {Erhard}, {Heyl}, {Hauke}, {Dalmonte},
  {Monz}, {Zoller} et~al.}}]{Martinez2016}
\bibinfo{author}{\bibfnamefont{E.~A.} \bibnamefont{{Martinez}}},
  \bibinfo{author}{\bibfnamefont{C.~A.} \bibnamefont{{Muschik}}},
  \bibinfo{author}{\bibfnamefont{P.}~\bibnamefont{{Schindler}}},
  \bibinfo{author}{\bibfnamefont{D.}~\bibnamefont{{Nigg}}},
  \bibinfo{author}{\bibfnamefont{A.}~\bibnamefont{{Erhard}}},
  \bibinfo{author}{\bibfnamefont{M.}~\bibnamefont{{Heyl}}},
  \bibinfo{author}{\bibfnamefont{P.}~\bibnamefont{{Hauke}}},
  \bibinfo{author}{\bibfnamefont{M.}~\bibnamefont{{Dalmonte}}},
  \bibinfo{author}{\bibfnamefont{T.}~\bibnamefont{{Monz}}},
  \bibinfo{author}{\bibfnamefont{P.}~\bibnamefont{{Zoller}}},
  \bibnamefont{et~al.}, \bibinfo{journal}{\nat} \textbf{\bibinfo{volume}{534}},
  \bibinfo{pages}{516} (\bibinfo{year}{2016}), \eprint{1605.04570}.

\bibitem[{\citenamefont{{Kasper}
  et~al.}(2016{\natexlab{a}})\citenamefont{{Kasper}, {Hebenstreit},
  {Oberthaler}, and {Berges}}}]{Kasper2015}
\bibinfo{author}{\bibfnamefont{V.}~\bibnamefont{{Kasper}}},
  \bibinfo{author}{\bibfnamefont{F.}~\bibnamefont{{Hebenstreit}}},
  \bibinfo{author}{\bibfnamefont{M.~K.} \bibnamefont{{Oberthaler}}},
  \bibnamefont{and} \bibinfo{author}{\bibfnamefont{J.}~\bibnamefont{{Berges}}},
  \bibinfo{journal}{Physics Letters B} \textbf{\bibinfo{volume}{760}},
  \bibinfo{pages}{742} (\bibinfo{year}{2016}{\natexlab{a}}),
  \eprint{1506.01238}.

\bibitem[{\citenamefont{{Kasper}
  et~al.}(2016{\natexlab{b}})\citenamefont{{Kasper}, {Hebenstreit},
  {Jendrzejewski}, {Oberthaler}, and {Berges}}}]{Kasper2016}
\bibinfo{author}{\bibfnamefont{V.}~\bibnamefont{{Kasper}}},
  \bibinfo{author}{\bibfnamefont{F.}~\bibnamefont{{Hebenstreit}}},
  \bibinfo{author}{\bibfnamefont{F.}~\bibnamefont{{Jendrzejewski}}},
  \bibinfo{author}{\bibfnamefont{M.~K.} \bibnamefont{{Oberthaler}}},
  \bibnamefont{and} \bibinfo{author}{\bibfnamefont{J.}~\bibnamefont{{Berges}}},
  \bibinfo{journal}{ArXiv e-prints}  (\bibinfo{year}{2016}{\natexlab{b}}),
  \eprint{1608.03480}.

\bibitem[{\citenamefont{{Schwinger}}(1951)}]{Schwinger1951}
\bibinfo{author}{\bibfnamefont{J.}~\bibnamefont{{Schwinger}}},
  \bibinfo{journal}{Physical Review} \textbf{\bibinfo{volume}{82}},
  \bibinfo{pages}{664} (\bibinfo{year}{1951}).

\bibitem[{\citenamefont{{Schmidt} et~al.}(1998)\citenamefont{{Schmidt},
  {Blaschke}, {R{\"o}pke}, {Smolyansky}, {Prozorkevich}, and
  {Toneev}}}]{Schmidt1998}
\bibinfo{author}{\bibfnamefont{S.}~\bibnamefont{{Schmidt}}},
  \bibinfo{author}{\bibfnamefont{D.}~\bibnamefont{{Blaschke}}},
  \bibinfo{author}{\bibfnamefont{G.}~\bibnamefont{{R{\"o}pke}}},
  \bibinfo{author}{\bibfnamefont{S.~A.} \bibnamefont{{Smolyansky}}},
  \bibinfo{author}{\bibfnamefont{A.~V.} \bibnamefont{{Prozorkevich}}},
  \bibnamefont{and} \bibinfo{author}{\bibfnamefont{V.~D.}
  \bibnamefont{{Toneev}}}, \bibinfo{journal}{International Journal of Modern
  Physics E} \textbf{\bibinfo{volume}{7}}, \bibinfo{pages}{709}
  (\bibinfo{year}{1998}), \eprint{hep-ph/9809227}.

\bibitem[{\citenamefont{{Kluger} et~al.}(1992)\citenamefont{{Kluger},
  {Eisenberg}, {Svetitsky}, {Cooper}, and {Mottola}}}]{Kluger1992}
\bibinfo{author}{\bibfnamefont{Y.}~\bibnamefont{{Kluger}}},
  \bibinfo{author}{\bibfnamefont{J.~M.} \bibnamefont{{Eisenberg}}},
  \bibinfo{author}{\bibfnamefont{B.}~\bibnamefont{{Svetitsky}}},
  \bibinfo{author}{\bibfnamefont{F.}~\bibnamefont{{Cooper}}}, \bibnamefont{and}
  \bibinfo{author}{\bibfnamefont{E.}~\bibnamefont{{Mottola}}},
  \bibinfo{journal}{\prd} \textbf{\bibinfo{volume}{45}}, \bibinfo{pages}{4659}
  (\bibinfo{year}{1992}).

\bibitem[{\citenamefont{{Hebenstreit} et~al.}(2011)\citenamefont{{Hebenstreit},
  {Alkofer}, and {Gies}}}]{Hebenstreit2011}
\bibinfo{author}{\bibfnamefont{F.}~\bibnamefont{{Hebenstreit}}},
  \bibinfo{author}{\bibfnamefont{R.}~\bibnamefont{{Alkofer}}},
  \bibnamefont{and} \bibinfo{author}{\bibfnamefont{H.}~\bibnamefont{{Gies}}},
  \bibinfo{journal}{Physical Review Letters} \textbf{\bibinfo{volume}{107}},
  \bibinfo{eid}{180403} (\bibinfo{year}{2011}), \eprint{1106.6175}.

\bibitem[{\citenamefont{{Kasper} et~al.}(2014)\citenamefont{{Kasper},
  {Hebenstreit}, and {Berges}}}]{Kasper2014}
\bibinfo{author}{\bibfnamefont{V.}~\bibnamefont{{Kasper}}},
  \bibinfo{author}{\bibfnamefont{F.}~\bibnamefont{{Hebenstreit}}},
  \bibnamefont{and} \bibinfo{author}{\bibfnamefont{J.}~\bibnamefont{{Berges}}},
  \bibinfo{journal}{\prd} \textbf{\bibinfo{volume}{90}}, \bibinfo{eid}{025016}
  (\bibinfo{year}{2014}), \eprint{1403.4849}.

\bibitem[{\citenamefont{{Kogut} and {Susskind}}(1975)}]{Kogut1975}
\bibinfo{author}{\bibfnamefont{J.}~\bibnamefont{{Kogut}}} \bibnamefont{and}
  \bibinfo{author}{\bibfnamefont{L.}~\bibnamefont{{Susskind}}},
  \bibinfo{journal}{\prd} \textbf{\bibinfo{volume}{11}}, \bibinfo{pages}{395}
  (\bibinfo{year}{1975}).

\bibitem[{\citenamefont{{Banks} et~al.}(1976)\citenamefont{{Banks}, {Susskind},
  and {Kogut}}}]{Banks1976}
\bibinfo{author}{\bibfnamefont{T.}~\bibnamefont{{Banks}}},
  \bibinfo{author}{\bibfnamefont{L.}~\bibnamefont{{Susskind}}},
  \bibnamefont{and} \bibinfo{author}{\bibfnamefont{J.}~\bibnamefont{{Kogut}}},
  \bibinfo{journal}{\prd} \textbf{\bibinfo{volume}{13}}, \bibinfo{pages}{1043}
  (\bibinfo{year}{1976}).

\bibitem[{\citenamefont{{Haegeman} et~al.}(2011)\citenamefont{{Haegeman},
  {Cirac}, {Osborne}, {Pi{\v z}orn}, {Verschelde}, and
  {Verstraete}}}]{Haegeman2011}
\bibinfo{author}{\bibfnamefont{J.}~\bibnamefont{{Haegeman}}},
  \bibinfo{author}{\bibfnamefont{J.~I.} \bibnamefont{{Cirac}}},
  \bibinfo{author}{\bibfnamefont{T.~J.} \bibnamefont{{Osborne}}},
  \bibinfo{author}{\bibfnamefont{I.}~\bibnamefont{{Pi{\v z}orn}}},
  \bibinfo{author}{\bibfnamefont{H.}~\bibnamefont{{Verschelde}}},
  \bibnamefont{and}
  \bibinfo{author}{\bibfnamefont{F.}~\bibnamefont{{Verstraete}}},
  \bibinfo{journal}{Physical Review Letters} \textbf{\bibinfo{volume}{107}},
  \bibinfo{eid}{070601} (\bibinfo{year}{2011}), \eprint{1103.0936}.

\bibitem[{\citenamefont{{Haegeman} et~al.}(2013)\citenamefont{{Haegeman},
  {Osborne}, and {Verstraete}}}]{Haegeman2013}
\bibinfo{author}{\bibfnamefont{J.}~\bibnamefont{{Haegeman}}},
  \bibinfo{author}{\bibfnamefont{T.~J.} \bibnamefont{{Osborne}}},
  \bibnamefont{and}
  \bibinfo{author}{\bibfnamefont{F.}~\bibnamefont{{Verstraete}}},
  \bibinfo{journal}{\prb} \textbf{\bibinfo{volume}{88}}, \bibinfo{eid}{075133}
  (\bibinfo{year}{2013}), \eprint{1305.1894}.

\bibitem[{\citenamefont{{Haegeman} et~al.}(2016)\citenamefont{{Haegeman},
  {Lubich}, {Oseledets}, {Vandereycken}, and {Verstraete}}}]{Haegeman2014a}
\bibinfo{author}{\bibfnamefont{J.}~\bibnamefont{{Haegeman}}},
  \bibinfo{author}{\bibfnamefont{C.}~\bibnamefont{{Lubich}}},
  \bibinfo{author}{\bibfnamefont{I.}~\bibnamefont{{Oseledets}}},
  \bibinfo{author}{\bibfnamefont{B.}~\bibnamefont{{Vandereycken}}},
  \bibnamefont{and}
  \bibinfo{author}{\bibfnamefont{F.}~\bibnamefont{{Verstraete}}},
  \bibinfo{journal}{Phys. Rev. B} \textbf{\bibinfo{volume}{94}},
  \bibinfo{pages}{165116} (\bibinfo{year}{2016}).

\bibitem[{\citenamefont{{Vidal}}(2007)}]{Vidal2007}
\bibinfo{author}{\bibfnamefont{G.}~\bibnamefont{{Vidal}}},
  \bibinfo{journal}{Physical Review Letters} \textbf{\bibinfo{volume}{98}},
  \bibinfo{eid}{070201} (\bibinfo{year}{2007}), \eprint{cond-mat/0605597}.

\bibitem[{\citenamefont{{Delfino} and {Viti}}(2016)}]{Delfino2016}
\bibinfo{author}{\bibfnamefont{G.}~\bibnamefont{{Delfino}}} \bibnamefont{and}
  \bibinfo{author}{\bibfnamefont{J.}~\bibnamefont{{Viti}}},
  \bibinfo{journal}{ArXiv e-prints}  (\bibinfo{year}{2016}),
  \eprint{1608.07612}.

\bibitem[{\citenamefont{Delfino}(2014)}]{Delfino2016b}
\bibinfo{author}{\bibfnamefont{G.}~\bibnamefont{Delfino}},
  \bibinfo{journal}{Journal of Physics A: Mathematical and Theoretical}
  \textbf{\bibinfo{volume}{47}}, \bibinfo{pages}{402001}
  (\bibinfo{year}{2014}),
  \urlprefix\url{http://stacks.iop.org/1751-8121/47/i=40/a=402001}.

\bibitem[{\citenamefont{{da Silva} et~al.}(2016)\citenamefont{{da Silva},
  {Lopez}, {Mas}, and {Serantes}}}]{daSilva2016}
\bibinfo{author}{\bibfnamefont{E.}~\bibnamefont{{da Silva}}},
  \bibinfo{author}{\bibfnamefont{E.}~\bibnamefont{{Lopez}}},
  \bibinfo{author}{\bibfnamefont{J.}~\bibnamefont{{Mas}}}, \bibnamefont{and}
  \bibinfo{author}{\bibfnamefont{A.}~\bibnamefont{{Serantes}}},
  \bibinfo{journal}{ArXiv e-prints}  (\bibinfo{year}{2016}),
  \eprint{1604.08765}.

\bibitem[{\citenamefont{Kormos et~al.}(2016)\citenamefont{Kormos, Collura,
  Takacs, and Calabrese}}]{Kormos2016}
\bibinfo{author}{\bibfnamefont{M.}~\bibnamefont{Kormos}},
  \bibinfo{author}{\bibfnamefont{M.}~\bibnamefont{Collura}},
  \bibinfo{author}{\bibfnamefont{G.}~\bibnamefont{Takacs}}, \bibnamefont{and}
  \bibinfo{author}{\bibfnamefont{P.}~\bibnamefont{Calabrese}},
  \bibinfo{journal}{Nat Phys} \textbf{\bibinfo{volume}{advance online
  publication}},  (\bibinfo{year}{2016}),
  \urlprefix\url{http://dx.doi.org/10.1038/nphys3934}.

\bibitem[{\citenamefont{Rakovszky et~al.}(2016)\citenamefont{Rakovszky,
  Mestyán, Collura, Kormos, and Takács}}]{Rakovszky2016}
\bibinfo{author}{\bibfnamefont{T.}~\bibnamefont{Rakovszky}},
  \bibinfo{author}{\bibfnamefont{M.}~\bibnamefont{Mestyán}},
  \bibinfo{author}{\bibfnamefont{M.}~\bibnamefont{Collura}},
  \bibinfo{author}{\bibfnamefont{M.}~\bibnamefont{Kormos}}, \bibnamefont{and}
  \bibinfo{author}{\bibfnamefont{G.}~\bibnamefont{Takács}},
  \bibinfo{journal}{Nuclear Physics B} \textbf{\bibinfo{volume}{911}},
  \bibinfo{pages}{805 } (\bibinfo{year}{2016}), ISSN \bibinfo{issn}{0550-3213},
  \urlprefix\url{//www.sciencedirect.com/science/article/pii/S0550321316302541}.

\bibitem[{\citenamefont{{Calabrese} and {Cardy}}(2005)}]{Calabrese2005}
\bibinfo{author}{\bibfnamefont{P.}~\bibnamefont{{Calabrese}}} \bibnamefont{and}
  \bibinfo{author}{\bibfnamefont{J.}~\bibnamefont{{Cardy}}},
  \bibinfo{journal}{Journal of Statistical Mechanics: Theory and Experiment}
  \textbf{\bibinfo{volume}{4}}, \bibinfo{pages}{04010} (\bibinfo{year}{2005}),
  \eprint{cond-mat/0503393}.

\bibitem[{\citenamefont{{Linden} et~al.}(2009)\citenamefont{{Linden},
  {Popescu}, {Short}, and {Winter}}}]{Linden2009}
\bibinfo{author}{\bibfnamefont{N.}~\bibnamefont{{Linden}}},
  \bibinfo{author}{\bibfnamefont{S.}~\bibnamefont{{Popescu}}},
  \bibinfo{author}{\bibfnamefont{A.~J.} \bibnamefont{{Short}}},
  \bibnamefont{and} \bibinfo{author}{\bibfnamefont{A.}~\bibnamefont{{Winter}}},
  \bibinfo{journal}{\pre} \textbf{\bibinfo{volume}{79}}, \bibinfo{eid}{061103}
  (\bibinfo{year}{2009}), \eprint{0812.2385}.

\bibitem[{\citenamefont{{Eisert} et~al.}(2015)\citenamefont{{Eisert},
  {Friesdorf}, and {Gogolin}}}]{Eisert2015}
\bibinfo{author}{\bibfnamefont{J.}~\bibnamefont{{Eisert}}},
  \bibinfo{author}{\bibfnamefont{M.}~\bibnamefont{{Friesdorf}}},
  \bibnamefont{and}
  \bibinfo{author}{\bibfnamefont{C.}~\bibnamefont{{Gogolin}}},
  \bibinfo{journal}{Nature Physics} \textbf{\bibinfo{volume}{11}},
  \bibinfo{pages}{124} (\bibinfo{year}{2015}), \eprint{1408.5148}.

\bibitem[{\citenamefont{{Rigol} et~al.}(2008)\citenamefont{{Rigol}, {Dunjko},
  and {Olshanii}}}]{Rigol2008}
\bibinfo{author}{\bibfnamefont{M.}~\bibnamefont{{Rigol}}},
  \bibinfo{author}{\bibfnamefont{V.}~\bibnamefont{{Dunjko}}}, \bibnamefont{and}
  \bibinfo{author}{\bibfnamefont{M.}~\bibnamefont{{Olshanii}}},
  \bibinfo{journal}{Nature} \textbf{\bibinfo{volume}{452}},
  \bibinfo{pages}{854} (\bibinfo{year}{2008}), \eprint{0708.1324}.

\bibitem[{\citenamefont{Deutsch}(1991)}]{Deutsch1991}
\bibinfo{author}{\bibfnamefont{J.~M.} \bibnamefont{Deutsch}},
  \bibinfo{journal}{Phys. Rev. A} \textbf{\bibinfo{volume}{43}},
  \bibinfo{pages}{2046} (\bibinfo{year}{1991}),
  \urlprefix\url{http://link.aps.org/doi/10.1103/PhysRevA.43.2046}.

\bibitem[{\citenamefont{{Srednicki}}(1994)}]{Srednicki1994}
\bibinfo{author}{\bibfnamefont{M.}~\bibnamefont{{Srednicki}}},
  \bibinfo{journal}{\pre} \textbf{\bibinfo{volume}{50}}, \bibinfo{pages}{888}
  (\bibinfo{year}{1994}), \eprint{cond-mat/9403051}.

\bibitem[{\citenamefont{{Tasaki}}(1998)}]{Tasaki1998}
\bibinfo{author}{\bibfnamefont{H.}~\bibnamefont{{Tasaki}}},
  \bibinfo{journal}{Physical Review Letters} \textbf{\bibinfo{volume}{80}},
  \bibinfo{pages}{1373} (\bibinfo{year}{1998}), \eprint{cond-mat/9707253}.

\bibitem[{\citenamefont{Rigol and Srednicki}(2012)}]{Rigol2012}
\bibinfo{author}{\bibfnamefont{M.}~\bibnamefont{Rigol}} \bibnamefont{and}
  \bibinfo{author}{\bibfnamefont{M.}~\bibnamefont{Srednicki}},
  \bibinfo{journal}{Phys. Rev. Lett.} \textbf{\bibinfo{volume}{108}},
  \bibinfo{pages}{110601} (\bibinfo{year}{2012}),
  \urlprefix\url{http://link.aps.org/doi/10.1103/PhysRevLett.108.110601}.

\bibitem[{\citenamefont{{Rigol}}(2009)}]{Rigol2009}
\bibinfo{author}{\bibfnamefont{M.}~\bibnamefont{{Rigol}}},
  \bibinfo{journal}{Physical Review Letters} \textbf{\bibinfo{volume}{103}},
  \bibinfo{eid}{100403} (\bibinfo{year}{2009}), \eprint{0904.3746}.

\bibitem[{\citenamefont{{Steinigeweg} et~al.}(2014)\citenamefont{{Steinigeweg},
  {Khodja}, {Niemeyer}, {Gogolin}, and {Gemmer}}}]{Steinigeweg2014}
\bibinfo{author}{\bibfnamefont{R.}~\bibnamefont{{Steinigeweg}}},
  \bibinfo{author}{\bibfnamefont{A.}~\bibnamefont{{Khodja}}},
  \bibinfo{author}{\bibfnamefont{H.}~\bibnamefont{{Niemeyer}}},
  \bibinfo{author}{\bibfnamefont{C.}~\bibnamefont{{Gogolin}}},
  \bibnamefont{and} \bibinfo{author}{\bibfnamefont{J.}~\bibnamefont{{Gemmer}}},
  \bibinfo{journal}{Physical Review Letters} \textbf{\bibinfo{volume}{112}},
  \bibinfo{eid}{130403} (\bibinfo{year}{2014}), \eprint{1311.0169}.

\bibitem[{\citenamefont{{Beugeling} et~al.}(2014)\citenamefont{{Beugeling},
  {Moessner}, and {Haque}}}]{Beugeling2014}
\bibinfo{author}{\bibfnamefont{W.}~\bibnamefont{{Beugeling}}},
  \bibinfo{author}{\bibfnamefont{R.}~\bibnamefont{{Moessner}}},
  \bibnamefont{and} \bibinfo{author}{\bibfnamefont{M.}~\bibnamefont{{Haque}}},
  \bibinfo{journal}{\pre} \textbf{\bibinfo{volume}{89}}, \bibinfo{eid}{042112}
  (\bibinfo{year}{2014}), \eprint{1308.2862}.

\bibitem[{\citenamefont{{Polkovnikov} et~al.}(2011)\citenamefont{{Polkovnikov},
  {Sengupta}, {Silva}, and {Vengalattore}}}]{Polkovnikov2011}
\bibinfo{author}{\bibfnamefont{A.}~\bibnamefont{{Polkovnikov}}},
  \bibinfo{author}{\bibfnamefont{K.}~\bibnamefont{{Sengupta}}},
  \bibinfo{author}{\bibfnamefont{A.}~\bibnamefont{{Silva}}}, \bibnamefont{and}
  \bibinfo{author}{\bibfnamefont{M.}~\bibnamefont{{Vengalattore}}},
  \bibinfo{journal}{Reviews of Modern Physics} \textbf{\bibinfo{volume}{83}},
  \bibinfo{pages}{863} (\bibinfo{year}{2011}), \eprint{1007.5331}.

\bibitem[{\citenamefont{{Riera} et~al.}(2012)\citenamefont{{Riera}, {Gogolin},
  and {Eisert}}}]{Riera2012}
\bibinfo{author}{\bibfnamefont{A.}~\bibnamefont{{Riera}}},
  \bibinfo{author}{\bibfnamefont{C.}~\bibnamefont{{Gogolin}}},
  \bibnamefont{and} \bibinfo{author}{\bibfnamefont{J.}~\bibnamefont{{Eisert}}},
  \bibinfo{journal}{Physical Review Letters} \textbf{\bibinfo{volume}{108}},
  \bibinfo{eid}{080402} (\bibinfo{year}{2012}), \eprint{1102.2389}.

\bibitem[{\citenamefont{{Berges} et~al.}(2004)\citenamefont{{Berges},
  {Bors{\'a}nyi}, and {Wetterich}}}]{Berges2004}
\bibinfo{author}{\bibfnamefont{J.}~\bibnamefont{{Berges}}},
  \bibinfo{author}{\bibfnamefont{S.}~\bibnamefont{{Bors{\'a}nyi}}},
  \bibnamefont{and}
  \bibinfo{author}{\bibfnamefont{C.}~\bibnamefont{{Wetterich}}},
  \bibinfo{journal}{Physical Review Letters} \textbf{\bibinfo{volume}{93}},
  \bibinfo{eid}{142002} (\bibinfo{year}{2004}), \eprint{hep-ph/0403234}.

\bibitem[{\citenamefont{{Ba{\~n}uls} et~al.}(2011)\citenamefont{{Ba{\~n}uls},
  {Cirac}, and {Hastings}}}]{Banuls2011}
\bibinfo{author}{\bibfnamefont{M.~C.} \bibnamefont{{Ba{\~n}uls}}},
  \bibinfo{author}{\bibfnamefont{J.~I.} \bibnamefont{{Cirac}}},
  \bibnamefont{and} \bibinfo{author}{\bibfnamefont{M.~B.}
  \bibnamefont{{Hastings}}}, \bibinfo{journal}{Physical Review Letters}
  \textbf{\bibinfo{volume}{106}}, \bibinfo{eid}{050405} (\bibinfo{year}{2011}),
  \eprint{1007.3957}.

\bibitem[{\citenamefont{{Mueller} et~al.}(2013)\citenamefont{{Mueller},
  {Adlam}, {Masanes}, and {Wiebe}}}]{Mueller2013}
\bibinfo{author}{\bibfnamefont{M.~P.} \bibnamefont{{Mueller}}},
  \bibinfo{author}{\bibfnamefont{E.}~\bibnamefont{{Adlam}}},
  \bibinfo{author}{\bibfnamefont{L.}~\bibnamefont{{Masanes}}},
  \bibnamefont{and} \bibinfo{author}{\bibfnamefont{N.}~\bibnamefont{{Wiebe}}},
  \bibinfo{journal}{ArXiv e-prints}  (\bibinfo{year}{2013}),
  \eprint{1312.7420}.

\bibitem[{\citenamefont{{Caux} and {Mossel}}(2011)}]{Caux2011}
\bibinfo{author}{\bibfnamefont{J.-S.} \bibnamefont{{Caux}}} \bibnamefont{and}
  \bibinfo{author}{\bibfnamefont{J.}~\bibnamefont{{Mossel}}},
  \bibinfo{journal}{Journal of Statistical Mechanics: Theory and Experiment}
  \textbf{\bibinfo{volume}{2}}, \bibinfo{pages}{02023} (\bibinfo{year}{2011}),
  \eprint{1012.3587}.

\bibitem[{\citenamefont{{Caux} and {Konik}}(2012)}]{Caux2012}
\bibinfo{author}{\bibfnamefont{J.-S.} \bibnamefont{{Caux}}} \bibnamefont{and}
  \bibinfo{author}{\bibfnamefont{R.~M.} \bibnamefont{{Konik}}},
  \bibinfo{journal}{Physical Review Letters} \textbf{\bibinfo{volume}{109}},
  \bibinfo{eid}{175301} (\bibinfo{year}{2012}), \eprint{1203.0901}.

\bibitem[{\citenamefont{{Caux} and {Essler}}(2013)}]{Caux2013}
\bibinfo{author}{\bibfnamefont{J.-S.} \bibnamefont{{Caux}}} \bibnamefont{and}
  \bibinfo{author}{\bibfnamefont{F.~H.~L.} \bibnamefont{{Essler}}},
  \bibinfo{journal}{Physical Review Letters} \textbf{\bibinfo{volume}{110}},
  \bibinfo{eid}{257203} (\bibinfo{year}{2013}), \eprint{1301.3806}.

\bibitem[{\citenamefont{{Gogolin} et~al.}(2011)\citenamefont{{Gogolin},
  {M{\"u}ller}, and {Eisert}}}]{Gogolin2011}
\bibinfo{author}{\bibfnamefont{C.}~\bibnamefont{{Gogolin}}},
  \bibinfo{author}{\bibfnamefont{M.~P.} \bibnamefont{{M{\"u}ller}}},
  \bibnamefont{and} \bibinfo{author}{\bibfnamefont{J.}~\bibnamefont{{Eisert}}},
  \bibinfo{journal}{Physical Review Letters} \textbf{\bibinfo{volume}{106}},
  \bibinfo{eid}{040401} (\bibinfo{year}{2011}), \eprint{1009.2493}.

\bibitem[{\citenamefont{{Cramer} et~al.}(2008)\citenamefont{{Cramer}, {Dawson},
  {Eisert}, and {Osborne}}}]{Cramer2008}
\bibinfo{author}{\bibfnamefont{M.}~\bibnamefont{{Cramer}}},
  \bibinfo{author}{\bibfnamefont{C.~M.} \bibnamefont{{Dawson}}},
  \bibinfo{author}{\bibfnamefont{J.}~\bibnamefont{{Eisert}}}, \bibnamefont{and}
  \bibinfo{author}{\bibfnamefont{T.~J.} \bibnamefont{{Osborne}}},
  \bibinfo{journal}{Physical Review Letters} \textbf{\bibinfo{volume}{100}},
  \bibinfo{eid}{030602} (\bibinfo{year}{2008}), \eprint{cond-mat/0703314}.

\bibitem[{\citenamefont{{Cassidy} et~al.}(2011)\citenamefont{{Cassidy},
  {Clark}, and {Rigol}}}]{Cassidy2011}
\bibinfo{author}{\bibfnamefont{A.~C.} \bibnamefont{{Cassidy}}},
  \bibinfo{author}{\bibfnamefont{C.~W.} \bibnamefont{{Clark}}},
  \bibnamefont{and} \bibinfo{author}{\bibfnamefont{M.}~\bibnamefont{{Rigol}}},
  \bibinfo{journal}{Physical Review Letters} \textbf{\bibinfo{volume}{106}},
  \bibinfo{eid}{140405} (\bibinfo{year}{2011}), \eprint{1008.4794}.

\bibitem[{\citenamefont{{Altland} and {Haake}}(2012)}]{Altland2012}
\bibinfo{author}{\bibfnamefont{A.}~\bibnamefont{{Altland}}} \bibnamefont{and}
  \bibinfo{author}{\bibfnamefont{F.}~\bibnamefont{{Haake}}},
  \bibinfo{journal}{Physical Review Letters} \textbf{\bibinfo{volume}{108}},
  \bibinfo{eid}{073601} (\bibinfo{year}{2012}), \eprint{1110.1270}.

\bibitem[{\citenamefont{{Vidmar} and {Rigol}}(2016)}]{Vidmar2016}
\bibinfo{author}{\bibfnamefont{L.}~\bibnamefont{{Vidmar}}} \bibnamefont{and}
  \bibinfo{author}{\bibfnamefont{M.}~\bibnamefont{{Rigol}}},
  \bibinfo{journal}{Journal of Statistical Mechanics: Theory and Experiment}
  \textbf{\bibinfo{volume}{6}}, \bibinfo{pages}{064007} (\bibinfo{year}{2016}),
  \eprint{1604.03990}.

\bibitem[{\citenamefont{{Marcuzzi} et~al.}(2013)\citenamefont{{Marcuzzi},
  {Marino}, {Gambassi}, and {Silva}}}]{Marcuzzi2013}
\bibinfo{author}{\bibfnamefont{M.}~\bibnamefont{{Marcuzzi}}},
  \bibinfo{author}{\bibfnamefont{J.}~\bibnamefont{{Marino}}},
  \bibinfo{author}{\bibfnamefont{A.}~\bibnamefont{{Gambassi}}},
  \bibnamefont{and} \bibinfo{author}{\bibfnamefont{A.}~\bibnamefont{{Silva}}},
  \bibinfo{journal}{Physical Review Letters} \textbf{\bibinfo{volume}{111}},
  \bibinfo{eid}{197203} (\bibinfo{year}{2013}), \eprint{1307.3738}.

\bibitem[{\citenamefont{{Essler} et~al.}(2014)\citenamefont{{Essler},
  {Kehrein}, {Manmana}, and {Robinson}}}]{Essler2014}
\bibinfo{author}{\bibfnamefont{F.~H.~L.} \bibnamefont{{Essler}}},
  \bibinfo{author}{\bibfnamefont{S.}~\bibnamefont{{Kehrein}}},
  \bibinfo{author}{\bibfnamefont{S.~R.} \bibnamefont{{Manmana}}},
  \bibnamefont{and} \bibinfo{author}{\bibfnamefont{N.~J.}
  \bibnamefont{{Robinson}}}, \bibinfo{journal}{\prb}
  \textbf{\bibinfo{volume}{89}}, \bibinfo{eid}{165104} (\bibinfo{year}{2014}),
  \eprint{1311.4557}.

\bibitem[{\citenamefont{{Geiger} et~al.}(2014)\citenamefont{{Geiger}, {Langen},
  {Mazets}, and {Schmiedmayer}}}]{Geiger2014}
\bibinfo{author}{\bibfnamefont{R.}~\bibnamefont{{Geiger}}},
  \bibinfo{author}{\bibfnamefont{T.}~\bibnamefont{{Langen}}},
  \bibinfo{author}{\bibfnamefont{I.~E.} \bibnamefont{{Mazets}}},
  \bibnamefont{and}
  \bibinfo{author}{\bibfnamefont{J.}~\bibnamefont{{Schmiedmayer}}},
  \bibinfo{journal}{New Journal of Physics} \textbf{\bibinfo{volume}{16}},
  \bibinfo{eid}{053034} (\bibinfo{year}{2014}), \eprint{1312.7568}.

\bibitem[{\citenamefont{{Abanin} et~al.}(2015)\citenamefont{{Abanin}, {De
  Roeck}, {Ho}, and {Huveneers}}}]{Abanin2015}
\bibinfo{author}{\bibfnamefont{D.~A.} \bibnamefont{{Abanin}}},
  \bibinfo{author}{\bibfnamefont{W.}~\bibnamefont{{De Roeck}}},
  \bibinfo{author}{\bibfnamefont{W.~W.} \bibnamefont{{Ho}}}, \bibnamefont{and}
  \bibinfo{author}{\bibfnamefont{F.}~\bibnamefont{{Huveneers}}},
  \bibinfo{journal}{ArXiv e-prints}  (\bibinfo{year}{2015}),
  \eprint{1510.03405}.

\bibitem[{\citenamefont{Lagendijk et~al.}(2009)\citenamefont{Lagendijk, van
  Tiggelen, and Wiersma}}]{Lagendijk2009}
\bibinfo{author}{\bibfnamefont{A.}~\bibnamefont{Lagendijk}},
  \bibinfo{author}{\bibfnamefont{B.}~\bibnamefont{van Tiggelen}},
  \bibnamefont{and} \bibinfo{author}{\bibfnamefont{D.}~\bibnamefont{Wiersma}},
  \bibinfo{journal}{Physics Today} \textbf{\bibinfo{volume}{62}},
  \bibinfo{pages}{24} (\bibinfo{year}{2009}).

\bibitem[{\citenamefont{{Nandkishore} and {Huse}}(2015)}]{Nandkishore2015}
\bibinfo{author}{\bibfnamefont{R.}~\bibnamefont{{Nandkishore}}}
  \bibnamefont{and} \bibinfo{author}{\bibfnamefont{D.~A.}
  \bibnamefont{{Huse}}}, \bibinfo{journal}{Annual Review of Condensed Matter
  Physics} \textbf{\bibinfo{volume}{6}}, \bibinfo{pages}{15}
  (\bibinfo{year}{2015}), \eprint{1404.0686}.

\bibitem[{\citenamefont{{Bauer} and {Nayak}}(2013)}]{Bauer2013}
\bibinfo{author}{\bibfnamefont{B.}~\bibnamefont{{Bauer}}} \bibnamefont{and}
  \bibinfo{author}{\bibfnamefont{C.}~\bibnamefont{{Nayak}}},
  \bibinfo{journal}{Journal of Statistical Mechanics: Theory and Experiment}
  \textbf{\bibinfo{volume}{9}}, \bibinfo{eid}{09005} (\bibinfo{year}{2013}),
  \eprint{1306.5753}.

\bibitem[{\citenamefont{{Serbyn} et~al.}(2015)\citenamefont{{Serbyn},
  {Papi{\'c}}, and {Abanin}}}]{Serbyn2015}
\bibinfo{author}{\bibfnamefont{M.}~\bibnamefont{{Serbyn}}},
  \bibinfo{author}{\bibfnamefont{Z.}~\bibnamefont{{Papi{\'c}}}},
  \bibnamefont{and} \bibinfo{author}{\bibfnamefont{D.~A.}
  \bibnamefont{{Abanin}}}, \bibinfo{journal}{Physical Review X}
  \textbf{\bibinfo{volume}{5}}, \bibinfo{eid}{041047} (\bibinfo{year}{2015}),
  \eprint{1507.01635}.

\bibitem[{\citenamefont{{Murg} et~al.}(2007)\citenamefont{{Murg}, {Verstraete},
  and {Cirac}}}]{Murg2007}
\bibinfo{author}{\bibfnamefont{V.}~\bibnamefont{{Murg}}},
  \bibinfo{author}{\bibfnamefont{F.}~\bibnamefont{{Verstraete}}},
  \bibnamefont{and} \bibinfo{author}{\bibfnamefont{J.~I.}
  \bibnamefont{{Cirac}}}, \bibinfo{journal}{\pra}
  \textbf{\bibinfo{volume}{75}}, \bibinfo{eid}{033605} (\bibinfo{year}{2007}),
  \eprint{cond-mat/0611522}.

\bibitem[{\citenamefont{{Corboz}
  et~al.}(2010{\natexlab{a}})\citenamefont{{Corboz}, {Or{\'u}s}, {Bauer}, and
  {Vidal}}}]{Corboz2009}
\bibinfo{author}{\bibfnamefont{P.}~\bibnamefont{{Corboz}}},
  \bibinfo{author}{\bibfnamefont{R.}~\bibnamefont{{Or{\'u}s}}},
  \bibinfo{author}{\bibfnamefont{B.}~\bibnamefont{{Bauer}}}, \bibnamefont{and}
  \bibinfo{author}{\bibfnamefont{G.}~\bibnamefont{{Vidal}}},
  \bibinfo{journal}{\prb} \textbf{\bibinfo{volume}{81}}, \bibinfo{eid}{165104}
  (\bibinfo{year}{2010}{\natexlab{a}}), \eprint{0912.0646}.

\bibitem[{\citenamefont{{Jordan} et~al.}(2008)\citenamefont{{Jordan},
  {Or{\'u}s}, {Vidal}, {Verstraete}, and {Cirac}}}]{Jordan2008}
\bibinfo{author}{\bibfnamefont{J.}~\bibnamefont{{Jordan}}},
  \bibinfo{author}{\bibfnamefont{R.}~\bibnamefont{{Or{\'u}s}}},
  \bibinfo{author}{\bibfnamefont{G.}~\bibnamefont{{Vidal}}},
  \bibinfo{author}{\bibfnamefont{F.}~\bibnamefont{{Verstraete}}},
  \bibnamefont{and} \bibinfo{author}{\bibfnamefont{J.~I.}
  \bibnamefont{{Cirac}}}, \bibinfo{journal}{Physical Review Letters}
  \textbf{\bibinfo{volume}{101}}, \bibinfo{eid}{250602} (\bibinfo{year}{2008}),
  \eprint{cond-mat/0703788}.

\bibitem[{\citenamefont{{Corboz}
  et~al.}(2010{\natexlab{b}})\citenamefont{{Corboz}, {Jordan}, and
  {Vidal}}}]{Corboz2010}
\bibinfo{author}{\bibfnamefont{P.}~\bibnamefont{{Corboz}}},
  \bibinfo{author}{\bibfnamefont{J.}~\bibnamefont{{Jordan}}}, \bibnamefont{and}
  \bibinfo{author}{\bibfnamefont{G.}~\bibnamefont{{Vidal}}},
  \bibinfo{journal}{\prb} \textbf{\bibinfo{volume}{82}}, \bibinfo{eid}{245119}
  (\bibinfo{year}{2010}{\natexlab{b}}), \eprint{1008.3937}.

\bibitem[{\citenamefont{{Kraus} et~al.}(2010)\citenamefont{{Kraus}, {Schuch},
  {Verstraete}, and {Cirac}}}]{Kraus2010}
\bibinfo{author}{\bibfnamefont{C.~V.} \bibnamefont{{Kraus}}},
  \bibinfo{author}{\bibfnamefont{N.}~\bibnamefont{{Schuch}}},
  \bibinfo{author}{\bibfnamefont{F.}~\bibnamefont{{Verstraete}}},
  \bibnamefont{and} \bibinfo{author}{\bibfnamefont{J.~I.}
  \bibnamefont{{Cirac}}}, \bibinfo{journal}{\pra}
  \textbf{\bibinfo{volume}{81}}, \bibinfo{eid}{052338} (\bibinfo{year}{2010}),
  \eprint{0904.4667}.

\bibitem[{\citenamefont{{Corboz} et~al.}(2014)\citenamefont{{Corboz}, {Rice},
  and {Troyer}}}]{Corboz2014}
\bibinfo{author}{\bibfnamefont{P.}~\bibnamefont{{Corboz}}},
  \bibinfo{author}{\bibfnamefont{T.~M.} \bibnamefont{{Rice}}},
  \bibnamefont{and} \bibinfo{author}{\bibfnamefont{M.}~\bibnamefont{{Troyer}}},
  \bibinfo{journal}{Physical Review Letters} \textbf{\bibinfo{volume}{113}},
  \bibinfo{eid}{046402} (\bibinfo{year}{2014}), \eprint{1402.2859}.

\bibitem[{\citenamefont{{Vanderstraeten}
  et~al.}(2015)\citenamefont{{Vanderstraeten}, {Mari{\"e}n}, {Verstraete}, and
  {Haegeman}}}]{Vanderstraeten2015b}
\bibinfo{author}{\bibfnamefont{L.}~\bibnamefont{{Vanderstraeten}}},
  \bibinfo{author}{\bibfnamefont{M.}~\bibnamefont{{Mari{\"e}n}}},
  \bibinfo{author}{\bibfnamefont{F.}~\bibnamefont{{Verstraete}}},
  \bibnamefont{and}
  \bibinfo{author}{\bibfnamefont{J.}~\bibnamefont{{Haegeman}}},
  \bibinfo{journal}{\prb} \textbf{\bibinfo{volume}{92}}, \bibinfo{eid}{201111}
  (\bibinfo{year}{2015}), \eprint{1507.02151}.

\bibitem[{\citenamefont{{Phien} et~al.}(2015)\citenamefont{{Phien}, {Bengua},
  {Tuan}, {Corboz}, and {Or{\'u}s}}}]{Phien2015}
\bibinfo{author}{\bibfnamefont{H.~N.} \bibnamefont{{Phien}}},
  \bibinfo{author}{\bibfnamefont{J.~A.} \bibnamefont{{Bengua}}},
  \bibinfo{author}{\bibfnamefont{H.~D.} \bibnamefont{{Tuan}}},
  \bibinfo{author}{\bibfnamefont{P.}~\bibnamefont{{Corboz}}}, \bibnamefont{and}
  \bibinfo{author}{\bibfnamefont{R.}~\bibnamefont{{Or{\'u}s}}},
  \bibinfo{journal}{\prb} \textbf{\bibinfo{volume}{92}}, \bibinfo{eid}{035142}
  (\bibinfo{year}{2015}), \eprint{1503.05345}.

\bibitem[{\citenamefont{{Corboz}}(2016)}]{Corboz2016}
\bibinfo{author}{\bibfnamefont{P.}~\bibnamefont{{Corboz}}},
  \bibinfo{journal}{\prb} \textbf{\bibinfo{volume}{94}}, \bibinfo{eid}{035133}
  (\bibinfo{year}{2016}), \eprint{1605.03006}.

\bibitem[{\citenamefont{{Vanderstraeten}
  et~al.}(2016)\citenamefont{{Vanderstraeten}, {Haegeman}, {Corboz}, and
  {Verstraete}}}]{Vanderstraeten2016b}
\bibinfo{author}{\bibfnamefont{L.}~\bibnamefont{{Vanderstraeten}}},
  \bibinfo{author}{\bibfnamefont{J.}~\bibnamefont{{Haegeman}}},
  \bibinfo{author}{\bibfnamefont{P.}~\bibnamefont{{Corboz}}}, \bibnamefont{and}
  \bibinfo{author}{\bibfnamefont{F.}~\bibnamefont{{Verstraete}}},
  \bibinfo{journal}{\prb} \textbf{\bibinfo{volume}{94}}, \bibinfo{eid}{155123}
  (\bibinfo{year}{2016}), \eprint{1606.09170}.

\bibitem[{\citenamefont{{Tagliacozzo} et~al.}(2014)\citenamefont{{Tagliacozzo},
  {Celi}, and {Lewenstein}}}]{Tagliacozzo2014}
\bibinfo{author}{\bibfnamefont{L.}~\bibnamefont{{Tagliacozzo}}},
  \bibinfo{author}{\bibfnamefont{A.}~\bibnamefont{{Celi}}}, \bibnamefont{and}
  \bibinfo{author}{\bibfnamefont{M.}~\bibnamefont{{Lewenstein}}},
  \bibinfo{journal}{Physical Review X} \textbf{\bibinfo{volume}{4}},
  \bibinfo{eid}{041024} (\bibinfo{year}{2014}), \eprint{1405.4811}.

\bibitem[{\citenamefont{{Haegeman} et~al.}(2015)\citenamefont{{Haegeman}, {Van
  Acoleyen}, {Schuch}, {Cirac}, and {Verstraete}}}]{Haegeman2015}
\bibinfo{author}{\bibfnamefont{J.}~\bibnamefont{{Haegeman}}},
  \bibinfo{author}{\bibfnamefont{K.}~\bibnamefont{{Van Acoleyen}}},
  \bibinfo{author}{\bibfnamefont{N.}~\bibnamefont{{Schuch}}},
  \bibinfo{author}{\bibfnamefont{J.~I.} \bibnamefont{{Cirac}}},
  \bibnamefont{and}
  \bibinfo{author}{\bibfnamefont{F.}~\bibnamefont{{Verstraete}}},
  \bibinfo{journal}{Physical Review X} \textbf{\bibinfo{volume}{5}},
  \bibinfo{eid}{011024} (\bibinfo{year}{2015}), \eprint{1407.1025}.

\bibitem[{\citenamefont{{Zohar} et~al.}(2015)\citenamefont{{Zohar}, {Burrello},
  {Wahl}, and {Cirac}}}]{Zohar2015}
\bibinfo{author}{\bibfnamefont{E.}~\bibnamefont{{Zohar}}},
  \bibinfo{author}{\bibfnamefont{M.}~\bibnamefont{{Burrello}}},
  \bibinfo{author}{\bibfnamefont{T.~B.} \bibnamefont{{Wahl}}},
  \bibnamefont{and} \bibinfo{author}{\bibfnamefont{J.~I.}
  \bibnamefont{{Cirac}}}, \bibinfo{journal}{Annals of Physics}
  \textbf{\bibinfo{volume}{363}}, \bibinfo{pages}{385} (\bibinfo{year}{2015}),
  \eprint{1507.08837}.

\bibitem[{\citenamefont{{Milsted} and {Osborne}}(2016)}]{Milsted2016}
\bibinfo{author}{\bibfnamefont{A.}~\bibnamefont{{Milsted}}} \bibnamefont{and}
  \bibinfo{author}{\bibfnamefont{T.~J.} \bibnamefont{{Osborne}}},
  \bibinfo{journal}{ArXiv e-prints}  (\bibinfo{year}{2016}),
  \eprint{1604.01979}.

\bibitem[{\citenamefont{{Zohar} et~al.}(2016)\citenamefont{{Zohar}, {Wahl},
  {Burrello}, and {Cirac}}}]{Zohar2016}
\bibinfo{author}{\bibfnamefont{E.}~\bibnamefont{{Zohar}}},
  \bibinfo{author}{\bibfnamefont{T.~B.} \bibnamefont{{Wahl}}},
  \bibinfo{author}{\bibfnamefont{M.}~\bibnamefont{{Burrello}}},
  \bibnamefont{and} \bibinfo{author}{\bibfnamefont{J.~I.}
  \bibnamefont{{Cirac}}}, \bibinfo{journal}{Annals of Physics}
  \textbf{\bibinfo{volume}{374}}, \bibinfo{pages}{84} (\bibinfo{year}{2016}),
  \eprint{1607.08115}.

\bibitem[{\citenamefont{{Haegeman} et~al.}(2012)\citenamefont{{Haegeman},
  {Pirvu}, {Weir}, {Cirac}, {Osborne}, {Verschelde}, and
  {Verstraete}}}]{Haegeman2012}
\bibinfo{author}{\bibfnamefont{J.}~\bibnamefont{{Haegeman}}},
  \bibinfo{author}{\bibfnamefont{B.}~\bibnamefont{{Pirvu}}},
  \bibinfo{author}{\bibfnamefont{D.~J.} \bibnamefont{{Weir}}},
  \bibinfo{author}{\bibfnamefont{J.~I.} \bibnamefont{{Cirac}}},
  \bibinfo{author}{\bibfnamefont{T.~J.} \bibnamefont{{Osborne}}},
  \bibinfo{author}{\bibfnamefont{H.}~\bibnamefont{{Verschelde}}},
  \bibnamefont{and}
  \bibinfo{author}{\bibfnamefont{F.}~\bibnamefont{{Verstraete}}},
  \bibinfo{journal}{\prb} \textbf{\bibinfo{volume}{85}}, \bibinfo{eid}{100408}
  (\bibinfo{year}{2012}), \eprint{1103.2286}.

\bibitem[{\citenamefont{Hatano and Suzuki}(2005)}]{Hatano2005}
\bibinfo{author}{\bibfnamefont{N.}~\bibnamefont{Hatano}} \bibnamefont{and}
  \bibinfo{author}{\bibfnamefont{M.}~\bibnamefont{Suzuki}},
  \emph{\bibinfo{title}{Quantum Annealing and Other Optimization Methods}}
  (\bibinfo{publisher}{Springer Berlin Heidelberg}, \bibinfo{address}{Berlin,
  Heidelberg}, \bibinfo{year}{2005}), chap. \bibinfo{chapter}{Finding
  Exponential Product Formulas of Higher Orders}, pp. \bibinfo{pages}{37--68},
  ISBN \bibinfo{isbn}{978-3-540-31515-5}.

\end{thebibliography}

\end{document}